\let\mypdfximage\pdfximage
\def\pdfximage{\immediate\mypdfximage}

\documentclass[journal,onecolumn,10pt]{IEEEtran1}

\makeatletter
\def\markboth#1#2{\def\leftmark{\@IEEEcompsoconly{\sffamily}\MakeUppercase{\protect#1}}%
\def\rightmark{\@IEEEcompsoconly{\sffamily}\MakeUppercase{\protect#2}}}
\makeatother

\let\labelindent\relax

\usepackage{times}
\usepackage{marvosym}
\usepackage{longtable}
\usepackage{graphics}
\usepackage{graphicx}
\usepackage{caption}
\usepackage[english]{babel}
\usepackage{subfig}
\usepackage{epsfig}
\usepackage{color}
\usepackage{multirow}
\usepackage{mathrsfs}
\usepackage{textcomp}
\usepackage{amsfonts}
\usepackage{amsmath}
\usepackage{amssymb}
\usepackage{nccmath}
\usepackage[numbers,square,sort&compress,comma]{natbib}
\usepackage{chapterbib}
\usepackage[ruled,vlined,linesnumbered,algosection]{algorithm2e}
\usepackage{setspace}
\usepackage{verbatim}
\usepackage{wrapfig}
\usepackage{dblfloatfix}
\usepackage{comment}
\usepackage[strict]{changepage}
\usepackage{ragged2e}
\usepackage{microtype}
\usepackage{enumitem}
\usepackage{nccmath}
\usepackage{hyperref}
\usepackage{doi}
\usepackage{afterpage}
\usepackage{multirow,bigdelim}
\usepackage{booktabs}
\usepackage{color}
\usepackage[outline]{contour}
\usepackage{xcolor}
\usepackage{colortbl}%
  
\usepackage{adjustbox}
\usepackage[nameinlink]{cleveref}
\usepackage{pdfpages}

\def\imagetop#1{\vtop{\null\hbox{#1}}}

\setlist{parsep=0pt,listparindent=\parindent}

\DeclareCaptionFont{singlespacing}{\setstretch{1}}
\numberwithin{figure}{section}
\renewcommand \thesection{\arabic{section}}
\numberwithin{equation}{section}

\hypersetup{bookmarks=false,bookmarksopen=false,pdfpagemode=empty,pdfstartview=}

\let\oldnl\nl
\newcommand{\nonl}{\renewcommand{\nl}{\let\nl\oldnl}}

\title{\singlespacing\sf\huge Estimating R\'{e}nyi's $\alpha$-Cross-Entropies in a Matrix-Based Way}
\author{Isaac J. Sledge, \emph{Member, IEEE} and Jos\'{e} C. Pr\'{i}ncipe, \emph{Life Fellow, IEEE}%
\thanks{\fontdimen2\font=1.55pt Isaac J. Sledge is the Senior Machine Learning Scientist and Dr. Delores M. Etter Assistant Secretary of the Navy Top Engineer with the Advanced Signal Processing and Automated Target Recognition Branch, Naval Surface Warfare Center, Panama City, FL, USA (email: isaac.j.sledge@navy.mil).  He the Chief Machine Learning Scientist with the Machine Intelligence Defense (MIND) lab at the Naval Sea Systems Command.}
\thanks{\fontdimen2\font=1.55pt Jos\'{e} C. Pr\'{i}ncipe is a Distinguished Professor and the Don D. and Ruth S. Eckis Chair with both the Department of Electrical and Computer Engineering and the Department of Biomedical Engineering, University of Florida, Gainesville, FL, USA (email: principe@ufl.edu).  He is the director of the Computational NeuroEngineering Laboratory (CNEL) at the University of Florida.\vspace{0.1cm}}
\thanks{The work of the authors was funded by grants N00014-19-WX-00636 (Marc Steinberg), N00014-21-WX-00476 (J. Tory Cobb), N00014-21-WX-00525 (Thomas McKenna), and N00014-21-WX-01348 (Marc Steinberg) from the US Office of Naval Research.  The first author was also supported by in-house laboratory independent research (ILIR) grant N00014-19-WX-00687 (Frank Crosby) from the US Office of Naval Research and a Naval Innovation in Science and Engineering (NISE) grant from NAVSEA.}%
}
\begin{document}
\bstctlcite{IEEEexample:BSTcontrol}

\maketitle
\RaggedRight\parindent=1.5em
\fontdimen2\font=2.1pt
\vspace{-1.55cm}\begin{abstract}\normalsize\singlespacing
\vspace{-0.25cm}{\small{\sf{\textbf{Abstract}}}}---Conventional information-theoretic quantities assume access to probability distributions.  Estimating such distributions is not trivial.  Here, we consider function-based formulations of cross entropy that sidesteps this a priori estimation requirement.  We propose three measures of R\'{e}nyi's $\alpha$-cross-entropies in the setting of reproducing-kernel Hilbert spaces.  Each measure has its appeals.  We prove that we can estimate these measures in an unbiased, non-parametric, and minimax-optimal way.  We do this via sample-constructed Gram matrices.  This yields matrix-based estimators of R\'{e}nyi's $\alpha$-cross-entropies.  These estimators satisfy all of the axioms that R\'{e}nyi established for divergences.  Our cross-entropies can thus be used for assessing distributional differences.  They are also appropriate for handling high-dimensional distributions, since the convergence rate of our estimator is independent of the sample dimensionality.

Python code for implementing these measures can be found at \href{https://github.com/isledge/MBRCE}{https://github.com/isledge/MBRCE}.

\end{abstract}%
\begin{IEEEkeywords}\normalsize\singlespacing
\vspace{-1.35cm}{{\small{\sf{\textbf{Index Terms}}}}---R\'{e}nyi's cross entropy, estimator, reproducing-kernel Hilbert space (RKHS), information-theoretic learning (ITL)}
\end{IEEEkeywords}
\IEEEpeerreviewmaketitle
\allowdisplaybreaks
\singlespacing

\vspace{-0.4cm}\subsection*{\small{\sf{\textbf{1.$\;\;\;$Introduction}}}}\addtocounter{section}{1}

Information-theoretic quantities, such as cross entropy, are defined over probability distributions \cite{StratonovichRL-book1975a,CoverTM-book2006a}.  These distributions are often not known in advance.  When the only information available is from a finite set of samples, then an appropriate distribution-estimation process must usually be employed \cite{ZivJ-jour1993a,DarbellayGA-jour1999a,WangQ-jour2005a,CaiH-jour2006a}.

Accurately estimating probability distributions is known to be highly challenging.  Improper estimation can greatly impede the use of the ensuing information-theoretic quantities, especially for learning \cite{PrincipeJC-book2010a}.

Our lab has shown that entropy-like and mutual-information-like quantities can be estimated in a principled way without knowledge of the underlying distributions \cite{GiraldoLGS-conf2013a,GiraldoLGS-jour2014a,YuS-jour2020a}.  This is done using the theory of functional operators in reproducing-kernel Hilbert spaces.

Here, we consider an operator-based framework for estimating cross-entropy-like quantities.  It characterizes distributional overlap.  Our framework avoids the intermediate step of explicit estimating distributions \cite{PalD-coll2010a,HeroAO-jour1999a,WangQ-jour2009a}.  Implicit distributional estimates are formed due to the nature of our framework, though.

More specifically, we assume access to a sets of available samples drawn from some unknown, arbitrary distributions.  We also suppose that these underlying distributions can be losslessly embedded in separable reproducing-kernel Hilbert spaces \cite{FukumizuK-coll2011a} by mean-element mappings \cite{SriperumbudurBK-conf2008a} (see \hyperref[sec2.1]{Section 2.1}).  We use either universal kernels \cite{ChristmannA-coll2011a} or characteristic kernels \cite{FukumizuK-coll2009a,NishiyamaY-jour2016a} to specify the reproducing-kernel Hilbert spaces.  The resulting mappings are provably guaranteed to be injective \cite{SriperumbudurBK-conf2010a,SriperumbudurBK-jour2011a}.  That is, different distributions can be distinguished within this Hilbert space (see \hyperref[sec2.1]{Section 2.1}).  This beneficial property permits defining measures of distance \cite{ChwialkowskiKP-coll2014a}, like cross entropies, using Hilbert-space operators.  Semi-metric \cite{GrettonA-coll2007a,SriperumbudurBK-conf2008a} and metric \cite{HeinM-conf2005a,FukumizuK-coll2008a,QuangMH-coll2014a,ChwialkowskiK-conf2014a,SejdinovicD-jour2013a} distances can also be considered between operators.

Direct access to the distributions is not required to assess distributional differences via operator-based cross-entropy measures.  By solving a corresponding pre-image problem \cite{KwokJTY-jour2004a}, the distributions can be recovered, at least in a point-estimate sense \cite{SongL-conf2008a,ChenY-conf2010a,HuszarF-conf2012a}.  Associated moments and interval measures can also be retrieved \cite{KanagawaM-conf2014a,KanagawaM-coll2016a}.  It is not our aim to do this here, though.  We simply specify a functional-operator-based cross-entropy measure.

We illustrate that the Hilbert-space operators can be empirically approximated using representations that rely on finite sums of Dirac measures centered at the samples \cite{SriperumbudurBK-jour2012a}.  These operator estimates are unbiased.  They also converge to the true operators at the minimax-optimal rate as a function of the sample size.  This convergence rate is independent of the sample dimensionality (see \hyperref[sec3.2]{Sections 3.2} and \hyperref[sec4.2]{4.2}).  Numerically computing these operators exactly is difficult, however.  They rely on access to an orthonormal basis for functional spaces.

The operators that we define have multi-linear forms that coincide with correlations of functions in a reproducing-kernel Hilbert space (See \hyperref[sec2.2]{Section 2.2}).  We refer to these covariance operators as Gram operators.  We show that the Gram operators can be replaced with easily-computable Gram matrices (see \hyperref[sec3.2]{Sections 3.2} and \hyperref[sec4.2]{4.2}).  The Gram matrices are obtainable from the application of either bounded, translation-invariant kernels \cite{SriperumbudurBK-jour2010a} or non-translation-invariant, strictly-positive-definite kernels \cite{FukumizuK-coll2010a} to a set of samples.  The Gram matrices posses spectral properties that are provably consistent with those of the empirical Gram operators.  Due to the form of our operator-based divergence functionals, we still obtain minimax optimality when using Gram matrices (see \hyperref[sec3.2]{Sections 3.2} and \hyperref[sec4.2]{4.2}).  This optimality allows us to create matrix-based notions of information-theoretic measures that act as though they had direct access to the true, but unknown, probability distributions from which the samples are drawn.  This is true even for finite-sample cases.  We thus do not need access to the functional-space basis.

We apply this formulation to generalize R\'{e}nyi's $\alpha$-cross-entropies \cite{RenyiA-coll1976a}.  R\'{e}nyi's cross-entropies are very general and contain many popular measures as limit cases \cite{vanErvenT-jour2014a}.  We develop three operator-based cross-entropy measures (see \hyperref[sec3.1]{Sections 3.1} and \hyperref[sec4.1]{4.1}).  They are parent quantities to notions of marginal, joint, and conditional R\'{e}nyi's $\alpha$-entropies \cite{TeixeiraA-jour2012a} that we previously defined in an operator-theoretic sense \cite{GiraldoLGS-jour2014a,YuS-jour2020a} (see \hyperref[secA.1]{Appendix A.1}).  The use of kernels that are additionally infinitely-divisible \cite{HornRA-jour1969a} is needed to recover these $\alpha$-entropies.  This is not a restrictive condition.  It requires only a simple modification of the Gram-matrix entries.

We prove that the corresponding matrix-based R\'{e}nyi's $\alpha$-cross-entropies satisfy the fundamental axioms that R\'{e}nyi originally considered for divergences (see \hyperref[secA.1]{Appendix A.1}).  They also have other beneficial properties (see \hyperref[secA.2]{Appendix A.2}).  These properties are present regardless of the sample-set size and dimensionality.  Our matrix-based functionals thus can behave as divergences even for small sample sets drawn from high-dimensional distributions.  However, a sufficient number of samples is still needed to ensure that the matrix-based-cross-entropy magnitudes are close enough to those from the operator-based cross-entropies.  

Throughout our presentation of these measures, we assume that our sample sets are vectorial in nature before being converted to Gram matrices.   This is done for the sake of presentation convenience.  Certain versions of our $\alpha$-cross-entropies can act as divergences for kernel-transformed representations that are inherently matrix-based, though.  Examples include adjacency matrices of undirected graphs along with square similarity and dissimilarity matrices.  Such data can be easily converted to Gram matrices via the application of kernels to the matrix entries.  This assumes axisymmetry of the underlying matrices, though.  Likewise, it assumes that the matrices are square.  

After defining each measure, we assess their empirical properties (see \hyperref[sec3.3]{Sections 3.3} and \hyperref[sec4.3]{4.3}).  We show that these measures empirically converge at a rate that is independent of the sample dimensionality, just as our theory predicts.  We also illustrate the importance of the kernel choice on the resulting cross entropies.

Our work here represents the first instance of operator-theoretic cross-entropy.  We show that there are some connections with our measures and those from quantum information theory (see \hyperref[sec5]{Section 5}).  The interpretation and utility of our measures are very different, though.  These distinctions permit investigators to use our estimators to assess population cross-entropies, optimize information-theoretic cost functions, and more.  The quantum variants cannot be employed in the same manner.  They are suited only for assessing information-theoretic quantities of random matrices.  Random matrices have characteristics that often differ greatly from those of the Gram matrices.

\subsection*{\small{\sf{\textbf{2.$\;\;\;$Preliminaries}}}}\addtocounter{section}{1}

The matrix-based cross-entropies that we define assess differences between probability distributions without needing direct access to them.  In this section, we outline the theory that makes this possible, which are operators posed in reproducing-kernel Hilbert spaces.  Such a functional space is appropriate for processing probability distributions.  This is because probability distributions are functions.

We start by recalling the notion of the mean element for univariate random variables (see \hyperref[def2.1]{Definition 2.1}).  Mean-element mappings are a principled way of characterizing probability measures.  These maps losslessly convert probability measures into points within a reproducing-kernel Hilbert space of functions (see \hyperref[prop2.1]{Proposition 2.1}).  Provided that the mapping is injective, then the Hilbert space separates points.  Unique distributions are assigned to unique points (see \hyperref[prop2.3]{Proposition 2.3}).  Distance measures can hence be defined within this Hilbert space to assess how much one probability measure is related to another (see \hyperref[prop2.2]{Proposition 2.2}).  Direct access to the underlying distributions is not required to do this for injective mappings (see \hyperref[prop2.4]{Proposition 2.4}).  This makes the mean-element map amenable for many applications where the true form of the distributions is either unknown or difficult to estimate in a non-parametric way.  Efficiently assessing cross entropy is one such application.

Following the univariate case, we consider the bivarate case.  We define the bivariate mean-element map (see \hyperref[def2.2]{Definition 2.2}) on tensor-product Hilbert spaces (see \hyperref[prop2.5]{Proposition 2.5}).  These spaces permit studying the interaction of samples from different distributions in a reproducing-kernel Hilbert space (see \hyperref[prop2.6]{Proposition 2.6}).  For the univariate case, injectivity of the mean-element map is guaranteed for characteristic kernels.  In the bivariate case, the matter is more complicated.  Products of characteristic kernels are not necessarily characteristic (see \hyperref[def2.3]{Definitions 2.3}--\hyperref[def2.4]{2.4} and \hyperref[prop2.7]{Proposition 2.7}).  Injectivity can be lost.  We thus require strict positive definiteness on the product kernel, which occurs whenever the kernel pairs are universal (see \hyperref[prop2.8]{Proposition 2.8}).  Many popular translation-varying and translation-invariant kernels are universal.  Consequently, kernel universality is not a very restrictive assumption in practice.

We illustrate that the mean-element maps are used in the construction of covariance operators on Hilbert spaces and tensor-product Hilbert spaces \cite{BakerC-jour1973a} (see \hyperref[def2.5]{Definitions 2.5}--\hyperref[def2.6]{2.6}).  We refer to these covariance operators as Gram operators.  Bilinear forms associated with the Gram operators correlate functions, and thus probability distributions, in reproducing-kernel Hilbert spaces.  In the next section, we show that the operators have empirical versions that can be constructed via distribution samples.  The empirical versions, which are Gram matrices, quantify relationships between sample pairs and hence the intra-distributional organization of samples.  That is, they provide insights into the distribution shape, which is needed for assessing cross entropy.  Covariance operators can also be defined to quantify inter-distributional shifts and therefore how much the distributions overlap (see \hyperref[def2.7]{Definition 2.7}).  This too helps with assessing cross entropy in certain cases.

An important result established in this section is that any process applied to the trace of the Gram operators will extend to the mean-element maps (see \hyperref[prop2.11]{Propositions 2.11} and \hyperref[prop2.14]{2.14}).  That is, the embeddings of probability distributions will be transformed.  

This is a powerful property that we exploit in our cross-entropy-like criteria.  We define these criteria in terms of Gram-operator trace.  These criteria provably satisfies foundational divergence axioms.  They hence quantify distributional differences.  Due to the mean-element-map relationship, access to the underlying probability distributions is not required when acting on the Gram-operator traces and their empirical estimates.  This property permits investigators to use our information-theoretic criteria as efficient drop-in estimators for a variety of problems.  No parametric assumptions are made about the distributions for appropriate kernel choices.  Arbitrary distributions can hence be characterized well and their differences assessed by our cross-entropy-like criteria.

Our formulation of the covariance and cross-covariance operators, and hence their traces, relies on access to a basis for the reproducing-kernel Hilbert space (See \hyperref[def2.5]{Definitions 2.5}--\hyperref[def2.6]{2.6} and \hyperref[prop2.13]{Proposition 2.13}).  Any orthonormal set is provably a basis (see \hyperref[prop2.10]{Proposition 2.10}).  The trace is independent of this basis choice.  However, it can be difficult to practically specify any such set depending on the selected kernel.  The basis may have countably-infinite components, in some cases (see \hyperref[prop2.9]{Propositions 2.9} and \hyperref[prop2.12]{2.12}).  In the next section, we sidestep this issue.  We rely on the connections between the covariance operators and the mean-element maps to show that, like the latter \cite{TolstikhinI-jour2017a}, the former can be estimated in an unbiased, minimax-optimal, non-parametric manner from the distributional samples.  We also have behaved spectral-difference bounds for eigenvalues of the Gram operators and empirical Gram matrices.  Given enough samples, the Gram matrices resemble the Gram operators.  Our formulation is thus well motivated.

Throughout, we assume access to a vector-based realization of matrices.  This enables us to present the background theory in an easy-to-understand manner.  This assumption is not practically required, though.  It is merely for the sake of convenience.  Only in a few instances are vector samples explicitly required.  We discuss these cases in the next section (see Section 3).

\phantomsection\label{sec2.1}
\subsection*{\small{\sf{\textbf{2.1.$\;\;\;$Distribution-Embedding Operators}}}}

{\small{\sf{\textbf{Univariate Operator Embeddings.}}}} We first recall some important properties of working with marginal probability measures in functional spaces that will be helpful throughout.

\phantomsection\label{def2.1}
\begin{itemize}
\item[] \-\hspace{0.5cm}{\small{\sf{\textbf{Definition 2.1: Univariate Mean-Element Map.}}}} Let $\mathcal{H}_{\kappa'}$ be a separable, reproducing-kernel Hilbert space \cite{AronszajnN-jour1950a} with a continuous reproducing kernel defined on the sample space $\kappa' \!:\! \mathcal{S} \!\times\! \mathcal{S} \!\to\! \mathbb{R}_{0,+}$.  Let $\varphi : \mathcal{S} \!\to\! \mathcal{H}_{\kappa'}$\\ \noindent be a measurable function which obeys $\kappa'(s,s') \!=\! \langle \varphi(s),\varphi(s') \rangle_{\mathcal{H}_{\kappa'}}$.  For a non-negative, non-zero-everywhere,\\ \noindent normalized kernel that is integrable, the mean element, $\mu_{p_{\mathcal{S}}}$, is defined via the Bochner integral
\begin{equation*}
\mathbb{E}_{(s,s')\sim p_{\mathcal{S}}}[\kappa'(s,s')] =\! \int\!\!\!\!\int_{\mathcal{S}} \langle \varphi(s),\varphi(s') \rangle_{\mathcal{H}_{\kappa'}} dp_{\mathcal{S}}(s,s')
\end{equation*}
where $p_{\mathcal{S}} \!\in\! \mathcal{M}^1_+(\mathcal{S})$ is a Borel probability measure over the topological sample space $\mathcal{S}$.
\end{itemize}
The mean element $\mu_{p_{\mathcal{S}}} \!=\! \mathbb{E}_{(s,s')\sim p_{\mathcal{S}}}[\kappa'(s,s')]$ can be viewed as an embedding of the measure in $\mathcal{H}_{\kappa'}$. That is, for\\ \noindent the set of all Borel probability measures, $\mu_{\kappa'} \!:\! \mathcal{M}^1_+(\mathcal{S}) \!\to\! \mathcal{H}_{\kappa'}$, with $\mu_{\kappa'}(p_{\mathcal{S}}) \!=\! \mu_{p_{\mathcal{S}}}$.  Unique distributions can\\ \noindent be mapped to unique points in the reproducing-kernel Hilbert space, for appropriate kernel choices.  This permits comparing distributions in this functional space.

Next, we outline the condition under which the embedding $\mu_{\kappa'}(p_{\mathcal{S}})$ exists and belongs to $\mathcal{H}_{\kappa'}$.

\phantomsection\label{prop2.1}
\begin{itemize}
\item[] \-\hspace{0.5cm}{\small{\sf{\textbf{Proposition 2.1: Univariate Riesz Representation Property.}}}} For a separable reproducing-kernel Hilbert space, with a continuous reproducing kernel defined on the samples, let $\kappa'(s,s') \!=\! \langle \varphi(s),\varphi(s') \rangle$, where $\varphi$\\ \noindent is measurable.  For a function $f \!\in\! \mathcal{H}_{\kappa'}$, if $\mathbb{E}_{(s,s')\sim p_{\mathcal{S}}}[\kappa'(s,s')^{1/2}] \!\in\! \mathbb{R}_{0,+}$, then we have that $\mu_{p_{\mathcal{S}}} \!\in\! \mathcal{H}_{\kappa'}$ and\\ \noindent $\mathbb{E}_{p_{\mathcal{S}}}[f(s)] \!=\! \langle f,\mu_{p_{\mathcal{S}}}\rangle_{\mathcal{H}_{\kappa'}}$.
\label{prop2.1}
\end{itemize}

From the equality $\mathbb{E}_{p_{\mathcal{S}}}[f(s)] \!=\! \langle f,\mu_{p_{\mathcal{S}}}\rangle_{\mathcal{H}_{\kappa'}}$, we simultaneously obtain the reproducing property of the expect-\\ \noindent ation operation in a reproducing-kernel Hilbert space.  That is, the functional expectation can be performed, with respect to the distribution, via an inner product of the function $f \!\in\! \mathcal{H}_{\kappa'}$ and the embedding $\mu_{p_{\mathcal{S}}} \!\in\! \mathcal{H}_{\kappa'}$.

There are two additional properties worth mentioning.

\phantomsection\label{prop2.2}
\begin{itemize}
\item[] \-\hspace{0.5cm}{\small{\sf{\textbf{Proposition 2.2: Distribution Distinguishability Property.}}}} Let $\mathcal{H}_{\kappa'}$ be a separable reproducing kernel Hilbert space with a continuous reproducing kernel defined on state-action pairs such that $\kappa'(s,s') \!=\! \langle \varphi(s),\varphi(s') \rangle$, where $\varphi$ is measurable.  Assume that this Hilbert space is dense on the space of all continuous, bounded functions.  We have that
\begin{itemize}
   \item[] \-\hspace{0.5cm}(i) The mean mapping $p_{\mathcal{S}} \!\mapsto\! \mu_{p_{\mathcal{S}}}$ is injective for all $p_{\mathcal{S}} \!\in\! \mathcal{M}^1_+(\mathcal{S})$, where $\mathcal{M}^1_+(\mathcal{S})$ is the set of all Borel\\ \noindent probability measures on the topological sample space.
   \item[] \-\hspace{0.5cm}(ii) $\langle\;\! \mu_{p_{\mathcal{S}}} \!-\! n^{-1}\sum_{i=1}^n \kappa'(s_i,s_i'),\cdot \rangle \!\leq\! 2n^{-1}\mathbb{E}_{s_i}\mathbb{E}_{\sigma_i}[\,\textnormal{sup}_{\|f\|_{\mathcal{H}_{\kappa'}} \leq 1}\,|\;\!\!\sum_{i=1}^m \sigma_if(s_i)|] \!+\! c_1$ is satisfied, with\vspace{-0.05cm}\\ \noindent probability at least $1 \!-\! \delta$, whenever $\|f\|_\infty \!\leq\! c_2$, $f \!\in\! \mathcal{H}_{\kappa'}$, where $c_1 \!=\! c_2(-\textnormal{log}(\delta)/n)^{1/2}$.
\end{itemize}
\end{itemize}
The first property implies that we can define distance measures and distance metrics between probability distributions in reproducing-kernel Hilbert spaces.  We will define three such measures in the next section.  The second property imparts that we do not need access to these distributions in order to estimate the distance-measure magnitudes.  This is possible provided that the Rademacher average is well behaved.  The second property is crucial for our purposes.  Without it, the theory we present in the next section would not be possible.

Injectivity of the mean-element mappings underlies much of our ability to operate on probability measures in reproducing-kernel Hilbert spaces.  We therefore characterize one of its properties.

\phantomsection\label{prop2.3}
\begin{itemize}
\item[] \-\hspace{0.5cm}{\small{\sf{\textbf{Proposition 2.3: Distributional Equality.}}}} Let $p_{\mathcal{S}}, q_{\mathcal{S}} \!\in\! \mathcal{M}^1_+(\mathcal{S})$.  If the mapping $p_{\mathcal{S}} \!\mapsto\! \mu_{p_{\mathcal{S}}}$ is injective for a\\ \noindent separable, reproducing- kernel Hilbert space, then $\mathbb{E}_{s \sim p_{\mathcal{S}}}[f(s)] \!=\! \mathbb{E}_{s \sim q_{\mathcal{S}}}[f(s)]$, $\forall f \!\in\! \mathcal{H}_{\kappa'}$, where 
\begin{equation*}
\mathbb{E}_{s \sim q_{\mathcal{S}}}[f(s)] =\! \int_{\mathcal{S}} f(s)dq_{\mathcal{S}}(s).
\end{equation*}
That is, the probability measures are equivalent, $p_{\mathcal{S}} \!=\! q_{\mathcal{S}}$.
\end{itemize}
Thus, $\langle p_{\mathcal{S}} \!-\! q_{\mathcal{S}},\cdot \rangle_{\mathcal{H}_{\kappa'}} \!=\! 0$ if and only if the measures are equal, $p_{\mathcal{S}} \!=\! q_{\mathcal{S}}$.  Kernels underlying injective mappings ensure that no information is lost when mapping distributions into a reproducing-kernel Hilbert space.  Absent this guarantee, it would not be appropriate to pose our function cross-entropies in such a functional space.

Several kernels, like the isotropic Gaussian function, yield injective mean mappings.  Kernels of this type are referred to as characteristic.  The following result characterizes when this occurs for arbitrary kernels.

\phantomsection\label{prop2.4}
\begin{itemize}
\item[] \-\hspace{0.5cm}{\small{\sf{\textbf{Proposition 2.4: Characteristic Kernel Condition.}}}} Let $\phi(\omega')$ be a continuous function on $\mathbb{R}^d$.  Let a kernel obey $\kappa'(s,s') \!=\! \langle\phi(s),\phi(s')\rangle_{\mathcal{H}_{\kappa'}}$.  If, for any $\xi \!\in\! \mathbb{R}^{d}$ and $\beta \!\in\! \mathbb{R}$, there exists $\tau,\tau_0 \!\in\! \mathbb{R}$, such that
\begin{equation*}
\int_{\mathbb{R}^d} \phi'(\tau(\omega' \!+\! \xi))^2 \phi'(\omega')^{-1} d\omega' \!<\! \infty,\;\;\; \phi'(\omega') =\! \int_{-\infty}^\infty \phi(\omega')e^{-2\pi i(s)\beta}ds,
\end{equation*}
for all $\tau_0 \!<\! \tau$, then $\mathcal{H}_{\kappa'}$ is dense in $\mathbb{L}^2(p'_{\mathcal{S}})$ where $p'_{\mathcal{S}} \!\in\! \mathcal{M}^1_+(\mathbb{R}^d)$.  Thus, $\kappa'$ is a characteristic kernel with respect\\ \noindent to the Borel sigma-field and defines an injective mean-element mapping.
\end{itemize}

We typically take the sample space to be some multi-dimensional real space.  It can also be the space of real-valued matrices.  In these settings, the characteristic kernels underlying the injective mean mappings are generalizations of characteristic functions.  That is, they generalize the property that characteristic functions uniquely determine a Borel probability measure on the reals.

\vspace{0.15cm}{\small{\sf{\textbf{Bivariate Operator Embeddings.}}}} We now extend the notion of embedding probability measures to the bivariate case so that they can be natively handled in a quotient Hilbert space.

We first show that it is possible to construct a quotient Hilbert space with a reproducing kernel.

\phantomsection\label{prop2.5}
\begin{itemize}
\item[] \-\hspace{0.5cm}{\small{\sf{\textbf{Proposition 2.5: Quotient Hilbert Space Existence.}}}} Let $\mathcal{H}_{\kappa_1'},\mathcal{H}_{\kappa_2'}$ be Hilbert spaces with reproducing kernels $\kappa_1',\kappa_2'$.  The tensor product of the vector spaces admits a functional completion $\mathcal{H}_{\kappa_1'} \!\otimes \mathcal{H}_{\kappa_2'}$ which is a quotient reproducing-kernel Hilbert space with a product reproducing kernel $(\kappa'_{1} \otimes \kappa_2')((s,s'),(a,a')) =$\\ \noindent $\kappa_{1}'(s,a) \otimes \kappa_{2}'(s',a')$, where $s,s' \!\in\! \mathcal{S}$ and $a,a' \!\in\! \mathcal{A}$.  Here, $\mathcal{S},\mathcal{A}$ are sample spaces.
\end{itemize}
Moreover, we can define a product measure, $\rho_{\mathcal{S} \times \mathcal{A}}$, on the sigma-algebras associated with $\mathcal{H}_{\kappa_1'},\mathcal{H}_{\kappa_2'}$.  The existence of this product measure, for the sigma-finite case, can be shown by Carath\'{e}odory's extension theorem.

As before, we can losslessly project product measures into quotient reproducing-kernel Hilbert spaces via an expectation operation.

\phantomsection\label{def2.2}
\begin{itemize}
\item[] \-\hspace{0.5cm}{\small{\sf{\textbf{Definition 2.2: Bivariate Mean-Element Map.}}}} Let $\mathcal{H}_{\kappa_1'} \!\otimes \mathcal{H}_{\kappa_2'}$ be a separable, reproducing-kernel Hilbert\\ \noindent space.  For a product measurable function $\varphi^\otimes \!:\! \mathcal{S} \!\times\! \mathcal{A} \!\to\! \mathcal{H}_{\kappa_1'} \!\otimes \mathcal{H}_{\kappa_2'}$ which obeys the following condition\vspace{-0.025cm}\\ \noindent $(\kappa'_{1} \otimes \kappa_2')((s,a),\cdot) \!=\! \langle \varphi^\otimes(s,a),\cdot \rangle_{\mathcal{H}_{\kappa'_1} \otimes \mathcal{H}_{\kappa'_2}}$, we have, for a non- negative, non-zero-everywhere kernel, that
\begin{equation*}
\mathbb{E}_{((s,a),(s'a'))\sim \rho_{\mathcal{S} \times \mathcal{A}}}[(\kappa'_{1} \otimes \kappa_2')((s,a),(s'a'))] =\! \int\!\!\!\!\int_{\mathcal{S} \times \mathcal{A}} \langle \varphi^\otimes(s,a),\varphi^\otimes(s',a') \rangle_{\mathcal{H}_{\kappa'_1} \otimes \mathcal{H}_{\kappa'_2}} d\rho_{\mathcal{S} \times \mathcal{A}}((s,a),(s',a'))
\end{equation*}
is the product mean element $\mu_{\rho_{\mathcal{S} \times \mathcal{A}}} \!=\! \mathbb{E}_{((s,a),\cdot)\sim \rho_{\mathcal{S} \times \mathcal{A}}}[\kappa'_{1, 2}((s,a),\cdot)]$, where $\rho_{\mathcal{S} \times \mathcal{A}} \!\in\! \mathcal{M}^1_+(\mathcal{S} \!\times\! \mathcal{A})$ is a Borel\\ \noindent probability measure over the product topological space.\vspace{-0.03cm}
\end{itemize}
However, we must impose $\mathbb{E}_{(s,s') \sim p_{\mathcal{S}}}[\kappa_1'(s,s')] \!\in\! \mathbb{R}_{0,+}$ and $\mathbb{E}_{(a,a') \sim q_{\mathcal{A}}}[\kappa_2'(a,a')] \!\in\! \mathbb{R}_{0,+}$,  where the Borel mea-\\ \noindent sures $p_{\mathcal{S}} \!\in\! \mathcal{M}^1_+(\mathcal{S})$ and $q_{\mathcal{A}} \!\in\! \mathcal{M}^1_+(\mathcal{A})$, so that the product kernel is Bochner $\rho_{\mathcal{S} \times \mathcal{A}}$-integrable.

The corresponding embedding $\mu_{\kappa'_1 \otimes\, \kappa'_2}(\rho_{\mathcal{S} \times \mathcal{A}}) \!=\! \mu_{\rho_{\mathcal{S} \times \mathcal{A}}}$ naturally exists in $\mathcal{H}_{\kappa'_1} \!\otimes \mathcal{H}_{\kappa'_2}$.

\phantomsection\label{prop2.6}
\begin{itemize}
\item[] \-\hspace{0.5cm}{\small{\sf{\textbf{Proposition 2.6: Bivariate Riesz Representation Property.}}}} Let $\mathcal{H}_{\kappa_1'} \!\otimes \mathcal{H}_{\kappa_2'}$ be a separable, quotient\\ \noindent reproducing-kernel Hilbert space, with a continuous reproducing kernel.  For $f \!\in\! \mathcal{H}_{\kappa_1'}$ and $g \!\in\! \mathcal{H}_{\kappa_2'}$, if\vspace{-0.025cm}\\ \noindent $\mathbb{E}_{((s,a),\cdot)\sim \rho_{\mathcal{S} \times \mathcal{A}}}[(\kappa'_{1} \otimes \kappa_2')((s,a),\cdot)^{1/2}] \!\in\! \mathbb{R}_{0,+}$, then we have that $\mu_{\rho_{\mathcal{S} \times \mathcal{A}}} \!\!\in\! \mathcal{H}_{\kappa_1'} \!\otimes \mathcal{H}_{\kappa_2'}$ and hence that\\ \noindent $\mathbb{E}_{\rho_{\mathcal{S} \times \mathcal{A}}}[f(s)g(a)] \!=\! \langle f \otimes g,\mu_{\rho_{\mathcal{S} \times \mathcal{A}}}\rangle_{\mathcal{H}_{\kappa_1'} \otimes \mathcal{H}_{\kappa_2'}}$.
\end{itemize}

Product characteristic kernels also exist for injective mappings.  The quotient reproducing-kernel Hilbert space is hence expressive enough to distinguish between distributions from their embeddings.

\phantomsection\label{def2.3}
\begin{itemize}
\item[] \-\hspace{0.5cm}{\small{\sf{\textbf{Definition 2.3: Product Kernel Characteristicness.}}}} A positive-definite product kernel $\kappa_1' \otimes \kappa_2'$ is characteristic to a set of probability measures if $\mathcal{M}_+^1(\mathcal{S} \!\times\! \mathcal{A}) \!\to\! \mathcal{H}_{\kappa_1'} \otimes \mathcal{H}_{\kappa_2'} \!:\! \rho_{\mathcal{S} \times \mathcal{A}} \!\mapsto\! \mu_{\rho_{\mathcal{S} \times \mathcal{A}}}$ is injective.
\end{itemize}

It has been shown that injective mean-element mappings are exactly the same as moment-generating functions of samples on either the marginal space or the quotient space.  The mean-element map contains information of all moments.  This property lends credence to the richness of a reproducing-kernel-Hilbert-space representation in losslessly describing the probability distributions and distinguishing them via their embeddings.  No parametric assumptions are made for appropriate kernel choices.  Arbitrary distributions can hence be modeled and processed when relying on constructs that employ the mean-element mappings.

It is important to observe that the product of two characteristic kernels, $\kappa_1',\kappa_2'$, does not necessarily yield a characteristic product kernel $\kappa_1' \otimes \kappa_2'$.  Rather, the product may only belong to a weaker class of kernels that do not guarantee injectivity \cite{SzaboZ-jour2018a}, which we outline below.

\phantomsection\label{def2.4}
\begin{itemize}
\item[] \-\hspace{0.5cm}{\small{\sf{\textbf{Definition 2.4: Product Kernel $i$-Characteristicness.}}}} Let $\kappa_1' \!:\! \mathcal{S} \!\times\! \mathcal{S} \!\to\! \mathbb{R}_{0,+}$ and $\kappa_2' \!:\! \mathcal{A} \!\times\! \mathcal{A} \!\to\! \mathbb{R}_{0,+}$ be\\ \noindent bounded kernels on topological spaces $\mathcal{S}$ and $\mathcal{A}$, respectively.  Let $\mathcal{J} \!\subseteq\! \mathbb{L}_2(\mathcal{S} \!\times\! \mathcal{A})$ such that $0 \!\in\! \mathcal{J}$.  The product kernel $\kappa_1' \!\otimes \kappa_2'$ is $\mathcal{J}$-integrally-strictly-positive-definite if and only if
\begin{equation*}
\int\!\!\!\!\int_{\mathcal{S} \times \mathcal{A}} (\kappa_1' \!\otimes \kappa_2')((s,a),(s',a'))df(s,a)df(s',a') \!>\!0,\;\; \forall f \!\in\! \mathcal{J}\backslash\{0\}.
\end{equation*}
If $\mathcal{J} \!=\! \{\rho_{\mathcal{S} \times \mathcal{A}} - p_{\mathcal{S}} \otimes p_{\mathcal{A}} : \rho_{\mathcal{S} \times \mathcal{A}} \!\in\! \mathcal{M}_1^+(\mathcal{S} \!\times\! \mathcal{A})\}$, then the tensor-product kernel is said to be $i$-characteristic. 
\end{itemize}
If $\mathcal{S}$ and $\mathcal{A}$ are second-countable, then, for characteristic kernels $\kappa_1',\kappa_2'$, their tensor product $\kappa_1' \otimes \kappa_2'$ is $i$-characteristic.  The converse is true, provided that the topological spaces are additionally Hausdorff.  

\phantomsection\label{prop2.7}
\begin{itemize}
\item[] \-\hspace{0.5cm}{\small{\sf{\textbf{Proposition 2.7: Product Kernel Non-Characteristicness.}}}} Let $\kappa_1' \otimes \kappa_2'$ be an arbitrary $i$-characteristic\\ \noindent tensor-product kernel.  This product kernel may not be characteristic, even if $\kappa_1',\kappa_2'$ are, individually, characteristic on their respective topological spaces $\mathcal{S}$, $\mathcal{A}$.
\end{itemize}

For injectivity to be present, we it is helpful to have that the product kernel be $c_0$-universal.  This is possible if both $\kappa_1',\kappa_2'$ are universal kernels.  Translation invariance of the kernels is not required \cite{FukumizuK-coll2010a} if we replace integrally-strictly positive definiteness with strict positive definiteness.  The isotropic-Gaussian kernel is universal, as are others like the exponential-inner-product kernel.

\phantomsection\label{prop2.8}
\begin{itemize}
\item[] \-\hspace{0.5cm}{\small{\sf{\textbf{Proposition 2.8: Product Kernel Strict Positive Definiteness.}}}} Let $\kappa_1' \!:\! \mathcal{S} \!\times\! \mathcal{S} \!\to\! \mathbb{R}_{0,+}$ and $\kappa_2' \!:\! \mathcal{A} \!\times\! \mathcal{A} \!\to\! \mathbb{R}_{0,+}$\\ \noindent be bounded, $c_0$-kernels on locally compact Polish spaces $\mathcal{S}$ and $\mathcal{A}$, respectively.  Let $\mathcal{J} \!\subseteq\! \mathcal{M}_b(\mathcal{S} \!\times\! \mathcal{A})$, such that $0 \!\in\! \mathcal{J}$, be a subset of the space of all finite signed measures on the product topological space.  The kernel $\kappa_1' \!\otimes \kappa_2'$ is $\mathcal{J}$-strictly-positive-definite if and only if
\begin{equation*}
\int\!\!\!\!\int_{\mathcal{S} \times \mathcal{A}} (\kappa_1' \!\otimes \kappa_2')((s,a),(s',a'))d\rho_{\mathcal{S} \times \mathcal{A}}((s,a),(s',a')) \!>\!0,\;\; \forall \rho_{\mathcal{S} \times \mathcal{A}} \!\in\! \mathcal{J}\backslash\{0\}.
\end{equation*}
If $\mathcal{J} \!=\! \mathcal{M}_b(\mathcal{S} \!\times\!\mathcal{A})$, the space of all finite signed measures, then the tensor-product kernel is said to be $c_0$-uni-\\ \noindent versal and hence characteristic. 
\end{itemize}
Without $c_0$-universality, the covariance operators that we define would not necessary correspond to transformations of embedded probability measures.  Throughout, we refer to $c_0$-universal kernels as just universal kernels.

\phantomsection\label{sec2.2}
\subsection*{\small{\sf{\textbf{2.2.$\;\;\;$Gram Operators}}}}

{\small{\sf{\textbf{Univariate Operators.}}}} We now implicitly use the mean-element map to define the notion of a covariance operator for both univariate and bivariate-product random variables.  We refer to this covariance operator as a Gram operator since it behaves akin to a conventional Gramian, but in a functional sense.  We will use this operator to mainly quantify marginal distribution shape.

We start with the Gram operator $\kappa \!:\! \mathcal{H}_{\kappa'} \!\to\! \mathcal{H}_{\kappa'}$ for the univariate product case.  We also define its trace, as our notions of cross-entropy rely on traces of univariate Gram operators.

\phantomsection\label{def2.5}
\begin{itemize}
\item[] \-\hspace{0.5cm}{\small{\sf{\textbf{Definition 2.5: Univariate Gram Operator.}}}} For a separable reproducing-kernel Hilbert space, with a continuous, characteristic reproducing kernel, let $\kappa'(s,s') \!=\! \langle \varphi(s),\varphi(s') \rangle_{\mathcal{H}_{\kappa'}}$, where $\varphi$ is measurable.  For $f,g \!\in\! \mathcal{H}_{\kappa'}$, the univariate Gram operator $\kappa \!:\! \mathcal{H}_{\kappa'} \!\to\! \mathcal{H}_{\kappa'}$ is given by the symmetric bilinear form
\begin{equation*}
\mathcal{K}(f,g) = \langle f,\kappa g\rangle =\! \int\!\!\!\!\int_{\mathcal{S}}\! \Bigg(\!\langle f,\varphi(s) \rangle_{\mathcal{H}_{\kappa'}} \langle \varphi(s'),g \rangle_{\mathcal{H}_{\kappa'}} \!\!\Bigg) dp_{\mathcal{S}}(s,s')
\end{equation*}
where $f(s) \!=\! \langle f,\varphi(s)\rangle_{\mathcal{H}_{\kappa'}}$ and thus $\mathcal{K}(f,g) \!=\! \mathbb{E}_{(s,s')\sim p_{\mathcal{S}}}[f(s)g(s')]$, for $s,s' \!\in\! \mathcal{S}$. 

We have that the trace of this operator is
\begin{equation*}
\textnormal{tr}(\kappa) = \sum_{j=1}^h \int\!\!\!\!\int_{\mathcal{S}}\! \Bigg(\!\langle \eta_j,\varphi(s) \rangle_{\mathcal{H}_{\kappa'}} \langle \varphi(s'),\eta_j \rangle_{\mathcal{H}_{\kappa'}} \!\!\Bigg) dp_{\mathcal{S}}(s,s'),
\end{equation*}
which follows from $\textnormal{tr}(\kappa) \!=\! \sum_{j=1}^h \mathcal{K}(\eta_j,\cdot)$.  Here, $\{\eta_j\}_{j=1}^h \!\subset\! \mathcal{H}_{\kappa'}$ is a complete orthonormal basis for $\mathcal{H}_{\kappa'}$.  
\end{itemize}
From this definition, we can view the Gram operator as an image point of an embedding of the probability measure from which samples are drawn.  This operator defines a bilinear form $\mathcal{K}(f,g) \!=\! \langle f,\kappa g\rangle$, with $f,g \!\in\! \mathcal{H}_{\kappa'}$, that corr-\\ \noindent esponds to the correlation of functions that belong to the reproducing-kernel Hilbert space $\mathcal{H}_{\kappa'}$ induced by $\kappa'$.

We have that $\kappa$ is a compact, trace-class operator.  The corresponding trace is finite.  The trace is also independent of the choice of basis, as any basis of a separable reproducing-kernel Hilbert space contains an orthonormal set.

\phantomsection\label{prop2.9}
\begin{itemize}
\item[] \-\hspace{0.5cm}{\small{\sf{\textbf{Proposition 2.9: Hilbert Space Separability.}}}} Let $\kappa'$ be a reproducing kernel defined for a univariate random variable.  A reproducing kernel Hilbert space $\mathcal{H}_{\kappa'}$ is separable if and only if its dimension is, at most, countable.  That Hilbert space's dimensionality is the cardinality of its basis.
\end{itemize}

\phantomsection\label{prop2.10}
\begin{itemize}
\item[] \-\hspace{0.5cm}{\small{\sf{\textbf{Proposition 2.10: Hilbert Space Basis Orthonormality.}}}} Let $\mathcal{H}_{\kappa'}$ be a separable reproducing-kernel Hilbert\\ \noindent space induced by a continuous reproducing kernel $\kappa'$.  If $\{\eta_j\}_{j=1}^h \!\subset\! \mathcal{H}_{\kappa'}$ is an orthonormal set, then there is a\\ \noindent basis that contains it.
\end{itemize}
Although a basis provably exists, specifying it can be difficult.  The trace operation assumes access to it.  We therefore consider an approximation strategy in the next section.

It is important to note that if there is no reproducing kernel, then the space is non-separable.  Non-separability complicates the Bochner integrability underlying the mean-element maps \cite{OwhadiH-jour2017a}.  Without the ability to losslessly embed probability distributions in a functional space, we will not have guarantees that our cross-entropy measures will behave well.

An important result is that the trace of the univariate Gram operator can be expressed as the norm of the univariate mean-element map.  Other $\alpha$-traces are also related, though their relationship with the mean-element mapping is more complicated.

\phantomsection\label{prop2.11}
\begin{itemize}
\item[] \-\hspace{0.5cm}{\small{\sf{\textbf{Proposition 2.11: Univariate Operator-Map Equivalence.}}}} Let $\mathcal{H}_{\kappa'}$ be a separable, reproducing-kernel Hilbert space with a continuous, characteristic reproducing kernel $\kappa'(s,s') \!=\! \langle \varphi(s),\varphi(s') \rangle_{\mathcal{H}_{\kappa'}}$, for a measurable $\varphi$.  We have that the relationship between the 2-trace of the univariate Gram operator, $\kappa$, and the univariate mean-element map $\mu_{\rho_{\mathcal{S}}}$ is
\begin{equation*}
\int\!\!\!\!\int_{\mathcal{S}}\! \Bigg(\!\langle \varphi(s),\kappa \varphi(s') \rangle_{\mathcal{H}_{\kappa'}} \!\!\Bigg) dp_{\mathcal{S}}(s,s') = \langle\hspace{0.01cm}\mu_{\rho_{\mathcal{S}}},\mu_{\rho_{\mathcal{S}}}\hspace{-0.02cm}\rangle_{\mathcal{H}_{\kappa'\kappa'}}.
\end{equation*}
Here, $\mathcal{H}_{\kappa'\kappa'}$ denotes the reproducing-kernel Hilbert space induced by the kernel $\kappa'(s,s')\kappa'(s,s')$.
\end{itemize}
This finding demonstrates that, in working with univariate Gram operators, we are implicitly dealing with the univariate mean-element maps.  Any measure defined in terms of these Gram operators will thus transform embeddings of the probability measures in a reproducing-kernel Hilbert space.

\vspace{0.15cm}{\small{\sf{\textbf{Bivariate Operators.}}}} We now define an operator $\gamma \!:\! \mathcal{H}_{\kappa_1'} \!\otimes \mathcal{H}_{\kappa_2'} \!\to\! \mathcal{H}_{\kappa_1'} \!\otimes \mathcal{H}_{\kappa_2'}$ for bivariate product random variables, where $\mathcal{H}_{\kappa_1'}$ and $\mathcal{H}_{\kappa_2'}$ are two reproducing-kernel Hilbert spaces.  This extends the notion of the Gram operator to a case that will prove useful when handling pairs of univariate operators. 

\phantomsection\label{def2.6}
\begin{itemize}
\item[] \-\hspace{0.5cm}{\small{\sf{\textbf{Definition 2.6: Bivariate Gram Operator.}}}} For separable reproducing-kernel Hilbert spaces $\mathcal{H}_{\kappa_1'}$ and $\mathcal{H}_{\kappa_2'}$, with continuous reproducing kernels defined on state-action pairs, let $\kappa'_1(s,s') \!=\! \langle \varphi(s),\varphi(s') \rangle_{\mathcal{H}_{\kappa_1'}}$ and\vspace{-0.05cm}\\ \noindent $\kappa'_2(a,a') \!=\! \langle \psi(a),\psi(a') \rangle_{\mathcal{H}_{\kappa_2'}}$, where $\varphi \!:\! \mathcal{S} \!\to\! \mathcal{H}_{\kappa_1'}$ and $\psi \!:\! \mathcal{A} \!\to\! \mathcal{H}_{\kappa_2'}$ are measurable.  Assume that the product\vspace{-0.025cm}\\ \noindent kernel is characteristic.  We define the bivariate Gram operator $\gamma$, which is over the quotient, completed Hilbert space, via the symmetric bilinear form
\begin{equation*}
\mathcal{G}(f,g) = \langle f,\gamma g\rangle_{\mathcal{H}_{\kappa_1'} \otimes \,\mathcal{H}_{\kappa_2'}} \!\!\,=\! \int\!\!\!\!\int_{\mathcal{S} \times \mathcal{A}} \Bigg(\!\langle f, \varphi^{\otimes}((s,a),\cdot) \rangle_{\mathcal{H}_{\kappa_1'} \otimes \,\mathcal{H}_{\kappa_2'}} \langle \varphi^{\otimes}((s,a),\cdot),g \rangle_{\mathcal{H}_{\kappa_1'} \otimes \,\mathcal{H}_{\kappa_2'}}\!\Bigg) d\rho_{\mathcal{S} \times \mathcal{A}}((s,a),\cdot).
\end{equation*}
Here, $f,g \!\in\! \mathcal{H}_{\kappa_1'} \!\otimes \mathcal{H}_{\kappa_2'}$.  The double integral is over the same joint topological space $\mathcal{S} \!\times\! \mathcal{A}$.

Using the bivariate mean-element $\mu_{\rho_{\mathcal{S} \times \mathcal{A}}}$, the corresponding trace of this operator is analogous to the univariate case,
\begin{equation*}
\textnormal{tr}(\gamma) = \sum_{q=1}^{r} \int\!\!\!\!\int_{\mathcal{S} \times \mathcal{A}}\! \Bigg(\!\langle \pi_q,\varphi^{\otimes}((s,a),\cdot) \rangle_{\mathcal{H}_{\kappa_1'} \!\otimes \mathcal{H}_{\kappa_2'}} \langle \varphi^{\otimes}((s,a),\cdot),\pi_q \rangle_{\mathcal{H}_{\kappa_1'} \!\otimes \mathcal{H}_{\kappa_2'}} \!\Bigg) d\rho_{\mathcal{S} \times \mathcal{A}}((s,a),\cdot).
\end{equation*}
Here, $\{\pi_q\}_{j=1}^r \!\subset\! \mathcal{H}_{\kappa_1'} \!\otimes \mathcal{H}_{\kappa_2'}$ is a complete orthonormal basis for the quotient, completed Hilbert space.
\end{itemize}

\phantomsection\label{prop2.12}
\begin{itemize}
\item[] \-\hspace{0.5cm}{\small{\sf{\textbf{Proposition 2.12: Quotient Hilbert Space Separability.}}}}  Let $\mathcal{H}_{\kappa_1'}$ and $\mathcal{H}_{\kappa_2'}$ be separable reproducing-kernel Hilbert spaces induced by a continuous reproducing kernels $\kappa_1'$ and $\kappa_1'$.  If $\mathcal{H}_{\kappa_1'}$ and $\mathcal{H}_{\kappa_2'}$ have orthonormal bases $\{\eta_j\}_{j=1}^{h}$ and $\{\vartheta_k\}_{k=1}^{p}$, respectively, then $\{\eta_j \otimes \vartheta_k\}_{j,k} \!\subset\! \mathcal{H}_{\kappa_1'} \!\otimes \mathcal{H}_{\kappa_2'}$, is an orthonormal set.  The Hilbert dimension of the tensor products is the product of the Hilbert dimensions.
\end{itemize}
For universal reproducing kernels, the Hilbert-space dimensionality is infinite.

We can extend the trace of the bivariate Gram operators to arbitrary positive powers, which we will need shortly.  A similar result holds in the univariate case.

\phantomsection\label{prop2.13}
\begin{itemize}
\item[] \-\hspace{0.5cm}{\small{\sf{\textbf{Proposition 2.13: Bivariate Gram Operator $\alpha$-Trace.}}}} For a separable, quotient reproducing-kernel Hilbert space, with a continuous reproducing kernel, let $(\kappa_1' \otimes \kappa_2')((s,\cdot),(a,\cdot)) \!=\! \langle \varphi^{\otimes}((s,a),\cdot),\cdot \rangle_{\mathcal{H}_{\kappa'}}$, where $\varphi^{\otimes}$ is\\ \noindent measurable.  Assume that the product kernel is characteristic.  For $\alpha \!\in\! \mathbb{R}_{0,+} \!\backslash \{1\}$, we have that the $\alpha$-trace of the bivariate Gram operator, $\gamma$, is
\begin{align*}
\textnormal{tr}(\gamma^\alpha) &= \sum_{q=1}^r \int\!\!\!\!\int_{\mathcal{S} \times \mathcal{A}}\! \Bigg(\!\langle \pi_q,\varphi^{\otimes}((s,a),\cdot) \rangle_{\mathcal{H}_{\kappa_1'} \otimes \mathcal{H}_{\kappa_2'}} \langle \varphi^{\otimes}((s,a),\cdot),\gamma^\alpha\pi_q \rangle_{\mathcal{H}_{\kappa_1'} \otimes \mathcal{H}_{\kappa_2'}} \!\!\Bigg) d\rho_{\mathcal{S} \times \mathcal{A}}((s,a),\cdot)\\
&= \int\!\!\!\!\int_{\mathcal{S} \times \mathcal{A}} \!\Bigg(\!\langle \varphi^{\otimes}((s,a),\cdot),\gamma^{\alpha-1}\varphi^{\otimes}((s,a),\cdot) \rangle_{\mathcal{H}_{\kappa_1'} \otimes \mathcal{H}_{\kappa_2'}}\!\Bigg) d\rho_{\mathcal{S} \times \mathcal{A}}((s,a),\cdot).
\end{align*}
The inner product is a positive-definite function that depends on the product probability measure.
\end{itemize}

As with the univariate case, the 2-trace of the bivariate Gram operator is connected to the bivariate mean-element mapping.  It is also equivalent to the square of the Hilbert-Schmidt norm \cite{FukumizuK-coll2008a}.

\phantomsection\label{prop2.14}
\begin{itemize}
\item[] \-\hspace{0.5cm}{\small{\sf{\textbf{Proposition 2.14: Bivariate Operator-Map Equivalence.}}}}  Let $\mathcal{H}_{\kappa_1'}$ and $\mathcal{H}_{\kappa_2'}$ be separable reproducing-kernel Hilbert spaces induced by a continuous reproducing kernels $\kappa_1'$ and $\kappa_1'$.  Assume that the product kernel $(\kappa_1' \otimes \kappa_2')((s,\cdot),(a,\cdot)) \!=\! \langle \varphi^{\otimes}((s,a),\cdot),\cdot \rangle_{\mathcal{H}_{\kappa'_1} \otimes \mathcal{H}_{\kappa'_2}}$ is characteristic.  We have that the relationship between the 2-trace of the bivariate Gram operator, $\gamma$, and the bivariate mean-element map $\mu_{\rho_{\mathcal{S} \times \mathcal{A}}}$ is
\begin{equation*}
\int\!\!\!\!\int_{\mathcal{S} \times \mathcal{A}} \!\Bigg(\!\langle \varphi^{\otimes}((s,a),\cdot),\gamma\varphi^{\otimes}((s,a),\cdot) \rangle_{\mathcal{H}_{\kappa_1'} \otimes \mathcal{H}_{\kappa_2'}}\!\Bigg) d\rho_{\mathcal{S} \times \mathcal{A}}((s,a),\cdot) = \langle\hspace{0.01cm}\mu_{\rho_{\mathcal{S} \times \mathcal{A}}},\mu_{\rho_{\mathcal{S} \times \mathcal{A}}}\hspace{-0.02cm}\rangle_{\mathcal{H}_{\kappa_1'\kappa_1'} \otimes \mathcal{H}_{\kappa_2'\kappa_2'}}.
\end{equation*}
Here, $\mathcal{H}_{\kappa_1'\kappa_1' \otimes \kappa_2'\kappa_2'}$ denotes the reproducing-kernel Hilbert space induced by the tensor-product kernel\\ \noindent $(\kappa_1'\kappa_1' \!\otimes\! \kappa_2'\kappa_2')((s,s'),(a,a'))$.
\end{itemize}

\noindent As a byproduct of the mean-element map relationship, we also obtain that the Gram operators are non-parametric, lossless characterizations of the interactions between probability measures.  The operators also contain information about all statistical moments.  Without these guarantees, then we could not be assured that working in such a functional space would preserve the necessary traits of the probability measures to facilitate proper analysis.

Lastly, we consider a bivariate operator that corresponds to cross-correlations of functions.  We will use this operator to quantify the overlap of marginal distributions.

\phantomsection\label{def2.7}
\begin{itemize}
\item[] \-\hspace{0.5cm}{\small{\sf{\textbf{Definition 2.7: Joint, Bivariate Gram Operator.}}}} For separable reproducing-kernel Hilbert spaces $\mathcal{H}_{\kappa_1'}$ and $\mathcal{H}_{\kappa_2'}$, with continuous, universal reproducing kernels defined on state-action pairs, let $\kappa'_1(s,s') \!=\! \langle \varphi(s),\varphi(s') \rangle_{\mathcal{H}_{\kappa_1'}}$\vspace{-0.025cm}\\ \noindent and $\kappa'_2(a,a') \!=\! \langle \psi(a),\psi(a') \rangle_{\mathcal{H}_{\kappa_2'}}$, where $\varphi \!:\! \mathcal{S} \!\times\! \mathcal{S} \!\to\! \mathcal{H}_{\kappa_1'}$ and $\psi \!:\! \mathcal{A} \!\times\! \mathcal{A} \!\to\! \mathcal{H}_{\kappa_2'}$ are measurable.  We define the\vspace{-0.01cm}\\ \noindent joint, bivariate Gram operator $\lambda$ via the bilinear form,\vspace{-0.15cm}
\begin{equation*}
\mathcal{L}(f,g) = \langle f,\lambda g\rangle_{\mathcal{H}_{\kappa_1'} \otimes^* \,\mathcal{H}_{\kappa_2'}} =\! \int_{\mathcal{S} \times \mathcal{A}} \Bigg(\!\langle f, \varphi(s) \rangle_{\mathcal{H}_{\kappa_1'}} \langle \varphi(s),\psi(a) \rangle_{\mathcal{H}_{\kappa_1'} \otimes \,\mathcal{H}_{\kappa_2'}} \langle \psi(a),g \rangle_{\mathcal{H}_{\kappa_2'}}\!\Bigg) d\rho_{\mathcal{S} \times \mathcal{A}}(s,a).
\end{equation*}
Here, $f \!\in\! \mathcal{H}_{\kappa_1'}$, $g \!\in\! \mathcal{H}_{\kappa_2'}$.
\end{itemize}

\subsection*{\small{\sf{\textbf{3.$\;\;\;$Bipartite Cross-Entropy}}}}\addtocounter{section}{1}

R\'{e}nyi, in \cite{RenyiA-coll1976a}, pursued an axiomatic investigation of Shannon's entropy.  He showed that a parameterized family of entropies could be constructed.  R\'{e}nyi also created bipartite relative entropies and bipartite cross-entropies, both of which are parameterized.  The latter of these is specified below.
\begin{itemize}
\item[] \-\hspace{0.5cm}{\small{\sf{\textbf{Definition 3.1: R\'{e}nyi's $\alpha$-Cross-Entropy}}}} Let $q_1,q_2 \!:\! \mathcal{S} \!\to\! \mathbb{R}_{0,+}$ be probability densities over a common topological sample space $\mathcal{S}$.  The $\alpha$-order R\'{e}nyi's cross-entropy, for $\alpha \!\in\! \mathbb{R}_{+} \!\backslash \{1\}$, is 
\begin{equation*}
H_\alpha(q_1\|q_2) = \frac{1}{\alpha \!-\! 1}\textnormal{log}\Bigg(\!\int_\mathcal{S} q_1^\alpha(s)q_2^{1-\alpha}(s) dp_\mathcal{S}(s)\!\Bigg).
\end{equation*}
Here, $p_{\mathcal{S}} \!\in\! \mathcal{M}^1_+(\mathcal{S})$ is a Borel probability measure over $\mathcal{S}$.
\end{itemize}
Intuitively, the $\alpha$-cross-entropy is a measure of distributional overlap.  

In this section, we extend R\'{e}nyi's $\alpha$-cross-entropies to take as input non-commutative Gram operators in reproducing-kernel Hilbert spaces.  We start with the most straightforward generalization, which we refer to as the bipartite, non-mirrored R\'{e}nyi's $\alpha$-cross-entropy (see \hyperref[def3.2]{Definition 3.2}).  This version provably satisfies the same axioms that R\'{e}nyi considered for divergences.  It therefore acts like a divergence.  

Unfortunately, this measure only satisfies the data-processing inequality on a small set of parameter values.  Outside of this set, any transformation from a given class of functions will artificially increase the distinguishability of the operators, leading to erroneous magnitudes.  We hence construct another relative entropy for non-commutative operators.  We refer to this as the bipartite, mirrored R\'{e}nyi's $\alpha$-cross-entropy (see \hyperref[def3.3]{Definition 3.3}).  It obeys this inequality on the full parameter-value range.  It too satisfies the fundamental axioms.  

Both the mirrored and non-mirrored cross-entropies rely on traces of either the univariate or bivariate Gram operators.  They hence rely on access to a basis for the reproducing-kernel Hilbert space.  As we noted, it can be difficult to explicitly specify a basis, depending on the kernel choice.  Here, we sidestep this issue by considering principled approximations of the Gram operators, which are empirical Gram matrices.  We do this for the univariate case (see \hyperref[def3.4]{Definition 3.4} and \hyperref[prop3.1]{Proposition 3.1}) and the bivariate case (see \hyperref[def3.5]{Definition 3.5} and \hyperref[prop3.2]{Proposition 3.2}).  This approximation leads us to define matrix-based estimators of cross-entropy (see \hyperref[alg:1]{Algorithms 3.1} and \hyperref[alg:1]{3.2}).  The Gram matrices are unbiased, non-parametric estimates of the Gram operators.  These matrices can be constructed through the application of kernels to pairwise distances between vector samples.  They can also be formed from the application of kernels to representations encoded by symmetric, non-negative matrices that are square.

We show that the spectral characteristics of the Gram matrices and Gram operators are consistent (see \hyperref[prop3.1]{Propositions 3.1} and \hyperref[prop3.2]{3.2}).  This holds for the univariate and bivariate cases.  We additionally bound the difference between the traces of the Gram operators and the traces of the Gram matrices (see \hyperref[prop3.1]{Propositions 3.1} and \hyperref[prop3.2]{3.2}).  These bounds are minimax optimal (see \hyperref[prop3.3]{Propositions 3.3} and \hyperref[prop3.4]{3.4}).  They hence are the best obtainable bounds for the worst possible condition.  Our matrix-based cross-entropies do not require direct access to the underlying probability distributions, just samples from them, to work well.  Such results indicate that the Gram matrices are suitable substitutes for the operators.  As well, the bounds are completely independent of the sample-set dimensionality (see \hyperref[prop3.3]{Propositions 3.3} and \hyperref[prop3.4]{3.4}).  This implies that our cross-entropies are appropriate for handling high-dimensional samples well.  Alternate estimators, like those that rely on approximating and tightening evidence lower bounds \cite{LiY-coll2016a}, can be rather sensitive to the sample dimensionality \cite{StoneCJ-jour1980a}.

It is important to note that proper matrix conditioning must be imposed for the Gram matrices, though, for them to be useful for assessing cross entropy.  This is because we will be, potentially, raising the Gram matrices to fractional powers, which will entail matrix inversion.  Poorly conditioned matrices will yield poor empirical results.

\phantomsection\label{sec3.1}
\subsection*{\small{\sf{\textbf{3.1.$\;\;\;$Operator-Based Cross-Entropies}}}}

{\small{\sf{\textbf{Bipartite, Non-Mirrored R\'{e}nyi's $\alpha$-Cross-Entropy.}}}} For non-commutative Gram operators, we specify an operator-based divergence measure that strongly resembles R\'{e}nyi's classical $\alpha$-divergences.  We refer to this function as the bipartite, non-mirrored case, as it contains only a single instance of two Gram operators in its primary term.

\phantomsection\label{def3.2}
\begin{itemize}
\item[] \-\hspace{0.5cm}{\small{\sf{\textbf{Definition 3.2: Bipartite, Non-Mirrored R\'{e}nyi's $\alpha$-Cross-Entropy.}}}} Let $\mathcal{H}_{\kappa_1'}$, $\mathcal{H}_{\kappa_2'}$ be separable reproducing-kernel Hilbert spaces, with continuous, universal reproducing kernels $\kappa_1',\kappa_2'$.  Let $\kappa_1 \!:\! \mathcal{H}_{\kappa_1'} \!\to\! \mathcal{H}_{\kappa_1'}$, $\kappa_2 \!:\! \mathcal{H}_{\kappa_2'} \!\to\! \mathcal{H}_{\kappa_2'}$ be the corresponding univariate Gram operators.  The bipartite, non-mirrored, operator-based R\'{e}nyi's $\alpha$-cross-entropy of $\kappa_1$ and $\kappa_2$, with respect to $\alpha \!\in\! \mathbb{R}_{0,+} \!\backslash \{1\}$, is
\begin{equation*}
C_\alpha(\kappa_1\|\kappa_2) = \frac{1}{\alpha \!-\! 1}\textnormal{log}\Bigg(\!\textnormal{tr}(\kappa_1^\alpha\kappa_2^{1-\alpha})\!\Bigg) \!-\! \frac{1}{\alpha \!-\! 1}\textnormal{log}\Bigg(\!\textnormal{tr}(\kappa_1)\!\Bigg),\; \textnormal{for } \textnormal{supp}(\kappa_2) \!\subseteq\! \textnormal{supp}(\kappa_1),
\end{equation*}
with $C_\alpha(\kappa_1\|\kappa_2) \!=\! \infty$ otherwise.  $\kappa_1^\alpha\kappa_2^{1-\alpha} \!=\! \gamma$ is, naturally, a bivariate Gram operator.  Here, $\textnormal{supp}(\kappa_*)$ is the\\ \noindent support of $\kappa_*$, i.e., the span of the eigenvectors of $\kappa_*$ corresponding to non-zero eigenvalues.  The variable $\kappa_*$ represents either $\kappa_1$ or $\kappa_2$.

\end{itemize}
In this cross-entropy measure, the arguments of the log terms can be viewed as transformations that transfer statistical properties of the measures to reproducing-kernel Hilbert spaces.  This follows from the theory established in the previous section.  The second log term acts as a normalization term.  It is consistent with what R\'{e}nyi proposed for incomplete distributions \cite{RenyiA-coll1976a}.

As we illustrate in the appendix (see \hyperref[secA.1]{Appendix A.1}), this operator-based divergence measure satisfies the five major axioms that R\'{e}nyi specified for divergence measures.  The sixth axiom, that of normalization, is trivial to demonstrate for Gram operators associated with single-Dirac-measure distributions.  This operator-based cross-entropy can thus behave like a divergence.  Several other properties are realized too, which are presented in the appendix.

\vspace{0.15cm}{\small{\sf{\textbf{Bipartite, Mirrored R\'{e}nyi's $\alpha$-Cross-Entropy.}}}} As we note in the appendix, an issue with the preceding measure is that it does not satisfy the data-processing inequality.  Certain transformations of the operators can artificially change the cross-entropy scores.  We hence consider an amended operator-based R\'{e}nyi's $\alpha$-cross-entropy that will not improperly alter the distributional difference.  We refer to this function as the mirrored case, due to the appearance of dual non-commutative operators in the primary term.

\phantomsection\label{def3.3}
\begin{itemize}
\item[] \-\hspace{0.5cm}{\small{\sf{\textbf{Definition 3.3: Bipartite, Mirrored R\'{e}nyi's $\alpha$-Cross-Entropy.}}}} Let $\mathcal{H}_{\kappa_1'}$, $\mathcal{H}_{\kappa_2'}$ be separable reproducing-kernel Hilbert spaces, with continuous, universal reproducing kernels $\kappa_1',\kappa_2'$.  Let $\kappa_1 \!:\! \mathcal{H}_{\kappa_1'} \!\to\! \mathcal{H}_{\kappa_1'}$, $\kappa_2 \!:\! \mathcal{H}_{\kappa_2'} \!\to\! \mathcal{H}_{\kappa_2'}$ be the corresponding Gram operators.  The mirrored, operator-based R\'{e}nyi's $\alpha$-divergence of $\kappa_1$ and $\kappa_2$, with respect to $\alpha \!\in\! \mathbb{R}_{+} \!\backslash \{1\}$, is
\begin{equation*}
C'_\alpha(\kappa_1\|\kappa_2) = \frac{1}{\alpha \!-\! 1}\textnormal{log}\Bigg(\!\textnormal{tr}\Bigg(\!\kappa_2^{\frac{1-\alpha}{2\alpha}} \kappa_1 \kappa_2^{\frac{1-\alpha}{2\alpha}}\!\Bigg)^{\!\!\alpha}\Bigg) \!-\! \frac{1}{\alpha \!-\! 1}\textnormal{log}\Bigg(\!\textnormal{tr}(\kappa_1)\!\Bigg),\; \textnormal{for } \textnormal{supp}(\kappa_2) \!\subseteq\! \textnormal{supp}(\kappa_1),
\end{equation*}
and $C'_\alpha(\kappa_1\|\kappa_2) \!=\! \infty$ otherwise.  $\kappa_2^{(1-\alpha)/2\alpha} \kappa_1 \kappa_2^{(1-\alpha)/2\alpha} \!=\! \gamma$ is a bivariate Gram operator.
\end{itemize}

This operator-based cross-entropy also satisfies the six fundamental axioms that R\'{e}nyi outlined for divergences, amongst other properties.  We provide more details in the appendix (see \hyperref[secA.1]{Appendix A.1}).

\phantomsection\label{sec3.2}
\subsection*{\small{\sf{\textbf{3.2.$\;\;\;$Empirical Estimation of Operator-Based Cross-Entropies}}}}

These cross-entropy measures assume access to the univariate Gram operators and hence the underlying probability distributions.  In practice, the distributions underlying the operators will only be known through random samples.  Therefore, the operators $\kappa$ and $\gamma$ can be estimated by $\hat{\kappa}$ and $\hat{\gamma}$ under the condition that $\mathbb{E}_{(s_i,s')\sim \hat{p}_{\mathcal{S}}}[\hat{\kappa}(s_i,s')] \!=\! \kappa$ and\vspace{-0.01cm}\\ \noindent $\mathbb{E}_{(s_i,a_i)\sim \hat{\rho}_{\mathcal{S} \times \mathcal{A}}}[\hat{\gamma}((s_i,a_i),\cdot)] \!=\! \gamma$.

We first show that the univariate Gram operators can be approximated.

\phantomsection\label{def3.4}
\begin{itemize}
\item[] \-\hspace{0.5cm}{\small{\sf{\textbf{Definition 3.4: Univariate Gram Operator Approximation.}}}} Let $\mathcal{H}_{\kappa'}$ be a separable reproducing-kernel Hilbert space, with a continuous, universal reproducing kernel.  An unbiased empirical estimate $\hat{\kappa}(f,g)$,\\ \noindent $f,g \!\in\! \mathcal{H}_{\kappa'}$, of the symmetric, self-adjoint operator $\kappa$ is
\begin{equation*}
\hat{\kappa}(f,g) = \langle f,\hat{\kappa} g \rangle_{\mathcal{H}_{\kappa'}} \!=\! \int\!\!\!\!\int_{\mathcal{S}} \langle f, \varphi(s_i) \rangle_{\mathcal{H}_{\kappa'}} \langle \varphi(s'),g \rangle_{\mathcal{H}_{\kappa'}} d\hat{p}_{\mathcal{S}}(s_i,s')
\end{equation*}
which evaluates to $n^{-1}\sum_{i=1}^n \langle f, \varphi(s_i) \rangle_{\mathcal{H}_{\kappa'}} \langle \varphi(s'),g \rangle_{\mathcal{H}_{\kappa'}}$, where $s_i \!\in\! \mathcal{S}$, $i \!\in\! \mathbb{N}_{1,n}$, are samples.  This corre-\\ \noindent sponds to replacing $p_{\mathcal{S}}$ with an empirical version $\hat{p}_{\mathcal{S}} \!\in\! \mathcal{M}^1_+(\mathcal{S})$ formed by a series of $n$ Dirac measures $\delta_{s_i}$,\\ \noindent $\hat{p}_{\mathcal{S}} \!=\! \sum_{i=1}^n \delta_{s_i}/n$.
\end{itemize}
As well, the bivariate Gram operators can be approximated.

\phantomsection\label{def3.5}
\begin{itemize}
\item[] \-\hspace{0.5cm}{\small{\sf{\textbf{Definition 3.5: Bivariate Gram Operator Approximation.}}}} Let $\mathcal{H}_{\kappa_1'}$, $\mathcal{H}_{\kappa_2'}$ be separable reproducing-kernel Hilbert spaces, with continuous, universal reproducing kernels $\kappa_1',\kappa_2'$.  An unbiased empirical estimate $\hat{\gamma}(f,g)$, $f,g \!\in\! \mathcal{H}_{\kappa_1'} \!\otimes\! \mathcal{H}_{\kappa_2'}$, of the symmetric, self-adjoint operator $\gamma$ is
\begin{equation*}
\hat{\gamma}(f,g) =\! \int\!\!\!\!\int_{\mathcal{S} \times \mathcal{A}} \langle f, \varphi^{\otimes}((s_i,a_i),\cdot) \rangle_{\mathcal{H}_{\kappa_1'} \otimes \,\mathcal{H}_{\kappa_2'}} \langle \varphi^{\otimes}((s_i,a_i),\cdot),g \rangle_{\mathcal{H}_{\kappa_1'} \otimes \,\mathcal{H}_{\kappa_2'}} d\hat{\rho}_{\mathcal{S} \times \mathcal{A}}((s_i,a_i),\cdot),
\end{equation*}
which evaluates to $n^{-1}\sum_{i=1}^n \langle f, \varphi^\otimes((s_i,a_i),\cdot) \rangle_{\mathcal{H}_{\kappa'_1} \otimes \mathcal{H}_{\kappa'_2}} \langle \varphi^\otimes((s_i,a_i),\cdot),g \rangle_{\mathcal{H}_{\kappa'}}$, where $s_i \!\in\! \mathcal{S}$, $a_i \!\in\! \mathcal{A}$,\vspace{-0.035cm}\\ \noindent $i \!\in\! \mathbb{N}_{1,n}$ are samples. This corresponds to replacing $\rho_{\mathcal{S} \times \mathcal{A}}$ with an empirical version \noindent $\hat{\rho}_{\mathcal{S} \times \mathcal{A}} \!\in\! \mathcal{M}^1_+(\mathcal{S} \!\times\! \mathcal{A})$ formed\vspace{-0.015cm}\\ \noindent by a series of $n$ Dirac measures.
\end{itemize}

Due to the connection between the reproducing kernels and the measurable feature maps, we can substitute, for the Gram operator $\kappa$, a non-negative, scaled Gram matrix $[\hat{\kappa}]_{i,j} \!\in\! \mathbb{R}_{0,+}$, $i,j \!\in\! \mathbb{N}_{1,n}$, constructed by all pairwise evalua-\\ \noindent tions of a normalized kernel $\kappa'$ to random samples, $s_i,s_j \!\in\! \mathcal{S}$, $i,j \!\in\! \mathbb{N}_{1,n}$.  Alternatively, for undirected graphs, a\\ \noindent kernel can be applied to the entries of square, symmetric adjacency matrices.  Not all kernels can be applied directly to all matrices, though.

\phantomsection\label{prop3.1}
\begin{itemize}
\item[] \-\hspace{0.5cm}{\small{\sf{\textbf{Proposition 3.1: Univariate Gram Operator and Gram Matrix Relationship.}}}} Let $[\hat{\kappa}]_{i,j} \!=\! \langle \varphi(s_i),\varphi(s_j) \rangle_{\mathcal{H}_{\kappa'}}$ be an empirical Gram matrix, $\hat{\kappa} \!\in\! \mathbb{R}_{0,+}^{n \times n}$, for a measurable feature map $\varphi$.  Assume that the corresponding kernel $\kappa'$ is continuous and universal.

Let $\hat{\kappa}$ be an unbiased empirical estimate of the univariate Gram operator $\kappa$.  $\hat{\kappa}$ has at most $n$ positive eigenvalues that satisfy $[\hat{\kappa}] \beta_i/n \!=\! \lambda^1_i\beta_i$, where $\lambda_i \!\in\! \mathbb{R}_+$ are eigenvalues and $\beta_i \!\in\! \mathbb{R}^n$ are eigenvectors.  As well, $n\lambda_i$ are all positive eigenvalues of $[\hat{\kappa}]$ under the assumption that the Gram matrices are normalized as $n^{-1}\sum_{i=1}^n [\hat{\kappa}]_{i,i} \!=\! 1$.

\end{itemize}
We impose a trace normalization constraint on the Gram matrices ensures that they will be infinitely divisible.  This permits us to construct notions of operator- and matrix-based entropy and mutual information via our cross-entropies.  We discuss this topic more in the appendix (see \hyperref[secA.1]{Appendix A.1}).

We can, similarly, substitute a non-negative Gram matrix $[\hat{\gamma}]_{i,j} \!\in\! \mathbb{R}_{0,+}$, $i,j \!\in\! \mathbb{N}_{1,n}$, for the bivariate Gram operator $\gamma$.  $[\hat{\gamma}]$ corresponds to the non-linear combination of univariate Gram matrices $[\hat{\kappa}]$.

\phantomsection\label{prop3.2}
\begin{itemize}
\item[] \-\hspace{0.5cm}{\small{\sf{\textbf{Proposition 3.2: Bivariate Gram Operator and Gram Matrix Relationship.}}}} Let $\hat{\gamma} \!\in\! \mathbb{R}_{0,+}^{n \times n}$ be an empirical Gram matrix, $[\hat{\gamma}]_{i,j} \!=\! \langle \varphi^\otimes(s_i,a_i),\varphi^\otimes(s_j,a_j) \rangle_{\mathcal{H}_{\kappa'_1} \otimes \mathcal{H}_{\kappa'_2}}$, for a measurable feature map $\varphi^\otimes$.  Assume that the\\ \noindent corresponding kernels $\kappa_1',\kappa_2'$ are continuous and universal.

Let $\hat{\gamma}$ be an unbiased empirical estimate of the univariate Gram operator $\gamma$.  $\hat{\gamma}$ has at most $n$ positive eigen-\\ \noindent values that satisfy $[\hat{\gamma}] \beta_i \!=\! \tau_i\beta_i$, where $\lambda_i \!\in\! \mathbb{R}_+$ are eigenvalues and $\beta_i \!\in\! \mathbb{R}^n$ are eigenvectors.  As well, $\tau_i$ are all positive eigenvalues of $[\hat{\gamma}]$.
\end{itemize}

Performing this substitution yields matrix-based $\alpha$-cross-entropies, which are stated in \hyperref[alg:1]{Algorithms 3.1} and \hyperref[alg:1]{3.2}.  Note that, due to the Gram-matrix scaling constraint, $n^{-1}\sum_{i=1}^n [\kappa_1]_{i,i} \!=\! 1$, the second log term evaluates to zero for both cross-entropies.

It is important to note that both the kernel choice and kernel parameters greatly influence the entries of the Gram matrices.  It hence influences the cross-entropy magnitudes.  We refer to \cite{FukumizuK-coll2010a,GrettonA-coll2012a} for discussions about kernel choices.  We also refer to \cite{GrettonA-coll2007a,GrettonA-coll2009a,GrettonA-jour2012a} for discussions about kernel parameters, at least in the context of using the $\alpha$-cross-entropies as a two-sample test \cite{EricM-coll2008a}.  For the general problem of learning kernels, see \cite{ArgyriouA-conf2006a,XuZ-coll2009a,KloftM-coll2010a}.

\begin{figure*}\label{alg:1}
\vspace{-0.5cm}
\hspace{-0.2cm}\begin{tabular}{c}
\imagetop{\parbox{0.465\linewidth}{
{\singlespacing\begin{algorithm}[H]
\DontPrintSemicolon
\SetAlFnt{\small} \SetAlCapFnt{\small}
\caption{Non-Mirrored Matrix-Based R\'{e}nyi's $\alpha$-Cross-Entropy $C_\alpha([\hat{\kappa}_1]\|[\hat{\kappa}_2])$}
\AlFnt{\small}\KwData{$\alpha \!\in\! \mathbb{R}_{0,+} \!\backslash\{1\}$:\! Order\! of\! the\! cross\! entropy.}
\AlFnt{\small}\KwIn{$[\hat{\kappa}_1],[\hat{\kappa}_2] \!\in\! \mathbb{R}_{0,+}^{n \times n}$:\! unit-trace-normalized,\! properly\! conditioned\! Gram\! matrices\! with\! the\! constraint\! $\textnormal{supp}([\hat{\kappa}_2]) \!\subseteq\! \textnormal{supp}([\hat{\kappa}_1])$.\vspace{0.2cm}}
\AlFnt{\small} Return\! $\displaystyle C_\alpha([\hat{\kappa}_1]\|[\hat{\kappa}_2]) \!=\! \frac{1}{\alpha \!-\! 1}\textnormal{log}\Bigg(\!\textnormal{tr}([\hat{\kappa}_1]^{\alpha} [\hat{\kappa}_2]^{1-\alpha})\!\Bigg)$.\;\vspace{0.05cm}
\end{algorithm}}}}

\hspace{0.01cm}

\imagetop{\parbox{0.51\linewidth}{
{\singlespacing\begin{algorithm}[H]
\DontPrintSemicolon
\SetAlFnt{\small} \SetAlCapFnt{\small}
\caption{Mirrored Matrix-Based R\'{e}nyi's $\alpha$-Cross-Entropy $C'_\alpha([\hat{\kappa}_1]\|[\hat{\kappa}_2])$}
\AlFnt{\small}\KwData{$\alpha \!\in\!  \mathbb{R}_{0,+} \!\backslash\{1\}$:\! Order\! of\! the\! cross\! entropy.}
\AlFnt{\small}\KwIn{$[\hat{\kappa}_1],[\hat{\kappa}_2] \!\in\! \mathbb{R}_{0,+}^{n \times n}$:\! unit-trace-normalized,\! properly\! conditioned\! Gram\! matrices\! with\! the\! constraint\! $\textnormal{supp}([\hat{\kappa}_2)] \!\subseteq\! \textnormal{supp}([\hat{\kappa}_1])$.\vspace{0.2cm}}
\AlFnt{\small} Return\! $\displaystyle C'_\alpha([\hat{\kappa}_1]\|[\hat{\kappa}_2]) \!=\! \frac{1}{\alpha \!-\! 1}\textnormal{log}\Bigg(\!\textnormal{tr}\Bigg(\![\hat{\kappa}_2]^{\frac{1-\alpha}{2\alpha}} [\hat{\kappa}_1] [\hat{\kappa}_2]^{\frac{1-\alpha}{2\alpha}}\!\Bigg)^{\!\!\alpha}\,\Bigg)$.\;\vspace{-0.3cm}
\end{algorithm}}}}
\end{tabular}\vspace{-0.6cm}
\end{figure*}

Replacing the empirical Gram operators with matrices is not arbitrary.  As we noted above, the spectral characteristics of the empirical Gram operators and Gram matrices are consistent.  The trace function therefore returns a proportionally equivalent result in either case, which lends credence to using easily computable Gram matrices.

\phantomsection\label{prop3.3}
\begin{itemize}
\item[] \-\hspace{0.5cm}{\small{\sf{\textbf{Proposition 3.3: Univariate Gram Matrix Convergence Properties.}}}} Let $\mathcal{H}_{\kappa'}$ be a separable reproducing-kernel Hilbert space.  Let $\hat{\kappa}$ be an approximate Gram operator and $[\hat{\kappa}] \!\in\! \mathbb{R}_{0,+}^{n \times n}$ be the corresponding normal-\\ \noindent ized Gram matrix.  Assume that the corresponding kernel $\kappa'$ is continuous and universal.  We have that, for $\alpha \!\in\! \mathbb{R}_+ \!\backslash \{1\}$,
\begin{itemize}
   \item[] \-\hspace{0.5cm}(i) $\textnormal{tr}(\hat{\kappa}) \!=\! \textnormal{tr}([\hat{\kappa}])$.
   \item[] \-\hspace{0.5cm}(ii) For the univariate operator $\kappa$, the inequality $|\textnormal{tr}(\kappa) \!-\! \textnormal{tr}(\hat{\kappa})| \!\leq\! c_1 (2\textnormal{log}(2/\delta)/n)^{1/2}$ is satisfied with\\ \noindent probability $1\!-\!\delta$, where the kernel magnitudes obey $\kappa' \!\leq\! c_1$, with $c_1 \!\in\! \mathbb{R}_+$.  Here, $\textnormal{tr}(\kappa) \!=\! \sum_{j=1}^h \langle \eta_j,\kappa \eta_j \rangle_{\mathcal{H}_{\kappa'_1}}$, where $\eta_j$ are elements of the orthonormal basis for $\mathcal{H}_{\kappa'}$.
  \item[] \-\hspace{0.5cm}(iii) The inequality $\textnormal{inf}_{\hat{\kappa}}\, \textnormal{sup}_{\kappa}\, \|\textnormal{tr}([\hat{\kappa}]) \!-\! \textnormal{tr}(\kappa)\|_{\mathcal{H}_{\kappa'}} \!\geq\! c_2(2c_3/n)^{1/2}$ is satisfied with non-zero probability,\\ \noindent where $c_2,c_3 \!\in\! \mathbb{R}_+$.  Hence, the trace of the univariate Gram matrix is minimax optimal.
\end{itemize}
\end{itemize}

\phantomsection\label{prop3.4}
\begin{itemize}
\item[] \-\hspace{0.5cm}{\small{\sf{\textbf{Proposition 3.4: Bivariate Gram Matrix Convergence Properties.}}}} Let $\mathcal{H}_{\kappa'_1}$, $\mathcal{H}_{\kappa'_2}$ be separable reproducing-kernel Hilbert spaces.  Let $\hat{\kappa_1},\hat{\kappa_2}$ be approximate univariate Gram operators and $[\hat{\kappa}_1],[\hat{\kappa}_2] \!\in\! \mathbb{R}_{0,+}^{n \times n}$ be the corresponding normalized Gram matrices.  Assume that the corresponding kernel $\kappa'_1,\kappa_2'$ are continuous and universal.  Moreover, assume that $\hat{\gamma}$ and $[\hat{\gamma}] \!\in\! \mathbb{R}_{0,+}^{n \times n}$ are, respectively, continuous combinations of $\hat{\kappa}_1,\hat{\kappa}_2$ and $[\hat{\kappa}_1],[\hat{\kappa}_2]$.  We have that
\begin{itemize}
   \item[] \-\hspace{0.5cm}(i) $\textnormal{tr}(\hat{\gamma}) \!=\! \textnormal{tr}([\hat{\gamma}])$, where $[\hat{\gamma}] \!=\! [\hat{\kappa}_1]^\alpha[\hat{\kappa}_2]^{1-\alpha}$ or $[\hat{\gamma}] \!=\! [\kappa_2]^{(1-\alpha)/2\alpha} [\kappa_1] [\kappa_2]^{(1-\alpha)/2\alpha}$, with $\hat{\gamma}$ having ana-\\ \noindent logous forms.  
   \item[] \-\hspace{0.5cm}(ii) For the bivariate operator $\gamma$, the inequality $|\textnormal{tr}(\gamma) \!-\! \textnormal{tr}(\hat{\gamma})| \!\leq\! \alpha c_1c_2c_3 (2\textnormal{log}(2/\delta)/n)^{1/2}$ is satisfied with probability $1 \!-\! \delta$.  We assume that the kernel magnitudes obey $\kappa'_1 \!\leq\! c_1$ and $\kappa'_2 \!\leq\! c_2$, with $c_1,c_2 \!\in\! \mathbb{R}_+$.  $c_3 \!\in\! \mathbb{R}_+$ takes the value $c_3 \!=\! 1$ for the non-mirrored case and $c_3 \!=\! \alpha$ for the mirrored case.  For both cross-entropies,\vspace{-0.02cm}\\ \noindent $\textnormal{tr}(\gamma) \!=\! \sum_{q=1}^r \langle \pi_q,\gamma \pi_q \rangle_{\mathcal{H}_{\kappa'_1} \otimes \mathcal{H}_{\kappa'_2}}$, where $\pi_q$ are elements of the orthonormal basis for $\mathcal{H}_{\kappa'_1} \!\otimes\! \mathcal{H}_{\kappa'_2}$.
  \item[] \-\hspace{0.5cm}(iii) The inequality $\textnormal{inf}_{\hat{\gamma}}\, \textnormal{sup}_{\gamma}\, \|\textnormal{tr}(\hat{\gamma}) \!-\! \textnormal{tr}(\gamma)\|_{\mathcal{H}_{\kappa'_1} \otimes \mathcal{H}_{\kappa'_2}} \!\geq\! c_4(2c_5/n)^{1/2}$ is satisfied with non-zero\vspace{-0.02cm}\\ \noindent probability, for constants $c_4,c_5 \!\in\! \mathbb{R}_+$.  Hence, the trace of the bivariate Gram matrix is minimax optimal.
\end{itemize}
\end{itemize}

We can thus deduce that Gram matrices, when constructed from universal kernels, provide a sufficiently good characterization of the empirical distributions.  This matrix-based cross-entropies hence can assess differences between probability distributions without needing direct access to those distributions.

It is important to notice that the convergence rate of the Gram matrices to the Gram operators is independent of the dimensionality of the samples.  It is, instead, primarily a function of the number of samples used in the estimation process.  This property makes our matrix-based notions of cross-entropy appealing for many applications.  Moreover, our measures avoid the concerns for plug-in density estimators \cite{AntosA-jour2001a}.  Namely, our measures do not have slow-rate-of-convergence issues for arbitrary distributions \cite{BirgeL-jour1986a,DevroyeL-jour1995a}.

\phantomsection\label{sec3.3}
\subsection*{\small{\sf{\textbf{3.3.$\;\;\;$Numerical Examples}}}}

\begin{figure*}
   \hspace{0.85cm}\includegraphics[width=6in]{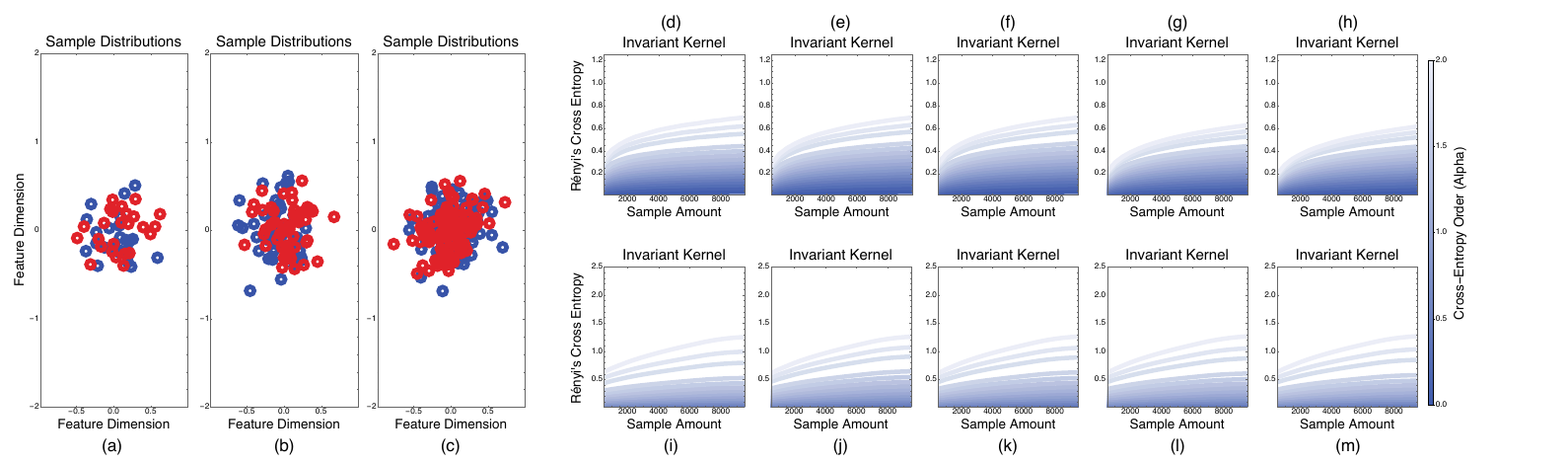}\vspace{-0.225cm}\\
   \caption[]{\fontdimen2\font=1.55pt\selectfont An overview of the sample-convergence properties for matrix-based cross-entropies.  (a)--(c) Two-dimensional scatter plots for samples drawn from two Gaussian distributions with equivalent means and variances.  Only the number of samples differs in the scatter plots, with the left-most plot having the fewest samples and the right-most plot having the most samples.  (d)--(h) plots of $C_\alpha([\hat{\kappa}_1]\|[\hat{\kappa}_2])$ as a function of the cross-entropy order, the number of samples, and the dimensionality of the distributions.  We consider (d) 2-dimensional, (e) 10-dimensional, (f) 25-dimensional, (g) 50-dimensional, and (h) 100-dimensional distributions.  Here we used a translation-invariant, Gaussian kernel, $\kappa'(s,s') \!=\! \textnormal{exp}(-\sigma \langle s \!-\! s',\cdot\rangle)$, $\sigma \!=\! 1$, for both $\kappa_1'$ and $\kappa_2'$.  (i)--(m) plots of $C_\alpha'([\hat{\kappa}_1]\|[\hat{\kappa}_2])$ as a function of the cross-entropy order, the number of samples, and the dimensionality of the distributions.  We consider the same distributional dimensionality as in (d)--(h).  We again used a translation-invariant, Gaussian kernel.  These results show that the convergence rate of the matrix-based cross-entropies is independent of the sample dimensionality.\vspace{-0.4cm}}
\label{fig:conv}
\end{figure*}

{\small{\sf{\textbf{Dimension Agnosticity Analysis.}}}} The theory that we outlined above posits that the convergence rate of the Gram-matrix-based cross-entropy to the Gram-operator-based cross-entropy is independent of the vector-sample dimensionality.  It is informative to show that this behavior occurs.  Examples of it are displayed in \hyperref[fig:conv]{Figure 3.1}.

In \hyperref[fig:conv]{Figure 3.1}, we plot cross-entropy magnitudes for our bipartite, mirrored and non-mirrored measures.  We draw vector samples from Gaussian distributions (red and blue) with near-equivalent variances and means.  The same number of samples are used for each distribution.  We vary the distribution dimensionality between two and a hundred.  Snapshots of the two-dimensional samples, for increasing sample amounts, are shown in \hyperref[fig:conv]{Figure 3.1}(a)--(c).  For each set of samples, we compute the pairwise sample distances and use the Gaussian kernel to form the Gram matrices.  We fix the kernel bandwidth to a constant, positive value.

Due to the high amount of distributional overlap, we expect that the cross-entropy magnitudes will be low.  They will not, however, necessarily be zero, since the sample sets are not equivalent.  The results in \hyperref[fig:conv]{Figure 3.1}(d)--(h) and (i)--(m) are aligned with these expectations, regardless of the sample dimensionality.  They also highlight the minimax convergence rate guarantees.  In each case, the cross-entropy magnitude grows proportionally to the square root of the number of samples.  The growth rate is additionally a function of the cross-entropy order.  This is anticipated, since the constant terms in the minimax bounds depend on this parameter.  Moreover, the cross-entropy curves are remarkably consistent, regardless of the dimensionality.

\begin{figure*}
   \hspace{0.35cm}\includegraphics[width=6in]{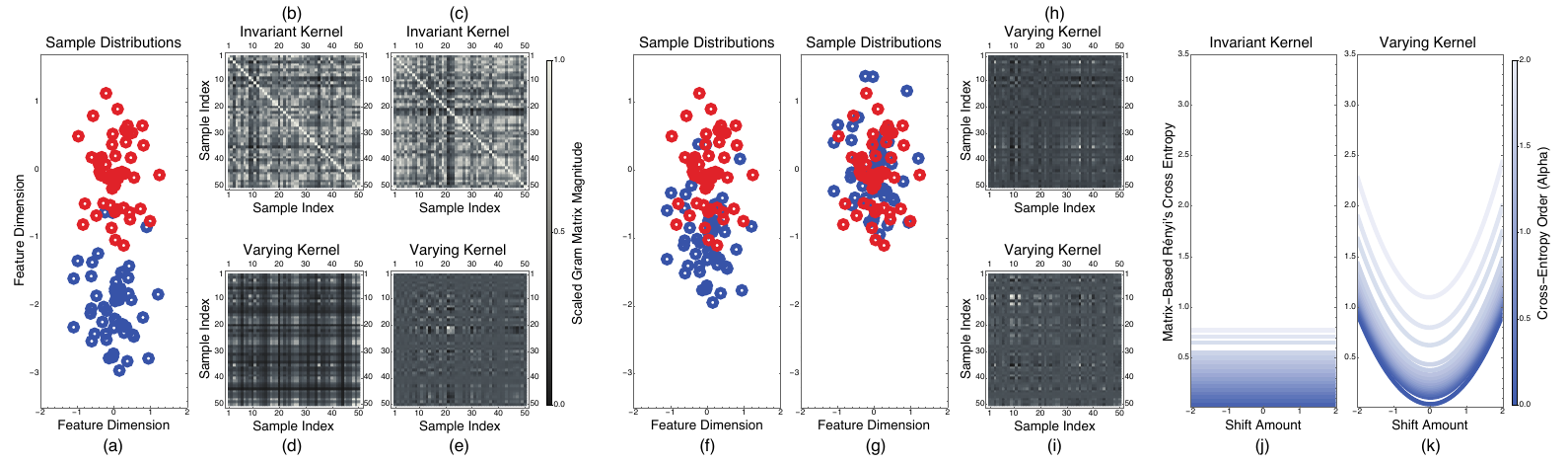}\vspace{-0.2cm}\\
   \caption[]{\fontdimen2\font=1.55pt\selectfont An overview of the importance of kernel choice when considering distributional shifts.  (a) A two-dimensional scatter plot for samples drawn from two Gaussian distributions with equivalent variances but different means.  (b)--(c) Empirical Gram matrices $\hat{\kappa}_1$ and $\hat{\kappa}_2$, respectively, for the blue and red distributions shown in (a).  Here a translation-invariant, Gaussian kernel was used, $\kappa'(s,s') \!=\! \textnormal{exp}(-\sigma \langle s \!-\! s',\cdot\rangle)$, $\sigma \!=\! 1$.  (d)--(e) Empirical Gram matrices $\hat{\kappa}_1$ and $\hat{\kappa}_2$, respectively, for the blue and red distributions shown in (a).  Here a translation-varying, exponential-inner- product kernel was used, $\kappa'(s,s') \!=\! \textnormal{exp}(\sigma s^\top\! s')$, $\sigma \!=\! 1$.  In (f)--(g), we shift the mean of the samples from the blue distribution in (a).  This shift is not captured by the translation-invariant kernel.  Only the Gram matrices for the translation-varying kernel change.  In (h) and (i), we plot the empirical Gram matrices $\hat{\kappa}_1$ associated with the blue distribution in (f)--(g).  The empirical Gram matrices $\hat{\kappa}_2$ associated with the red distribution in (f)--(g) stay the same, since the distribution does not change.  (j)--(k) Plots of the matrix-based R\'{e}nyi's $\alpha$-cross-entropy $C_\alpha(\hat{\kappa}_1\|\hat{\kappa}_2)$ as a function of the mean shift amount of the blue distribution.  The $\alpha$-cross-entropy is constant for the translation-invariant kernel.  It changes for the translation- varying kernel.  The lowest $\alpha$-cross-entropy magnitude is achieved when the means of the two distributions are equivalent, as shown in (g).  This result aligns with our expectation.  It hence demonstrates that translation-varying kernels should be employed when using either $C_\alpha(\hat{\kappa}_1\|\hat{\kappa}_2)$ or $C_\alpha'(\hat{\kappa}_1\|\hat{\kappa}_2)$ if distributional offsets are to be captured in addition to distributional shape changes.\vspace{-0.4cm}}
\label{fig:transinv}
\end{figure*}

\begin{figure*}
   \hspace{0.35cm}\includegraphics[width=6in]{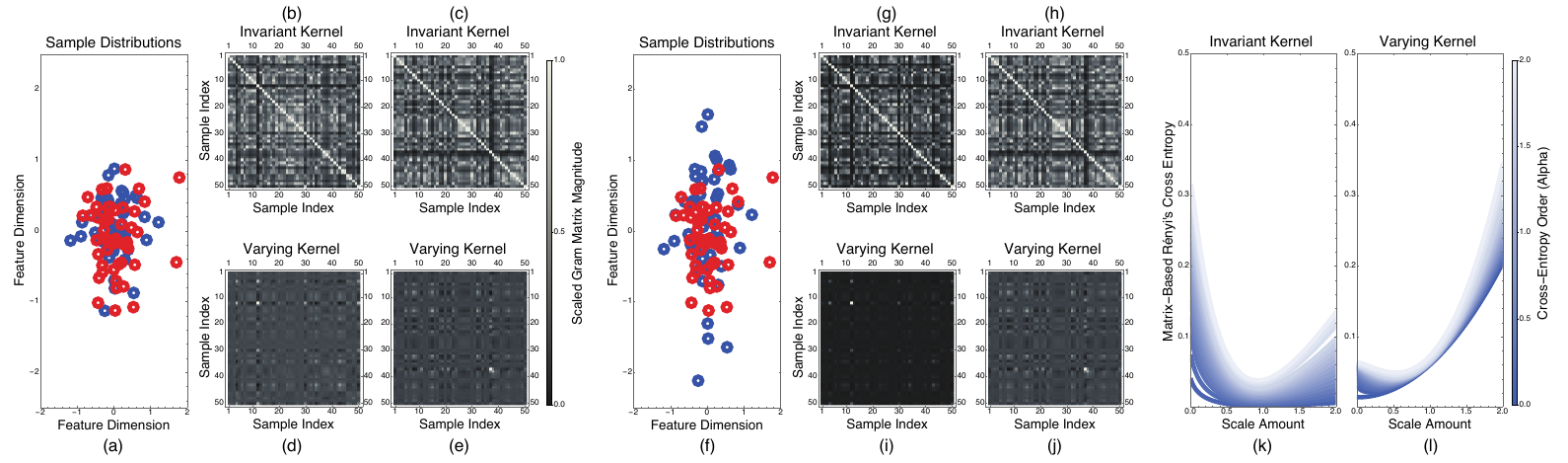}\vspace{-0.2cm}\\
   \caption[]{\fontdimen2\font=1.55pt\selectfont An overview of cross-entropy estimation when considering distributional shape changes.  (a) A two-dimensional scatter plot for samples drawn from two Gaussian distributions with equivalent variances and means.  (b)--(c) Empirical Gram matrices $\hat{\kappa}_1$ and $\hat{\kappa}_2$, respectively, for the blue and red distributions shown in (a).  Here a translation-invariant, Gaussian kernel was used, $\kappa'(s,s') \!=\! \textnormal{exp}(-\sigma \langle s \!-\! s',\cdot\rangle)$, $\sigma \!=\! 1$.  (d)--(e) Empirical Gram matrices $\hat{\kappa}_1$ and $\hat{\kappa}_2$, respectively, for the blue and red distributions shown in (a).  Here a translation-varying, exponential-inner-product kernel was used, $\kappa'(s,s') \!=\! \textnormal{exp}(\sigma s^\top\! s')$, $\sigma \!=\! 1$.  In (f), we increase the variance of the samples from the blue distribution in (a).  This scale change is captured by the empirical Gram matrices in (g) and (i).  The Gram matrices in (h) and (j) naturally do not change from (c) and (e), since the red distribution is not modified.  (k)--(l) Plots of the matrix-based R\'{e}nyi's $\alpha$-cross-entropy $C_\alpha(\hat{\kappa}_1\|\hat{\kappa}_2)$ as a function of the variance scale amount of the blue distribution.  The results for $C'_\alpha(\hat{\kappa}_1\|\hat{\kappa}_2)$ are similar to those for $C_\alpha(\hat{\kappa}_1\|\hat{\kappa}_2)$ for this example and hence not plotted.  Regardless of which kernel is employed, the cross-entropies reach a minimal value near the unit-scale case.  This result aligns well with our expectations.  It hence demonstrates that both translation-invariant and translation-varying kernels can be used to quantify changes in the distribution shape.\vspace{-0.4cm}}
\label{fig:transvar}
\end{figure*}

\vspace{0.15cm}{\small{\sf{\textbf{Distribution Assessment Analysis.}}}} It is also informative to demonstrate how well the cross-entropy measures can assess distributional overlap when employing different kernels.  We consider two examples for vector samples drawn from Gaussian distributions.  These examples are presented in \hyperref[fig:transinv]{Figures 3.2}--\hyperref[fig:transvar]{3.3}.

In \hyperref[fig:transinv]{Figures 3.2}, we consider the case of distributional mean shift.  Samples for one distribution (red) are held fixed while the mean for the samples for the other distribution (blue) are shifted.  The same number of samples are used for each distribution.  Snapshots are shown in \hyperref[fig:transinv]{Figures 3.2}(a), (f), and (g).  Both distributions have the same variance.  We hence expect that the matrix-based, bipartite, mirrored $\alpha$-cross-entropy should reach a minimum value whenever the distributional means are equivalent.  The cross-entropy should gradually rise as the means diverge and hence the distributions have increasingly minimal overlap.  This behavior is observed in \hyperref[fig:transinv]{Figure 3.2}(k) when using translation-varying kernels, like the exponential inner-product kernel.  It does not, however, emerge for translation-invariant kernels, like the popular Gaussian kernel, which can be seen in \hyperref[fig:transinv]{Figure 3.2}(j).  In both cases, we plot the cross-entropy as a function of the mean shift amount and cross-entropy order.

There is a simple explanation for this discrepancy between expected and observed behavior.  The bipartite, non-mirrored cross-entropy does not account for the cross-correlation of samples from both distributions.  Rather, it relies on Gram matrices that correspond to autocorrelations of samples from only a single distribution.  If the kernel is insensitive to uniform translations of the samples, then the Gram matrices, like those in \hyperref[fig:transinv]{Figures 3.2}(b) and (c), will remain constant.  The cross-entropy measure will stay the same too.  This is due to the properties imparted by the kernel on the inner product of the reproducing-kernel Hilbert space.  Only for a translation-varying kernel can the autocorrelation Gram matrices potentially change for this type of distributional adjustment.  This is illustrated in \hyperref[fig:transinv]{Figures 3.2}(e), (h), and (i).

In \hyperref[fig:transvar]{Figure 3.3}, we consider distributional variance changes.  Samples for one distribution (red) are held fixed while the variance for the samples from the other distribution (blue) are adjusted uniformly for each dimension.  Snapshots of this process are shown in \hyperref[fig:transvar]{Figures 3.3}(a) and (f).  Both distributions have the same mean.  We therefore anticipate that the matrix-based bipartite $\alpha$-cross-entropy should achieve a maximal value when the variance is zero and the samples from one distribution collapse to a delta function.  It should monotonically decrease as the variance is raised until the distributional spreads are equal.  Further increases in variance should yield monotonic increases in the cross-entropy.  This occurs in \hyperref[fig:transvar]{Figure 3.3}(k) for the translation-invariant Gaussian kernel.  A similar trend is observed in \hyperref[fig:transvar]{Figure 3.3}(k) for the translation-varying exponential-inner-product kernel.  However, the mirrored cross-entropy does does not achieve the highest value for the zero-variance case.  In both cases, we plot the cross-entropy as a function of the variance scaling amount and the cross-entropy order.

This example again illustrates the influence of the kernel choice on the estimated cross-entropy.  For the translation-varying kernel that we employed, properties of the inner product are such that samples which are far from the population mean are given a greater emphasis than those closer to it.  This creates a few high-magnitude entries in the autocorrelation-based Gram matrices as the variance scale is increased.  \hyperref[fig:transvar]{Figures 3.3}(d) and (i) show this.  These high-magnitude entries dominate the cross-entropy calculation, leading to increasing cross-entropy values as the scale of one distribution is continuously raised past the equivalence point.  The translation-invariant kernel that we use does not share this property. 

Regardless of the kernel choice, it is important to notice that several theoretical properties of the matrix-based cross-entropies are upheld in these examples.  We clearly have non-negativity and order monotonicity.

Another observation from these experiments is that the cross-entropy measure cannot intrinsically detect when there is no distributional overlap.  The measure will simply continue to increase as the samples are shifted farther apart.  This is not a defect of our measure.  It is simply a byproduct of handling continuous distributions using finite samples.  Moreover, it may be possible that, after drawing additional samples, distributional overlap is achieved, even if the sample means are far apart.  This can occur, for instance, with distributions that have infinite support.  It is thus not always appropriate to conclude that there is no distributional overlap.

\phantomsection\label{sec4}
\subsection*{\small{\sf{\textbf{4.$\;\;\;$Tripartite Cross-Entropy}}}}\addtocounter{section}{1}

The bipartite cross-entropies that we define have issues that prevent their use in certain circumstances.  For instance, they assume that the sizes of the empirical Gram matrices are equivalent.  They hence codify the relationships between distributions described by the same number of samples.  

In some instances, investigators may wish to assess cross-entropies for differing-cardinality sets.  We therefore introduce a third cross-entropy measure that overcomes this issue (see \hyperref[def4.1]{Definition 4.1}).  It is based on our lab's notion of the cross-information potential \cite{XuJ-jour2008a}.  We refer to this measure as the tripartite $\alpha$-cross-entropy.  This measure utilizes a joint, bivariate operator that performs cross-correlations of functions.  It also relies on univariate operators.

This measure offers other advantages.  Due to the use of a joint, bivariate operator, translation-invariant kernels can be employed within the tripartite $\alpha$-cross-entropy to assess distributional shape and shift changes.  In contrast, both bipartite $\alpha$-cross-entropies can only quantify distributional shape differences when using translation-invariant kernels.  This stems from properties of the non-joint, bivariate operators.  Translation-varying kernels are needed, for these bipartite measures, to quantify distributional shifts and hence how much they overlap.  Some translation-varying kernels may alter the properties of the inner-product in a way that yields unexpected cross-entropy behaviors, however.  Translation-invariant kernels, like the Gaussian kernel, tend yield cross-entropy magnitudes that align better with intuition.  We hence would prefer to use them whenever possible. 

We must note, though, that due to its construction, the measure cannot be readily applied to purely matrix- or graph-based modalities.  Rather, it relies on access to vector-based samples, which are required to empirically approximate the cross-correlation operator using Gram matrices.  Such vector-based samples can, however, usually be uncovered via multi-dimensional scaling.  Moreover, the pairwise distances between the vector-based samples often resemble well the original matrix entries.  This is thus not an overly restrictive condition.

For the bipartite cross-entropies, we rely on the spectral consistency of the approximate Gram operators and Gram matrices to initially motivate our empirical formulation.  For the tripartite cross-entropies, we only have spectral consistency in some cases (see \hyperref[def4.2]{Definition 4.2} and \hyperref[prop4.1]{Proposition 4.1}).  This is because we may have differing-cardinality sets, which preclude eigenanalyses.  We therefore primarily rely on minimax optimality of the measure (see \hyperref[prop4.2]{Proposition 4.2}) to motivate its use.  Our minimax optimality guarantee is limited currently to just radial universal kernels, though.  While the set of such kernels does encompass many popular choices, such as the Gaussian kernel, it does not include all universal kernels.  It is an open problem as to if this constraint can be relaxed while still maintaining a convergence rate that is independent of the sample dimensionality.

\phantomsection\label{sec4.1}
\subsection*{\small{\sf{\textbf{4.1.$\;\;\;$Operator-Based Cross-Entropies}}}}

For non-commutative Gram operators, we specify the tripartite R\'{e}nyi's $\alpha$-cross-entropy below.  We refer to it as such since it has three operator-based arguments.

\phantomsection\label{def4.1}
\begin{itemize}
\item[] \-\hspace{0.5cm}{\small{\sf{\textbf{Definition 4.1: Tripartite R\'{e}nyi's $\alpha$-Cross-Entropy.}}}} Let $\mathcal{H}_{\kappa_1'}$, $\mathcal{H}_{\kappa_2'}$ be separable reproducing-kernel Hilbert spaces, with continuous, universal reproducing kernels $\kappa_1',\kappa_2'$.  Let $\kappa_1 \!:\! \mathcal{H}_{\kappa_1'} \!\to\! \mathcal{H}_{\kappa_1'}$, $\kappa_2 \!:\! \mathcal{H}_{\kappa_2'} \!\to\! \mathcal{H}_{\kappa_2'}$.  Let\\ \noindent $\kappa_{1,2} \!:\! \mathcal{H}_{\kappa_1'} \!\otimes^*\! \mathcal{H}_{\kappa_2'} \!\to\! \mathcal{H}_{\kappa_1'} \!\otimes^*\! \mathcal{H}_{\kappa_2'}$.  The tripartite R\'{e}nyi's $\alpha$-cross-entropy of $\kappa_1$, $\kappa_{1,2}$, and $\kappa_2$, for $\alpha \!\in\! \mathbb{R}_{+} \!\backslash \{1\}$, is
\begin{equation*}
C''_\alpha(\kappa_1\|\kappa_{1,2}\|\kappa_2) = \frac{1}{\alpha \!-\! 1}\textnormal{log}\Bigg(\!\mathbb{E}_{(s,s') \sim p_\mathcal{S}}[\kappa_1] \!+\! \mathbb{E}_{(a,a') \sim p_\mathcal{A}}[\kappa_2] \!-\! 2\mathbb{E}_{(s,a) \sim \rho_{\mathcal{S} \times \mathcal{A}}}[\kappa_{1,2}] \!\Bigg) \!+\! \frac{1}{\alpha \!-\! 1}\textnormal{log}\Bigg(\!\textnormal{tr}(\kappa_1^\alpha)\!\Bigg),
\end{equation*}
for $\textnormal{supp}(\kappa_2) \!\subseteq\! \textnormal{supp}(\kappa_1)$, and $C''_\alpha(\kappa_1\|\kappa_{1,2}\|\kappa_2) \!=\! \infty$ otherwise.  $\kappa_{1,2} \!=\! \lambda$ is a joint, bivariate Gram operator. 
\end{itemize}
We take inspiration from Shannon's cross-entropy for this cross-entropy measure.  The first term, an operator-based cross-information potential, acts like a divergence.  The second term is an operator-based entropy. 

Much like the bipartite measures, the tripartite version satisfies the six axioms that R\'{e}nyi proposed for divergences.  It thus, theoretically, behaves as a divergence.  Practically, it does too.  We elaborate on these properties and others in the appendix (see \hyperref[secA.1]{Appendix A.1}).

\phantomsection\label{sec4.2}
\subsection*{\small{\sf{\textbf{4.2.$\;\;\;$Empirical Estimation of Operator-Based Cross-Entropies}}}}

As with the previous two measures, we approximate the operators.  We have already considered the univariate case, so all that remains is the joint, bivariate case.  The associated operator $\lambda$ can be estimated by $\hat{\lambda}$ subject to the\\ \noindent constraint $\mathbb{E}_{s_i \sim \hat{p}_{\mathcal{S}}, a_j \sim \hat{p}_{\mathcal{A}} }[\hat{\lambda}(s_i,a_j)] \!=\! \lambda$.

\phantomsection\label{def4.2}
\begin{itemize}
\item[] \-\hspace{0.5cm}{\small{\sf{\textbf{Definition 4.2: Joint, Bivariate Gram Operator Approximation.}}}} An unbiased empirical estimate $\hat{\lambda}(f,g)$, $f \!\in\! \mathcal{H}_{\kappa_1'}$, $g \!\in\! \mathcal{H}_{\kappa_2'}$, of the self-adjoint operator $\lambda$ is\vspace{-0.1cm}
\begin{equation*}
\hat{\gamma}(f,g) =\! \int_{\mathcal{S}}\!\int_{\mathcal{A}} \Bigg(\langle f, \varphi(s_i) \rangle_{\mathcal{H}_{\kappa_1'}} \langle \varphi(s_i),\psi(a_j) \rangle_{\mathcal{H}_{\kappa_1'} \otimes \,\mathcal{H}_{\kappa_2'}} \langle \psi(a_j),g \rangle_{\mathcal{H}_{\kappa_2'}}\!\Bigg) dp_{\mathcal{S}}(s_i)dp_{\mathcal{A}}(a_j),
\end{equation*}
which evaluates to $n^{-1}m^{-1}\sum_{i=1}^n\sum_{j=1}^m \langle f, \varphi(s_i) \rangle_{\mathcal{H}_{\kappa'_1}} \langle \varphi(s_i),\psi(a_j) \rangle_{\mathcal{H}_{\kappa_1'} \otimes \,\mathcal{H}_{\kappa_2'}} \langle \psi(a_j),g \rangle_{\mathcal{H}_{\kappa'_1}}$.  Here, $s_i \!\in\! \mathcal{S}$\\ \noindent and $a_j \!\in\! \mathcal{A}$, $i \!\in\! \mathbb{N}_{1,n}$ and $j \!\in\! \mathbb{N}_{1,n}$ are samples. 
\end{itemize}

We can again substitute, for the Gram operator $\lambda$, a non-negative Gram matrix $[\hat{\lambda}]_{i,j} \!\in\! \mathbb{R}_{0,+}$, $i \!\in\! \mathbb{N}_{1,n}$, $j \!\in\! \mathbb{N}_{1,m}$.  This provides a matrix-based estimate of cross-entropy, which is outlined in \hyperref[alg:1]{Algorithms 4.1}.  

Access to vector-based samples is required to form the joint, bivariate Gram matrix.  For data that naturally exist as either graphs or matrices, an embedding to a metric space will need to found first.  If no metric-space embedding of the matrix- or graph-based samples is possible, then the joint operator cannot be formed.  Only our bipartite cross-entropy measures can be applied.

We previously motivated the use of Gram matrices by claiming that they share spectral properties with the corresponding Gram operators.  Below, we do the same for the joint, bivariate Gram operator in the case where $m \!=\! n$.  

\phantomsection\label{prop4.1}
\begin{itemize}
\item[] \-\hspace{0.5cm}{\small{\sf{\textbf{Proposition 4.1: Joint, Bivariate Gram Operator and Gram Matrix Relationship.}}}} Let $\hat{\lambda} \!\in\! \mathbb{R}_{0,+}^{n \times n}$ be an\vspace{-0.025cm}\\ \noindent empirical Gram matrix, $[\hat{\lambda}]_{i,j} \!=\! \langle \varphi(s_i),\psi(a_j) \rangle_{\mathcal{H}_{\kappa'_1} \otimes \mathcal{H}_{\kappa'_2}}$, for measurable feature maps $\varphi,\psi$.  Assume that the\\ \noindent corresponding kernels $\kappa_1',\kappa_2'$ are continuous and universal.

Let $\hat{\lambda}$ be an unbiased empirical estimate of the univariate Gram operator $\lambda$.  $\hat{\lambda}$ has at most $n$ positive eigen-\\ \noindent values that satisfy $[\hat{\lambda}] \beta_i \!=\! \tau_i\beta_i$, where $\tau_i \!\in\! \mathbb{R}_+$ are eigenvalues and $\beta_i \!\in\! \mathbb{R}^n$ are eigenvectors.  As well, $\tau_i$ are all\\ \noindent positive eigenvalues of $[\hat{\lambda}]$.
\end{itemize}

\setcounter{algocf}{0}
\begin{figure*}\label{alg:3}
\vspace{-0.5cm}
\hspace{-0.2cm}
\begin{tabular}{c}
\imagetop{\parbox{1\linewidth}{
{\singlespacing\begin{algorithm}[H]
\DontPrintSemicolon
\SetAlFnt{\small} \SetAlCapFnt{\small}
\caption{Tripartite Matrix-Based R\'{e}nyi's $\alpha$-Cross-Entropy $C_\alpha''([\hat{\kappa}_1]\|[\hat{\kappa}_{1,2}]\|[\hat{\kappa}_2])$}
\AlFnt{\small}\KwData{$\alpha \!\in\! \mathbb{R}_{0,+} \!\backslash\{1\}$:\! Order\! of\! the\! divergence.}
\AlFnt{\small}\KwIn{$[\hat{\kappa}_1] \!\in\! \mathbb{R}_{0,+}^{n \times n}$,\! $[\hat{\kappa}_{1,2}] \!\in\! \mathbb{R}_{0,+}^{n \times m}$,\! $[\hat{\kappa}_2] \!\in\! \mathbb{R}_{0,+}^{m \times m}$: properly\! conditioned\! Gram\! matrices\! with\! the\! constraint\vspace{-0.05cm}\newline $\textnormal{supp}([\hat{\kappa}_2]) \!\subseteq\! \textnormal{supp}([\hat{\kappa}_1])$.\vspace{0.2cm}}
\AlFnt{\small} Return\! $\displaystyle C''_\alpha([\hat{\kappa}_1]\|[\hat{\kappa}_{1,2}]\|[\hat{\kappa}_2]) \!=\! \frac{1}{\alpha \!-\! 1}\textnormal{log}\Bigg(\!\mathbb{E}_{(s,s') \sim p_\mathcal{S}}[[\hat{\kappa}_1]] \!+\! \mathbb{E}_{(a,a') \sim p_\mathcal{A}}[[\hat{\kappa}_2]] \!-\! 2\mathbb{E}_{(s,a) \sim \rho_{\mathcal{S} \times \mathcal{A}}}[[\hat{\kappa}_{1,2}]] \!\Bigg) \!+\! \frac{1}{\alpha \!-\! 1}\textnormal{log}\Bigg(\!\textnormal{tr}([\hat{\kappa}_1]^\alpha)\!\Bigg)$.\;
\end{algorithm}}}}
\end{tabular}\vspace{-0.6cm}
\end{figure*}

For the bipartite cross-entropy measures, we have spectral consistency results that justify our use of Gram matrices.  For the tripartite measure, we do not necessarily have them, though.  This is because an eigenproblem cannot be solved for the joint, bivariate Gram matrix, given that the matrix may be square.  We therefore just rely on minimax optimality of the divergence term to motivate this approximation.

\phantomsection\label{prop4.2}
\begin{itemize}
\item[] \-\hspace{0.5cm}{\small{\sf{\textbf{Proposition 4.2: Tripartite Cross-Entropy Convergence Properties.}}}} Let $\mathcal{H}_{\kappa_1'},\mathcal{H}_{\kappa_2'}$ be separable reproducing-kernel Hilbert spaces with continuous, radial, universal reproducing kernels.  Let ${\kappa}_1$, ${\kappa}_{1,2}$, and ${\kappa}_2$ be Gram oper-\\ \noindent ators and $[\hat{\kappa}_1] \!\in\! \mathbb{R}_{0,+}^{n \times n}$, $[\hat{\kappa}_{1,2}] \!\in\! \mathbb{R}_{0,+}^{n \times m}$, and $[\hat{\kappa}_2] \!\in\! \mathbb{R}_{0,+}^{m \times m}$ be the corresponding Gram matrices, with $n,m \!\in\! \mathbb{N}$.  We have that
\begin{itemize}
   \item[] \-\hspace{0.5cm}(i) $\textnormal{inf}_{[\hat{\kappa}_1]}\, \textnormal{sup}_{{\kappa}_1}\, |\mathbb{E}_{s_i,s_j \sim p_{\mathcal{S}}} [[\hat{\kappa}_1]] \!-\! \mathbb{E}_{s,s' \sim p_{\mathcal{S}}}[\kappa_1]| \!\geq\! c_1 n^{-1/2}$ with non-zero probability, where $c_1 \!\in\! \mathbb{R}_+$.
   \item[] \-\hspace{0.5cm}(ii) $\textnormal{inf}_{[\hat{\kappa}_{1,2}]}\, \textnormal{sup}_{{\kappa}_{1,2}}\, |\mathbb{E}_{s_i,a_j \sim p_{\mathcal{S} \times \mathcal{A}}} [[\hat{\kappa}_{1,2}]] \!-\! \mathbb{E}_{s,a \sim p_{\mathcal{S} \times \mathcal{A}}}[\kappa_{1,2}]| \!\geq\! c_2 (n^{-1/2} \!+\! m^{-1/2})$ with non-zero prob-\\ \noindent ability, where $c_2 \!\in\! \mathbb{R}_+$.
   \item[] \-\hspace{0.5cm}(iii) $\textnormal{inf}_{[\hat{\kappa}_2]}\, \textnormal{sup}_{{\kappa}_2}\, |\mathbb{E}_{a_i,a_j \sim p_{\mathcal{A}}} [[\hat{\kappa}_1]] \!-\! \mathbb{E}_{a,a' \sim p_{\mathcal{A}}}[\kappa_2]| \!\geq\! c_3 m^{-1/2}$ with non-zero probability, where $c_3 \!\in\! \mathbb{R}_+$.
\end{itemize}
This holds regardless of if $m \!<\!n$ or $n \!\leq\! m$.  Therefore, the Gram-matrix-based version of the tripartite R\'{e}nyi's $\alpha$-cross entropy converges at a minimax-optimal rate for $\alpha \!\in\! \mathbb{R}_{+} \!\backslash \{1\}$.
\end{itemize}

Again, this claim shows that we can estimate cross-entropy-like quantities without direct access to the underlying probability distributions.

As with our bipartite cross-entropy measures, the tripartite version converges at a rate that is independent of the dimensionality of the sample dimensionality when using radial kernels.  For non-radial kernels, we obtain a sub-optimal convergence rate that depends on the sample dimensionality.

\phantomsection\label{sec4.3}
\subsection*{\small{\sf{\textbf{4.3.$\;\;\;$Numerical Examples}}}}

\setcounter{figure}{0}
\begin{figure*}
   \hspace{0.35cm}\includegraphics[width=6in]{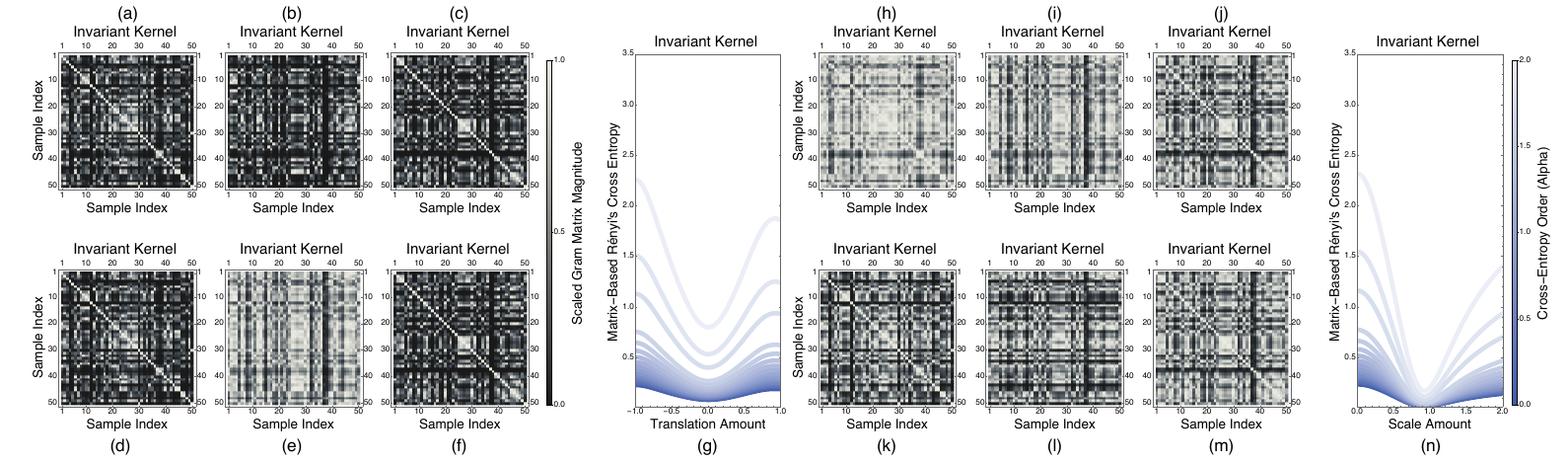}\vspace{-0.2cm}\\
   \caption[]{\fontdimen2\font=1.55pt\selectfont An overview of the cross-entropy estimation when considering distributional shape changes and shifts.  In (a)--(c), we plot the Gram matrices $\hat{\kappa}_1$, $\hat{\kappa}_{1,2}$, and $\hat{\kappa}_2$, respectively, obtained for samples from two Gaussian distributions with equivalent means and variances.  Here a translation-invariant, Gaussian kernel was used, $\kappa'(s,s') \!=\! \textnormal{exp}(-\sigma \langle s \!-\! s',\cdot\rangle)$, $\sigma \!=\! 1$.  We then shifted the means of both distributions so that they were further apart.  As shown in (d) and (f), the intra-distributional Gram matrices, $\hat{\kappa}_1$ and $\hat{\kappa}_2$, remain the same.  The inter-distributional Gram matrix, $\hat{\kappa}_{1,2}$, changes to reflect this mean shift.  In (g), we find that the $\alpha$-cross-entropy $C_\alpha(\hat{\kappa}_{1}\|\hat{\kappa}_{1,2}\|\hat{\kappa}_{2})$ changes according to the shift amount.  This result aligns well with our expectations.  In (h)--(n), we consider the case where the variance for one of the distributions changes.  Increasing the variance leads to changes in both $\hat{\kappa}_{1}$ and $\hat{\kappa}_{1,2}$, which can be seen by comparing (h) and (i) to (k) and (l).  The Gram matrix $\hat{\kappa}_{2}$ remains the same in (j) and (m), since it is not modified.  In (n), we find that the $\alpha$-cross-entropy $C_\alpha(\hat{\kappa}_{1}\|\hat{\kappa}_{1,2}\|\hat{\kappa}_{2})$ changes according to the variance scale amount.  This result aligns well with our expectations.  The plots in (g) and (n) also show that the $\alpha$-cross-entropy is monotonically increasing with respect to $\alpha$.  Taken together, these results indicate that $C_\alpha(\hat{\kappa}_{1}\|\hat{\kappa}_{1,2}\|\hat{\kappa}_{2})$ overcomes the issues associated with $C_\alpha(\hat{\kappa}_{1}\|\hat{\kappa}_{2})$ and $C_\alpha'(\hat{\kappa}_{1}\|\hat{\kappa}_{2})$ when translation-invariant kernels are employed. \vspace{-0.4cm}}
\label{fig:tripartite}
\end{figure*}

{\small{\sf{\textbf{Distribution Assessment Analysis.}}}} While the bipartite cross-entropy measures can assess both distributional shift and scale changes, their magnitudes do not always align well with intuition depending on the kernel type.  We show that the tripartite measure does not share this shortcoming.  Examples are presented in \hyperref[fig:tripartite]{Figure 4.1}.

In \hyperref[fig:tripartite]{Figure 4.1}, we consider both distributional mean and variance shifts.  We draw samples from two-dimensional Gaussian distributions.  For the mean-shift experiments, samples for one distribution are held fixed while the mean for the samples for the other distribution are shifted.  The distribution variances are the same.  For the variance-shift experiments, we artificially scale the distributional spread for one distribution.  The means are kept the same.  The same number of samples are drawn from each distribution.  As with the experiments for the bipartite measures, for the mean-shift experiments, we expect the tripartite measure to achieve a minimum whenever the distributional means are equal.  The tripartite cross-entropy should monotonically increase as the means diverge and hence the distributions have increasingly minimal overlap.  As well, for the variance-scale experiments, we expect the tripartite measure to achieve a minimum when the variances are equivalent and then monotonically increase according to the scale amount.  These behaviors are present in \hyperref[fig:tripartite]{Figure 4.1}(g) and (n) when using the translation-invariant Gaussian kernel.  

For both the mean-shift and variance-scale experiments, the observed behavior is a byproduct of using a joint, bivariate Gram matrix within the measure.  This matrix corresponds to cross-correlations of functions.  It hence captures inter-distributional sample-distance changes, even for translation-invariant kernels, as highlighted in \hyperref[fig:tripartite]{Figures 4.1}(b) and (e).  Moreover, such a matrix provides ancillary details about distributional shape differences, again due to capturing inter-distributional distance changes.  This can be seen in \hyperref[fig:tripartite]{Figures 4.1}(i) and (l).  The univariate Gram matrices, in contrast, correspond to autocorrelations of functions and hence intra-distributional sample distances.  Such matrices do not change with global translations, as shown in \hyperref[fig:tripartite]{Figures 4.1}(a) and (c) along with \hyperref[fig:tripartite]{Figures 4.1}(d) and (e), when using translation-invariant kernels, which is due to properties of the induced inner product.  These matrices are only sensitive to local distance modifications and hence distributional shape changes.  This is apparent from \hyperref[fig:tripartite]{Figures 4.1}(h) and (j) along with \hyperref[fig:tripartite]{Figures 4.1}(k) and (m).

\phantomsection\label{sec5}
\subsection*{\small{\sf{\textbf{5$\;\;\;$Comparisons}}}}\addtocounter{section}{1}

The matrix-based $\alpha$-cross-entropies that we have defined resemble quantum-information-theoretic counterparts.  Our non-mirrored $\alpha$-cross-entropy is analogous to Petz's original formulation of a quantum R\'{e}nyi relational entropy for stochastic matrices \cite{PetzD-jour1986a,HiaiF-jour2011a}.  Our mirrored variant is also related to the minimal divergence of M\"{u}ller-Lennert et al. \cite{MullerLennertM-jour2013a} for matrices.  A variety of properties have been demonstrated for these quantum measures \cite{PetzD-jour2003a,BeigiS-jour2013a,FrankRL-jour2013a,SutterD-jour2016a} and many have shown how to use these measures to generalize classical notions from information theory to the quantum case \cite{TomamhicelM-jour2009a,MosonyiM-jour2011a}.  We refer readers to \cite{RuskaiMB-jour2002a,ZhangL-jour2014a,FawziO-jour2015a,ErkerP-jour2015a} for additional work on quantum information theory.

There is, however, a fundamental distinction between our work and theirs.  In the quantum setting, the operators correspond to positive-semi-definite density matrices.  Entries of these matrices are mixtures of state vectors that a system may assume.  These divergences hence assess differences between quantum states in a Hilbert space.  In our case, the operators quantify correlations between functions.  We construct empirical versions of these operators, in the form of Gram matrices, which correspond to mappings of the finite sample set to a separable, reproducing-kernel Hilbert space of functions.  Due to the connections of Gram matrices with mean-element maps, we are assured that our cross-entropies are assessing differences between probability measures without needing direct knowledge of them.  This behavior is not present in the quantum-information-theoretic case.  This is because the random matrices are not formed from the application of kernels to the samples and hence do not characterize embeddings of distributions.  As a consequence of this distinction, the matrix-based quantum divergences cannot be applied for statistical testing \cite{GrettonA-jour2011a} in the non-quantum case \cite{MosonyiM-jour2015a}.  Our matrix-based cross-entropies can.  Our measures can also be readily extended to multiple variables, which offers similar functionality to \cite{SejdinovicD-coll2013a}.  

Our matrix-based cross-entropies thus are an entirely different formulation that have only superficial connections to the quantum relative entropies.  As well, our measures do not appear to stem from any other operator-theoretic concepts, such as either delta divergences, log-determinant divergences \cite{SraS-coll2012a}, or log-Hilbert-Schmidt metrics \cite{MnihHQ-coll2014a}, all of which are defined for Gram matrices.  Our contributions thus appear to provide the first matrix-based, unbiased, non-parametric estimators for R\'{e}nyi's cross-entropy.  Additionally, our measures can be transformed to specify R\'{e}nyi-divergence-like quantities, along with quantities associated with special cases, like the Kullback-Leibler divergence.  Our measure does not require distributional assumptions to assess distributional overlap.  It thus is suitable for general statistical testing of arbitrary distributions.  Previous definitions of non-quantum, matrix-based divergences have typically been limited to embeddings of Gaussian distributions \cite{MnihHQ-jour2017a}.  No convergence and convergence-rate guarantees have been provided.  Moreover, the measures defined in \cite{MnihHQ-jour2017a} can only handle sample sets with equivalent cardinalities.  While our bipartite measures share the same trait, our tripartite measure does not.

\subsection*{\small{\sf{\textbf{6.$\;\;\;$Conclusions}}}}\addtocounter{section}{1}

In this paper, we propose a functional-operator formulation of R\'{e}nyi's $\alpha$-cross-entropy.  We approximate the underlying operators using Gram-matrix-based representations.  This yields R\'{e}nyi's $\alpha$-cross-entropies suitable for kernelized matrix- and graph-based data.  The $\alpha$-cross-entropies can also be applied to vector-based data that are converted into Gram matrices through the application of a kernel to vector pairs.

We prove that the matrix-based $\alpha$-cross-entropies satisfy the axiomatic properties that R\'{e}nyi established for divergences.  Our criteria thus, in theory, can act as divergences.  It is appropriate to consider them as divergences for matrix and graph samples that do not have metric-space vector realizations.  For vector samples converted into Gram matrices, though, our bipartite $\alpha$-cross-entropies should not be viewed as divergences, even though their expressions resemble the classical case.  This is because characteristics of our $\alpha$-cross-entropies do not permit effective statistical testing in this case.  They, for instance, can only compare matrices of the same size.  As well, certain kernel choices may not simultaneously quantify shape and positional differences in the functional-space-embedded distributions.

Due to these shortcomings, we also offer a tripartite, matrix-based R\'{e}nyi's $\alpha$-cross-entropy.  This criterion innately overcomes all of the aforementioned issues of our bipartite $\alpha$-cross-entropies, regardless of if the kernels are either translation varying or translation invariant.  It hence is more appealing for statistical testing.  This cross-entropy, however, can only be applied for graphs and matrices that have a vector realization.  If no realization exists, then our $\alpha$-cross-entropies must be used.  This is because our $\alpha$-divergence requires the approximation of a joint cross-covariance operator between sample sets, which can only be constructed if vector samples are available.  Our bipartite $\alpha$-cross entropies, in contrast, require only the approximation of marginal covariance operators that are obtained from each set independently.  The approximations for the marginal operators can be formed for either vector samples or matrix-based representations.  

A major advantage of our cross-entropy measures is that the convergence rate for the operator approximation depends only on the number of samples, not the sample dimensionality.  This property allows practitioners to consider assessing distributional overlap for high-dimensional samples where existing plug-in estimators, like Parzen windows, would be largely ineffective.  For the bipartite measures, this convergence-rate guarantee is available for all universal kernels and arbitrary distributions.  For the tripartite measure, we only have a dimensionally-agnostic error rate for radial universal kernels applied to samples from arbitrary distributions.  In either case, we have minimax optimality.  We thus avoid slow-convergence-rate issues that are prevalent for plug-in density estimators.

\renewcommand*{\bibfont}{\raggedright}
\renewcommand\bibsection{\subsection*{\small{\sf{\textbf{References}}}}}
{\singlespacing\fontsize{9.75}{10}\selectfont \bibliography{sledgebib} \bibliographystyle{IEEEtran}}

\begin{thebibliography}{10}
\providecommand{\url}[1]{#1}
\csname url@samestyle\endcsname
\providecommand{\newblock}{\relax}
\providecommand{\bibinfo}[2]{#2}
\providecommand{\BIBentrySTDinterwordspacing}{\spaceskip=0pt\relax}
\providecommand{\BIBentryALTinterwordstretchfactor}{4}
\providecommand{\BIBentryALTinterwordspacing}{\spaceskip=\fontdimen2\font plus
\BIBentryALTinterwordstretchfactor\fontdimen3\font minus
  \fontdimen4\font\relax}
\providecommand{\BIBforeignlanguage}[2]{{%
\expandafter\ifx\csname l@#1\endcsname\relax
\typeout{** WARNING: IEEEtran.bst: No hyphenation pattern has been}%
\typeout{** loaded for the language `#1'. Using the pattern for}%
\typeout{** the default language instead.}%
\else
\language=\csname l@#1\endcsname
\fi
#2}}
\providecommand{\BIBdecl}{\relax}
\BIBdecl

\bibitem{StratonovichRL-book1975a}
R.~L. Stratonovich, \emph{Information Theory}.\hskip 1em plus 0.5em minus
  0.4em\relax Moscow, Soviet Union: Sovetskoe Radio, 1975.

\bibitem{CoverTM-book2006a}
T.~M. Cover and J.~A. Thomas, \emph{Elements of Information Theory}.\hskip 1em
  plus 0.5em minus 0.4em\relax New York, NY, USA: John Wiley and Sons, 2006.

\bibitem{ZivJ-jour1993a}
\BIBentryALTinterwordspacing
J.~Ziv and N.~Merhav, ``A measure of relative entropy between individual
  sequences with application to universal classification,'' \emph{IEEE
  Transactions on Information Theory}, vol.~39, no.~4, pp. 1270--1279, 1933.
  Available: \url{http://dx.doi.org/10.1109/18.243444}
\BIBentrySTDinterwordspacing

\bibitem{DarbellayGA-jour1999a}
\BIBentryALTinterwordspacing
G.~A. Darbellay and I.~Vajda, ``Estimation of the information by an adaptive
  partitioning of the observation space,'' \emph{IEEE Transactions on
  Information Theory}, vol.~45, no.~4, pp. 1315--1321, 1999. Available:
  \url{http://dx.doi.org/10.1109/18.761290}
\BIBentrySTDinterwordspacing

\bibitem{WangQ-jour2005a}
\BIBentryALTinterwordspacing
Q.~Wang, S.~R. Kulkarni, and S.~Verdu, ``Divergence estimation of continuous
  distributions based on data-dependent partitions,'' \emph{IEEE Transactions
  on Information Theory}, vol.~51, no.~9, pp. 3064--3074, 2005. Available:
  \url{http://dx.doi.org/10.1109/TIT.2005.853314}
\BIBentrySTDinterwordspacing

\bibitem{CaiH-jour2006a}
\BIBentryALTinterwordspacing
H.~Cai, S.~R. Kulkarni, and S.~Verdu, ``Universal divergence estimation for
  finite-alphabet sources,'' \emph{IEEE Transactions on Information Theory},
  vol.~52, no.~8, pp. 3456--3475, 2006. Available:
  \url{http://dx.doi.org/10.1109/TIT.2006.878182}
\BIBentrySTDinterwordspacing

\bibitem{PrincipeJC-book2010a}
J.~C. {Pr\'{i}ncipe}, \emph{Information Theoretic Learning}.\hskip 1em plus
  0.5em minus 0.4em\relax New York City, NY, USA: Springer-Verlag, 2010.

\bibitem{GiraldoLGS-conf2013a}
\BIBentryALTinterwordspacing
L.~G. {Sanchez Giraldo} and J.~C. {Pr\'{i}ncipe}, ``Information theoretic
  learning with infinitely divisible kernels,'' in \emph{Proceedings of the
  International Conference on Learning Representations {(ICLR)}}, Scottsdale,
  AZ, USA, May 2-4 2013, pp. 1--11. Available:
  \url{https://arxiv.org/abs/1301.3551}
\BIBentrySTDinterwordspacing

\bibitem{GiraldoLGS-jour2014a}
\BIBentryALTinterwordspacing
L.~G. {Sanchez Giraldo}, M.~Rao, and J.~C. {Pr\'{i}ncipe}, ``Measures of
  entropy from data using infinitely divisible kernels,'' \emph{IEEE
  Transactions on Information Theory}, vol.~61, no.~1, pp. 535--548, 2014.
  Available: \url{http://dx.doi.org/10.1109/TIT.2014.2370058}
\BIBentrySTDinterwordspacing

\bibitem{YuS-jour2020a}
\BIBentryALTinterwordspacing
S.~Yu, L.~G. {S\'{a}nchez Giraldo}, R.~Jenssen, and J.~C. {Pr\'{i}ncipe},
  ``Multivariate extensions of matrix-based {R\'{e}nyi's} $\alpha$-order
  entropy functional,'' \emph{IEEE Transactions on Pattern Analysis and Machine
  Intelligence}, vol.~42, no.~11, pp. 2960--2966, 2020. Available:
  \url{http://dx.doi.org/10.1109/TPAMI.2019.2932976}
\BIBentrySTDinterwordspacing

\bibitem{PalD-coll2010a}
D.~{P\'{a}l}, B.~{P\'{o}czos}, and C.~{Szepesv\'{a}ri}, ``Estimation of
  {R\'{e}nyi} entropy and mutual information based on generalized
  nearest-neighbor graphs,'' in \emph{Advances in Neural Information Processing
  Systems {(NIPS)}}, J.~Lafferty, C.~Williams, J.~Shawe-Taylor, R.~Zemel, and
  A.~Culotta, Eds.\hskip 1em plus 0.5em minus 0.4em\relax Red Hook, NY, USA:
  Curran Associates, 2010, pp. 1849--1857.

\bibitem{HeroAO-jour1999a}
\BIBentryALTinterwordspacing
A.~O. Hero and O.~J.~J. Michel, ``Asymptotic theory of greedy approximations to
  minimal k-point random graphs,'' \emph{IEEE Transactions on Information
  Theory}, vol.~46, no.~6, pp. 1921--1938, 1999. Available:
  \url{http://dx.doi.org/10.1109/18.782114}
\BIBentrySTDinterwordspacing

\bibitem{WangQ-jour2009a}
\BIBentryALTinterwordspacing
Q.~Wang, S.~R. Kulkarni, and S.~Verdu, ``Divergence estimation for
  multidimensional densities via $k$-nearest-neighbor distances,'' \emph{IEEE
  Transactions on Information Theory}, vol.~55, no.~5, pp. 2392--2405, 2009.
  Available: \url{http://dx.doi.org/10.1109/TIT.2009.2016060}
\BIBentrySTDinterwordspacing

\bibitem{FukumizuK-coll2011a}
K.~Fukumizu, G.~R.~G. Lanckriet, and B.~K. Sriperumbudur, ``Learning in
  {Hilbert} vs. {Banach} spaces: {A} measure embedding viewpoint,'' in
  \emph{Advances in Neural Information Processing Systems {(NIPS)}},
  J.~Shawe-Taylor, R.~Zemel, P.~Bartlett, F.~Pereira, and K.~Q. Weinberger,
  Eds.\hskip 1em plus 0.5em minus 0.4em\relax Red Hook, NY, USA: Curran
  Associates, 2011, pp. 1773--1781.

\bibitem{SriperumbudurBK-conf2008a}
B.~K. Sriperumbudur, A.~Gretton, K.~Fukumizu, G.~R.~G. Lanckriet, and {B.
  Sch\"{o}lkopf}, ``Injective {Hilbert} space embeddings of probability
  measures,'' in \emph{Proceedings of the Conference on Learning Theory
  {(COLT)}}, Helsinki, Finland, July 9-12 2008, pp. 111--122.

\bibitem{ChristmannA-coll2011a}
A.~Christmann and I.~Steinwart, ``Universal kernels on non-standard input
  spaces,'' in \emph{Advances in Neural Information Processing Systems
  {(NIPS)}}, J.~Lafferty, C.~Williams, J.~Shawe-Taylor, R.~Zemel, and
  A.~Culotta, Eds.\hskip 1em plus 0.5em minus 0.4em\relax Red Hook, NY, USA:
  Curran Associates, 2011, pp. 406--414.

\bibitem{FukumizuK-coll2009a}
K.~Fukumizu, A.~Gretton, G.~R.~G. Lanckriet, B.~{Sch\"{o}lkopf}, and B.~K.
  Sriperumbudur, ``Characteristic kernels on groups and semigroups,'' in
  \emph{Advances in Neural Information Processing Systems {(NIPS)}}, D.~Koller,
  D.~Schuurmans, Y.~Bengio, and L.~Bottou, Eds.\hskip 1em plus 0.5em minus
  0.4em\relax Red Hook, NY, USA: Curran Associates, 2009, pp. 473--480.

\bibitem{NishiyamaY-jour2016a}
Y.~Nishiyama and K.~Fukumizu, ``Characteristic kernels and infinitely divisible
  distributions,'' \emph{Journal of Machine Learning Research}, vol.~17, no.~1,
  pp. 1--28, 2016.

\bibitem{SriperumbudurBK-conf2010a}
B.~K. Sriperumbudur, K.~Fukumizu, and G.~R.~G. Lanckriet, ``On the relation
  between universality, characteristic kernels and {RKHS} embedding of
  measures,'' in \emph{Proceedings of the International Conference on
  Artificial Intelligence and Statistics {(AISTATS)}}, Sardinia, Italy, May
  13-15 2010, pp. 773--780.

\bibitem{SriperumbudurBK-jour2011a}
------, ``Universality, characteristic kernels and {RKHS} embedding of
  measures,'' \emph{Journal of Machine Learning Research}, vol.~12, no.~1, pp.
  2389--5410, 2011.

\bibitem{ChwialkowskiKP-coll2014a}
K.~P. Chwialkowski, D.~Sejdinovic, and A.~Gretton, ``A wild bootstrap for
  degenerate kernel tests,'' in \emph{Advances in Neural Information Processing
  Systems {(NIPS)}}, Z.~Ghahramani, M.~Welling, C.~Cortes, N.~Lawrence, and
  K.~Q. Weinberger, Eds.\hskip 1em plus 0.5em minus 0.4em\relax Red Hook, NY,
  USA: Curran Associates, 2014, pp. 3608--3616.

\bibitem{GrettonA-coll2007a}
A.~Gretton, K.~Borgwardt, M.~Rasch, B.~{Sch\"{o}lkopf}, and A.~J. Smola, ``A
  kernel method for the two-sample problem,'' in \emph{Advances in Neural
  Information Processing Systems {(NIPS)}}, B.~{Sch\"{o}lkopf}, J.~Platt, and
  T.~Hoffman, Eds.\hskip 1em plus 0.5em minus 0.4em\relax Red Hook, NY, USA:
  Curran Associates, 2007, pp. 513--520.

\bibitem{HeinM-conf2005a}
M.~Hein and O.~Bousquet, ``Hilbertian metrics and positive definite kernels on
  probability measures,'' in \emph{Proceedings of the International Conference
  on Artificial Intelligence and Statistics {(AISTATS)}}, Bridgetown, Barbados,
  January 6-8 2005, pp. 136--143.

\bibitem{FukumizuK-coll2008a}
K.~Fukumizu, A.~Gretton, X.~Sun, and B.~{Sch\"{o}lkopf}, ``Kernel measures of
  conditional dependence,'' in \emph{Advances in Neural Information Processing
  Systems {(NIPS)}}, J.~Platt, D.~Koller, Y.~Singer, and S.~T. Roweis,
  Eds.\hskip 1em plus 0.5em minus 0.4em\relax Red Hook, NY, USA: Curran
  Associates, 2008, pp. 489--496.

\bibitem{QuangMH-coll2014a}
M.~H. Quang, M.~S. Biagio, and V.~Murino, ``{Log-Hilbert-Schmidt} metric
  between positive definite operators on {Hilbert} spaces,'' in \emph{Advances
  in Neural Information Processing Systems {(NIPS)}}, Z.~Ghahramani,
  M.~Welling, C.~Cortes, N.~Lawrence, and K.~Q. Weinberger, Eds.\hskip 1em plus
  0.5em minus 0.4em\relax Red Hook, NY, USA: Curran Associates, 2014, pp.
  388--396.

\bibitem{ChwialkowskiK-conf2014a}
\BIBentryALTinterwordspacing
K.~Chwialkowski and A.~Gretton, ``A kernel independence test for random
  processes,'' in \emph{Proceedings of the International Conference on Machine
  Learning {(ICML)}}, Beijing, China, June 21-26 2014, pp. 1422--1430.
  Available: \url{https://arxiv.org/abs/1402.4501}
\BIBentrySTDinterwordspacing

\bibitem{SejdinovicD-jour2013a}
\BIBentryALTinterwordspacing
D.~Sejdinovic, B.~K. Sriperumbudur, A.~Gretton, and K.~Fukumizu, ``Equivalence
  of distance-based and {RKHS}-based statistics in hypothesis testing,''
  \emph{Annals of Statistics}, vol.~41, no.~5, pp. 2263--2291, 2013. Available:
  \url{http://dx.doi.org/10.1214/13-AOS1140}
\BIBentrySTDinterwordspacing

\bibitem{KwokJTY-jour2004a}
\BIBentryALTinterwordspacing
J.~T.-Y. Kwok and I.~W.-H. Tsang, ``The pre-image problem in kernel methods,''
  \emph{IEEE Transactions on Neural Networks}, vol.~15, no.~6, pp. 1517--1525,
  2004. Available: \url{http://dx.doi.org/10.1109/TNN.2004.837781}
\BIBentrySTDinterwordspacing

\bibitem{SongL-conf2008a}
\BIBentryALTinterwordspacing
L.~Song, X.~Zhang, A.~J. Smola, A.~Gretton, and B.~{Sch\"{o}lkopf}, ``Tailoring
  density estimation via reproducing kernel moment matching,'' in
  \emph{Proceedings of the International Conference on Machine Learning
  {(ICML)}}, Helsinki, Finland, July 5-9 2008, pp. 992--999. Available:
  \url{http://dx.doi.org/10.1145/1390156.1390281}
\BIBentrySTDinterwordspacing

\bibitem{ChenY-conf2010a}
\BIBentryALTinterwordspacing
Y.~Chen, M.~Welling, and A.~J. Smola, ``Super-samples from kernel herding,'' in
  \emph{Proceedings of the Conference on Uncertainty in Artificial Intelligence
  {(UAI)}}, Catalina Island, CA, USA, July 8-11 2010, pp. 109--116. Available:
  \url{https://arxiv.org/abs/1203.3472}
\BIBentrySTDinterwordspacing

\bibitem{HuszarF-conf2012a}
\BIBentryALTinterwordspacing
F.~{Husz\'{a}r} and D.~Duvenaud, ``Optimally-weighted herding is {Bayesian}
  quadrature,'' in \emph{Proceedings of the Conference on Uncertainty in
  Artificial Intelligence {(UAI)}}, Catalina Island, CA, USA, August 15-17
  2012, pp. 377--386. Available: \url{https://arxiv.org/abs/1204.1664}
\BIBentrySTDinterwordspacing

\bibitem{KanagawaM-conf2014a}
M.~Kanagawa and K.~Fukumizu, ``Recovering distributions from {Gaussian RKHS}
  embeddings,'' in \emph{Proceedings of the International Conference on
  Artificial Intelligence and Statistics {(AISTATS)}}, Reykjavik, Iceland,
  April 22-25 2014, pp. 457--465.

\bibitem{KanagawaM-coll2016a}
M.~Kanagawa, B.~K. Sriperumbudur, and K.~Fukumizu, ``Convergence guarantees for
  kernel-based quadrature rules in misspecified settings,'' in \emph{Advances
  in Neural Information Processing Systems {(NIPS)}}, D.~Lee, M.~Sugiyama,
  U.~Luxburg, I.~Guyon, and R.~Garnett, Eds.\hskip 1em plus 0.5em minus
  0.4em\relax Red Hook, NY, USA: Curran Associates, 2016, pp. 3296--3304.

\bibitem{SriperumbudurBK-jour2012a}
\BIBentryALTinterwordspacing
B.~K. Sriperumbudur, K.~Fukumizu, A.~Gretton, B.~{Sch\"{o}lkopf}, and G.~R.~G.
  Lanckriet, ``On the empirical estimation of integral probability metrics,''
  \emph{Electronic Journal of Statistics}, vol.~6, no.~1, pp. 1550--1599, 2012.
  Available: \url{http://dx.doi.org/10.1214/12-EJS722}
\BIBentrySTDinterwordspacing

\bibitem{SriperumbudurBK-jour2010a}
B.~K. Sriperumbudur, A.~Gretton, K.~Fukumizu, B.~{Sch\"{o}lkopf}, and G.~R.~G.
  Lanckriet, ``Hilbert space embeddings and metrics on probability measures,''
  \emph{Journal of Machine Learning Research}, vol.~11, no.~1, pp. 1517--1561,
  2012.

\bibitem{FukumizuK-coll2010a}
K.~Fukumizu, A.~Gretton, G.~R.~G. Lanckriet, B.~{Sch\"{o}lkopf}, and B.~K.
  Sriperumbudur, ``Kernel choice and classifiability for {RKHS} embeddings of
  probability distributions,'' in \emph{Advances in Neural Information
  Processing Systems {(NIPS)}}, Y.~Bengio, D.~Schuurmans, J.~Lafferty,
  C.~Williams, and A.~Culotta, Eds.\hskip 1em plus 0.5em minus 0.4em\relax Red
  Hook, NY, USA: Curran Associates, 2010, pp. 1750--1758.

\bibitem{RenyiA-coll1976a}
A.~{R\'{e}nyi}, ``On measures of entropy and information,'' in \emph{Selected
  Papers of {Alfr\'{e}d R\'{e}nyi}}, P.~{Tur\'{a}n}, Ed.\hskip 1em plus 0.5em
  minus 0.4em\relax Budapest, Turkey: {Akad\'{e}miai Kaid\'{o}}, 1976, vol.~2,
  pp. 565--580.

\bibitem{vanErvenT-jour2014a}
\BIBentryALTinterwordspacing
T.~{van Erven} and P.~{Harremo\"{e}s}, ``{R\'{e}nyi} divergence and
  {Kullback-Leibler} divergence,'' \emph{IEEE Transactions on Information
  Theory}, vol.~60, no.~7, pp. 3797--3820, 2014. Available:
  \url{http://dx.doi.org/10.1109/TIT.2014.2320500}
\BIBentrySTDinterwordspacing

\bibitem{TeixeiraA-jour2012a}
\BIBentryALTinterwordspacing
A.~Teixeira, A.~Matos, and L.~Antunes, ``Conditional {R\'{e}nyi} entropies,''
  \emph{IEEE Transactions on Information Theory}, vol.~58, no.~7, pp.
  4273--4277, 2012. Available: \url{http://dx.doi.org/10.1109/TIT.2012.2192713}
\BIBentrySTDinterwordspacing

\bibitem{HornRA-jour1969a}
\BIBentryALTinterwordspacing
R.~A. Horn, ``The theory of infinitely divisible matrices and kernels,''
  \emph{Transactions of the American Mathematical Society}, vol. 136, pp.
  269--286, 1969. Available: \url{http://dx.doi.org/10.2307/1994714}
\BIBentrySTDinterwordspacing

\bibitem{BakerC-jour1973a}
\BIBentryALTinterwordspacing
C.~Baker, ``Joint measures and cross-covariance operators,'' \emph{Transactions
  of the American Mathematical Society}, vol. 186, pp. 273--289, 1973.
  Available: \url{http://dx.doi.org/10.2307/1996566}
\BIBentrySTDinterwordspacing

\bibitem{TolstikhinI-jour2017a}
I.~Tolstikhin, B.~K. Sriperumbudur, and K.~Muandet, ``Minimax estimation of
  kernel mean embeddings,'' \emph{Journal of Machine Learning Research},
  vol.~18, no.~1, pp. 1--47, 2017.

\bibitem{AronszajnN-jour1950a}
\BIBentryALTinterwordspacing
N.~Aronszajn, ``Theory of reproducing kernels,'' \emph{Transactions of the
  American Mathematical Society}, vol.~68, no.~3, pp. 337--404, 1950.
  Available: \url{http://dx.doi.org/10.2307/1990404}
\BIBentrySTDinterwordspacing

\bibitem{SzaboZ-jour2018a}
\BIBentryALTinterwordspacing
Z.~{Szab\'{o}} and B.~K. Sriperumbudur, ``Characteristic and universal tensor
  product kernels,'' \emph{Journal of Machine Learning Research}, vol.~18,
  no.~1, pp. 1--29, 2018. Available: \url{https://arxiv.org/abs/1708.08157}
\BIBentrySTDinterwordspacing

\bibitem{OwhadiH-jour2017a}
\BIBentryALTinterwordspacing
H.~Owhadi and C.~Scovel, ``Separability of reproducing kernel spaces,''
  \emph{Proceedings of the American Mathematical Society}, vol. 145, no.~5, pp.
  2131--2138, 2017. Available: \url{http://doi.org/10.1090/proc/13354}
\BIBentrySTDinterwordspacing

\bibitem{LiY-coll2016a}
Y.~Li and R.~E. Turner, ``{R\'{e}nyi} divergence variational inference,'' in
  \emph{Advances in Neural Information Processing Systems {(NIPS)}}, D.~Lee,
  M.~Sugiyama, U.~Luxburg, I.~Guyon, and R.~Garnett, Eds.\hskip 1em plus 0.5em
  minus 0.4em\relax Red Hook, NY, USA: Curran Associates, 2016, pp. 1073--1081.

\bibitem{StoneCJ-jour1980a}
\BIBentryALTinterwordspacing
C.~J. Stone, ``Optimal rates of convergence for nonparametric estimators,''
  \emph{Annals of Statistics}, vol.~8, no.~6, pp. 1348--1360, 1980. Available:
  \url{http://dx.doi.org/10.1214/aos/1176345206}
\BIBentrySTDinterwordspacing

\bibitem{GrettonA-coll2012a}
A.~Gretton, D.~Sejdinovic, H.~Strathmann, S.~Balakrishnan, M.~Pontil,
  K.~Fukumizu, and B.~K. Sriperumbudur, ``Optimal kernel choice for large-scale
  two-sample tests,'' in \emph{Advances in Neural Information Processing
  Systems {(NIPS)}}, F.~Pereira, C.~J.~C. Burges, L.~Bottou, and K.~Q.
  Weinberger, Eds.\hskip 1em plus 0.5em minus 0.4em\relax Red Hook, NY, USA:
  Curran Associates, 2012, pp. 1205--1213.

\bibitem{GrettonA-coll2009a}
A.~Gretton, K.~Fukumizu, Z.~Harchaoui, and B.~K. Sriperumbudur, ``A fast,
  consistent kernel two-sample test,'' in \emph{Advances in Neural Information
  Processing Systems {(NIPS)}}, Y.~Bengio, D.~Schuurmans, J.~Lafferty,
  C.~Williams, and A.~Culotta, Eds.\hskip 1em plus 0.5em minus 0.4em\relax Red
  Hook, NY, USA: Curran Associates, 2009, pp. 673--681.

\bibitem{GrettonA-jour2012a}
A.~Gretton, K.~Borgwardt, M.~Rasch, {B. Sch\"{o}lkopf}, and A.~J. Smola, ``A
  kernel two-sample test,'' \emph{Journal of Machine Learning Research},
  vol.~13, no.~1, pp. 723--773, 2012.

\bibitem{EricM-coll2008a}
M.~Eric, F.~Bach, and Z.~Harchaoui, ``Testing for homogeneity with kernel
  {Fisher} discriminant analysis,'' in \emph{Advances in Neural Information
  Processing Systems {(NIPS)}}, J.~Platt, D.~Koller, Y.~Singer, and S.~T.
  Roweis, Eds.\hskip 1em plus 0.5em minus 0.4em\relax Red Hook, NY, USA: Curran
  Associates, 2008, pp. 609--616.

\bibitem{ArgyriouA-conf2006a}
\BIBentryALTinterwordspacing
A.~Argyriou, R.~Hauser, C.~A. Micchelli, and M.~Pontil, ``A {DC}-programming
  algorithm for kernel selection,'' in \emph{Proceedings of the International
  Conference on Machine Learning {(ICML)}}, Pittsburgh, PA, USA, June 25-29
  2006, pp. 41--48. Available: \url{http://dx.doi.org/10.1145/1143844.1143850}
\BIBentrySTDinterwordspacing

\bibitem{XuZ-coll2009a}
M.~Kloft, U.~Brefeld, P.~Laskov, K.-R. {M\"{u}ller}, A.~Zien, and
  S.~Sonnenburg, ``An extended level method for efficient multiple kernel
  learning,'' in \emph{Advances in Neural Information Processing Systems
  {(NIPS)}}, D.~Koller, D.~Schuurmans, Y.~Bengio, and L.~Bottou, Eds.\hskip 1em
  plus 0.5em minus 0.4em\relax Red Hook, NY, USA: Curran Associates, 2009, pp.
  1825--1832.

\bibitem{KloftM-coll2010a}
------, ``Efficient and accurate {$\ell_p$}-norm multiple kernel learning,'' in
  \emph{Advances in Neural Information Processing Systems {(NIPS)}}, Y.~Bengio,
  D.~Schuurmans, J.~Lafferty, C.~Williams, and A.~Culotta, Eds.\hskip 1em plus
  0.5em minus 0.4em\relax Red Hook, NY, USA: Curran Associates, 2010, pp.
  997--1005.

\bibitem{AntosA-jour2001a}
\BIBentryALTinterwordspacing
A.~Antos and I.~Kontoyiannis, ``Convergence properties of functional estimates
  for discrete distributions,'' \emph{Random Structures and Algorithms},
  vol.~19, no. 3-4, pp. 163--193, 2001. Available:
  \url{http://dx.doi.org/10.1002/rsa.10019}
\BIBentrySTDinterwordspacing

\bibitem{BirgeL-jour1986a}
\BIBentryALTinterwordspacing
L.~{Birg\'{e}}, ``On estimating a density using {Hellinger} distance and some
  other strange facts,'' \emph{Probability Theory and Related Fields}, vol.
  710, no.~1, pp. 271--291, 1986. Available:
  \url{http://dx.doi.org/10.1007/BF00332312}
\BIBentrySTDinterwordspacing

\bibitem{DevroyeL-jour1995a}
\BIBentryALTinterwordspacing
L.~Devroye, ``Another proof of a slow convergence result of {Birg\'{e}},''
  \emph{Statistics and Probability Letters}, vol.~23, no.~1, pp. 63--67, 1995.
  Available: \url{http://dx.doi.org/10.1016/0167-7152(94)00095-P}
\BIBentrySTDinterwordspacing

\bibitem{XuJ-jour2008a}
\BIBentryALTinterwordspacing
J.-W. Xu, A.~R.~C. Paiva, I.~Park, and J.~C. {Pr\'{i}ncipe}, ``A reproducing
  kernel {Hilbert} space framework for information-theoretic learning,''
  \emph{IEEE Transactions on Signal Processing}, vol.~56, no.~12, pp.
  5891--5902, 2008. Available: \url{http://dx.doi.org/10.1109/TSP.2008.2005085}
\BIBentrySTDinterwordspacing

\bibitem{PetzD-jour1986a}
\BIBentryALTinterwordspacing
D.~Petz, ``Quasi-entropies for finite quantum systems,'' \emph{Reports on
  Mathematical Physics}, vol.~23, no.~1, pp. 57--65, 1986. Available:
  \url{http://dx.doi.org/10.1016/0034-4877(86)90067-4}
\BIBentrySTDinterwordspacing

\bibitem{HiaiF-jour2011a}
\BIBentryALTinterwordspacing
F.~Hiai, M.~Mosonyi, D.~Petz, and C.~{B\'{e}ny}, ``Quantum $f$-divergences and
  error correction,'' \emph{Reviews in Mathematical Physics}, vol.~23, no.~7,
  pp. 691--747, 2011. Available:
  \url{http://dx.doi.org/10.1142/S0129055X11004412}
\BIBentrySTDinterwordspacing

\bibitem{MullerLennertM-jour2013a}
\BIBentryALTinterwordspacing
L.~{M\"{u}ller-Lennert}, D.~Dupuis, O.~Szehr, S.~Fehr, and M.~Tomamichel, ``On
  quantum {R\'{e}nyi} entropies: {A} new generalization and some properties,''
  \emph{Journal of Mathematical Physics}, vol.~54, no.~1, pp. 122\,203(1--20),
  2013. Available: \url{http://dx.doi.org/10.1063/1.4838856}
\BIBentrySTDinterwordspacing

\bibitem{PetzD-jour2003a}
\BIBentryALTinterwordspacing
D.~Petz, ``Monotonicity of quantum relative entropy revisited,'' \emph{Reviews
  of Mathematical Physics}, vol.~15, no.~1, pp. 79--91, 2003. Available:
  \url{http://dx.doi.org/10.1142/S0129055X03001576}
\BIBentrySTDinterwordspacing

\bibitem{BeigiS-jour2013a}
\BIBentryALTinterwordspacing
S.~Beigi, ``Sandwiched {R\'{e}nyi} divergence satisfies data processing
  inequality,'' \emph{Journal of Mathematical Physics}, vol.~54, no.~1, pp.
  122\,202(1--12), 2013. Available: \url{http://dx.doi.org/10.1063/1.4838855}
\BIBentrySTDinterwordspacing

\bibitem{FrankRL-jour2013a}
\BIBentryALTinterwordspacing
R.~L. Frank and E.~H. Lieb, ``Montonicity of a relative {R\'{e}nyi} entropy,''
  \emph{Journal of Mathematical Physics}, vol.~54, no.~1, pp. 122\,201(1--5),
  2013. Available: \url{http://dx.doi.org/10.1063/1.4838835}
\BIBentrySTDinterwordspacing

\bibitem{SutterD-jour2016a}
\BIBentryALTinterwordspacing
D.~Sutter, M.~Tomamichel, and A.~W. Harrow, ``Strengthened monotonicity of
  relative entropy via pinched {Petz} recovery map,'' \emph{IEEE Transactions
  on Information Theory}, vol.~62, no.~5, pp. 2907--2913, 2016. Available:
  \url{http://dx.doi.org/10.1109/TIT.2016.2545680}
\BIBentrySTDinterwordspacing

\bibitem{TomamhicelM-jour2009a}
\BIBentryALTinterwordspacing
M.~Tomamichel, R.~Colbeck, and R.~Renner, ``A fully quantum asymptotic
  equipartition property,'' \emph{IEEE Transactions on Information Theory},
  vol.~55, no.~12, pp. 5840--5847, 2009. Available:
  \url{http://dx.doi.org/10.1109/TIT.2009.2032797}
\BIBentrySTDinterwordspacing

\bibitem{MosonyiM-jour2011a}
\BIBentryALTinterwordspacing
M.~Mosonyi and F.~Hiai, ``On the quantum {R\'{e}nyi} relative entropies and
  related capacity formulas,'' \emph{IEEE Transactions on Information Theory},
  vol.~57, no.~4, pp. 2474--2487, 2011. Available:
  \url{http://dx.doi.org/10.1109/TIT.2011.2110050}
\BIBentrySTDinterwordspacing

\bibitem{RuskaiMB-jour2002a}
\BIBentryALTinterwordspacing
M.~B. Ruskai, ``Inequalities for quantum entropy: {A} review with conditions
  for equality,'' \emph{Journal of Mathematical Physics}, vol.~43, no.~1, pp.
  4358(1--18), 2002. Available: \url{http://dx.doi.org/10.1063/1.1497701}
\BIBentrySTDinterwordspacing

\bibitem{ZhangL-jour2014a}
\BIBentryALTinterwordspacing
L.~Zhang and J.~Wu, ``A lower bound of quantum conditional mutual
  information,'' \emph{Journal of Physics A: Mathematical and Theoretical},
  vol.~47, no.~1, pp. 415\,303(1--11), 2014. Available:
  \url{http://dx.doi.org/10.1088/1751-8113/47/41/415303}
\BIBentrySTDinterwordspacing

\bibitem{FawziO-jour2015a}
\BIBentryALTinterwordspacing
O.~Fawzi and R.~Renner, ``Quantum conditional mutual information and
  approximate {Markov} chains,'' \emph{Communications in Mathematical Physics},
  vol. 340, no.~1, pp. 575--611, 2015. Available:
  \url{http://dx.doi.org/10.1007/s00220-015-2466-x}
\BIBentrySTDinterwordspacing

\bibitem{ErkerP-jour2015a}
\BIBentryALTinterwordspacing
P.~Erker, ``How not to {R\'{e}nyi}-generalize the quantum conditional mutual
  information,'' \emph{Journal of Physics A: Mathematical and Theoretical},
  vol.~48, no.~1, pp. 275\,303(1--9), 2015. Available:
  \url{http://dx.doi.org/10.1088/1751-8113/48/27/275303}
\BIBentrySTDinterwordspacing

\bibitem{GrettonA-jour2011a}
A.~Gretton and L.~{Gy\"{o}rfi}, ``Consistent nonparametric tests of
  independence,'' \emph{Journal of Machine Learning Research}, vol.~11, no.~1,
  pp. 1391--1423, 2011.

\bibitem{MosonyiM-jour2015a}
\BIBentryALTinterwordspacing
M.~Mosonyi and T.~Ogawa, ``Quantum hypothesis testing and the operational
  interpretation of the quantum {R\'{e}nyi} relative entropies,''
  \emph{Communications in Mathematical Physics}, vol. 334, no.~1, pp.
  1617--1648, 2015. Available:
  \url{http://dx.doi.org/10.1007/s00220-014-2248-x}
\BIBentrySTDinterwordspacing

\bibitem{SejdinovicD-coll2013a}
D.~Sejdinovic, A.~Gretton, and W.~Bergsma, ``A kernel test for three-variable
  interactions,'' in \emph{Advances in Neural Information Processing Systems
  {(NIPS)}}, C.~J.~C. Burges, L.~Bottou, M.~Welling, Z.~Ghahramani, and K.~Q.
  Weinberger, Eds.\hskip 1em plus 0.5em minus 0.4em\relax Red Hook, NY, USA:
  Curran Associates, 2013, pp. 637--646.

\bibitem{SraS-coll2012a}
S.~Sra, ``A new metric on the manifold of kernel matrices with application to
  matrix geometric means,'' in \emph{Advances in Neural Information Processing
  Systems {(NIPS)}}, F.~Pereira, C.~J.~C. Burges, L.~Bottou, and K.~Q.
  Weinberger, Eds.\hskip 1em plus 0.5em minus 0.4em\relax Red Hook, NY, USA:
  Curran Associates, 2012, pp. 144--152.

\bibitem{MnihHQ-coll2014a}
H.~Q. Minh, M.~S. Siagio, and V.~Murino, ``Log-{Hilbert-Schmidt} metric between
  positive definite operators on {Hilbert} spaces,'' in \emph{Advances in
  Neural Information Processing Systems {(NIPS)}}, Z.~Ghahramani, M.~Welling,
  C.~Cortes, N.~Lawrence, and K.~Q. Weinberger, Eds.\hskip 1em plus 0.5em minus
  0.4em\relax Red Hook, NY, USA: Curran Associates, 2014, pp. 388--396.

\bibitem{MnihHQ-jour2017a}
\BIBentryALTinterwordspacing
H.~Q. Minh, ``Infinite-dimensional log-determinant divergences between positive
  definite trace class operators,'' \emph{Linear Algebra and its Applications},
  vol. 528, no.~1, pp. 331--383, 2017. Available:
  \url{http://dx.doi.org/10.1016/j.laa.2016.09.018}
\BIBentrySTDinterwordspacing

\end{thebibliography}

\clearpage\newpage

\subsection*{\small{\sf{\textbf{Appendix A}}}}

\phantomsection\label{secA.1}
\subsection*{\small{\sf{\textbf{A.1.$\;\;\;$Matrix-based R\'{e}nyi's $\alpha$-Cross-Entropy Properties}}}}

In what follows, we demonstrate that our matrix-based R\'{e}nyi's $\alpha$-cross-entropies possess the same characteristics as the classical R\'{e}nyi's $\alpha$-divergences despite not directly having access to probability distributions.  Their quantities thus have similar interpretations.

Throughout, we assume that $\kappa_1,\kappa_2$ are non-zero, positive-semi-definite, univariate Gram operators for reproducing-kernel Hilbert spaces $\mathcal{H}_{\kappa_1'}$ and $\mathcal{H}_{\kappa_2'}$.  We also assume that $[\hat{\kappa}_1],[\hat{\kappa}_2] \!\in\! \mathbb{R}^{n \times n}_{0,+}$ are corresponding normalized Gram\vspace{-0.025cm}\\ \noindent matrices.  Likewise, $\kappa_{1,2}$ is a joint, bivariate Gram operator that is non-zero and $[\hat{\kappa}_{1,2}] \!\in\! \mathbb{R}^{n \times n}_{0,+}$ is the associated Gram matrix.  We sometimes relax the assumption that the Gram matrices are square.

{\small{\sf{\textbf{Non-negativity.}}}} First, we show that the matrix-based $\alpha$-cross-entropies are non-negative in practical settings.  The cross-entropies are zero when the two arguments are equivalent. 
\begin{itemize}
\item[] \-\hspace{0.5cm}{\small{\sf{\textbf{Proposition A.1.}}}} If $\textnormal{tr}([\hat{\kappa}_1]) \!\geq\! \textnormal{tr}([\hat{\kappa}_2])$, then $C_\alpha([\hat{\kappa}_1]\|[\hat{\kappa}_2]) \!\geq\! 0$.  If $[\hat{\kappa}_1] \!=\! [\hat{\kappa}_2]$, and hence $\textnormal{tr}([\hat{\kappa}_1]) \!=\! \textnormal{tr}([\hat{\kappa}_2])$,\\ \noindent then $C_\alpha([\hat{\kappa}_1]\|[\hat{\kappa}_2]) \!=\! 0$.  The same results hold in the mirrored case.  Non-negativity similarly holds in the\vspace{-0.02cm}\\ \noindent tripartite case for square and non-square Gram matrices, whereas nullity holds only for square matrices.
\end{itemize}
\noindent A stronger statement, that the $\alpha$-cross-entropy is zero if and only if $[\hat{\kappa}_1] \!=\! [\hat{\kappa}_2]$, with $[\hat{\kappa}_1], [\hat{\kappa}_2] \!\succeq\! 0$ and $\textnormal{tr}([\hat{\kappa}_1]) \!\geq$\\ \noindent $\textnormal{tr}([\hat{\kappa}_2])$, is also possible.  Demonstrating this claim relies on an application of the inequalities in Proposition A.9.

For matrices that are not positive (semi-)definite, there is no guarantee that the matrix-based $\alpha$-cross-entropies are non-negative.  Fortunately, this cannot occur in practice, as the cross-entropy arguments are Gram matrices.
\begin{itemize}
\item[] \-\hspace{0.5cm}{\small{\sf{\textbf{Proposition A.2.}}}} If $[\hat{\kappa}_1] \!-\! [\hat{\kappa}_2] \!\succeq\! 0$, then we have that $C_\alpha([\hat{\kappa}_1]\|[\hat{\kappa}_2]) \!\geq\! 0$.  If, however, $[\hat{\kappa}_1] \!-\! [\hat{\kappa}_2] \!\preceq\! 0$, then $C_\alpha([\hat{\kappa}_1]\|[\hat{\kappa}_2]) \!\leq\! 0$.  The same results hold in the mirrored and tripartite cases.
\end{itemize}

{\small{\sf{\textbf{Invariance.}}}} We can also demonstrate that the matrix-based $\alpha$-cross-entropies are conserved under unitary transformations; they hence preserve inner products.
\begin{itemize}
\item[] \-\hspace{0.5cm}{\small{\sf{\textbf{Proposition A.3.}}}} We have $C_\alpha([\hat{\kappa}_1]\|[\hat{\kappa}_2]) \!=\! C_\alpha(U[\hat{\kappa}_1]U^*\|U[\hat{\kappa}_2]U^*)$ for any unitary operator $U \!\!\in\! \mathbb{R}^{n \times n}$.  The same results hold in the mirrored and tripartite cases.
\end{itemize}
The non-mirrored $\alpha$-cross-entropies also satisfy $C_\alpha([\hat{\kappa}_1]\|[\hat{\kappa}_2]) \!=\! C_\alpha(V[\hat{\kappa}_1]V^*\|V[\hat{\kappa}_2]V^*)$ for any isometry matrix\\ \noindent $V \!\!\in\! \mathbb{R}^{n \times n}$; again the same result holds in the mirrored and tripartite cases.  Isometry invariance naturally implies\vspace{-0.015cm}\\ \noindent unitary invariance.  Without isometry invariance, uniformly translating the arguments, for instance, would alter the cross-entropies, thereby complicating ensuing inferences.  

Scaling the arguments can induce a change in the divergence, as we would expect. 
\begin{itemize}
\item[] \-\hspace{0.5cm}{\small{\sf{\textbf{Proposition A.4.}}}} For $\rho_1,\rho_2 \!\in\! \mathbb{R}_+$, and $\alpha \!\in\! \mathbb{R}_+\!\backslash \{1\}$, $C_\alpha(\rho_1[\hat{\kappa}_1]\|\rho_2[\hat{\kappa}_2]) \!=\! C_\alpha([\hat{\kappa}_1]\|[\hat{\kappa}_2]) \!+\! \textnormal{log}(\rho_1/\rho_2)$. The\\ \noindent  same result hold in the mirrored case.  It similarly holds in the tripartite case for both square and non-square univariate Gram matrices.
\end{itemize}

{\small{\sf{\textbf{Additivity.}}}} Additivity for tensor products is also well respected.
\begin{itemize}
\item[] \-\hspace{0.5cm}{\small{\sf{\textbf{Proposition A.5.}}}} Let $[\hat{\kappa}_3],[\hat{\kappa}_4] \!\in\! \mathbb{R}_{0,+}^{n \times n}$, be normalized, univariate Gram matrices.  We have that\\ \noindent $C_\alpha'([\hat{\kappa}_1] \!\otimes [\hat{\kappa}_3] \|[\hat{\kappa}_2] \otimes [\hat{\kappa}_4]) \!=\! C_\alpha([\hat{\kappa}_1] \| [\hat{\kappa}_2]) \!+\! C_\alpha([\hat{\kappa}_3] \| [\hat{\kappa}_4])$ with the additional constraint that $\textnormal{supp}([\hat{\kappa}_4]) \!\subseteq\! \textnormal{supp}([\hat{\kappa}_3])$.  The same result holds in the mirrored case.  It does too for the tripartite case for both square and non-square univariate Gram matrices.
\end{itemize}
This result can be extended to both finite and countable additivity.  

Tensor summation of matrix arguments yields a generalized mean.
\begin{itemize}
\item[] \-\hspace{0.5cm}{\small{\sf{\textbf{Proposition A.6.}}}} Let $[\hat{\kappa}_3],[\hat{\kappa}_4] \!\in\! \mathbb{R}_{0,+}^{n \times n}$, be univariate, normalized Gram matrices.  There exists a continuous, strictly monotonic function $g$ where, for $\alpha \!\in\! \mathbb{R}_+ \!\backslash \{1\}$,
\begin{equation*}
C_\alpha([\hat{\kappa}_1] \oplus [\hat{\kappa}_3] \| [\hat{\kappa}_2] \oplus [\hat{\kappa}_4]) \!=\! g^{-1}\Bigg(\frac{\textnormal{tr}([\hat{\kappa}_1])}{\textnormal{tr}([\hat{\kappa}_1] \!+\! [\hat{\kappa}_2])}\Bigg(\!g(C_\alpha([\hat{\kappa}_1]\|[\hat{\kappa}_2])) + g(C_\alpha([\hat{\kappa}_3]\|[\hat{\kappa}_4]))\!\Bigg)\!\!\Bigg),
\end{equation*}
with the additional constraint that $\textnormal{supp}([\hat{\kappa}_4]) \!\subseteq\! \textnormal{supp}([\hat{\kappa}_3])$.  The same result holds in the mirrored case.  It does too for the tripartite case for both square and non-square matrices. 
\end{itemize}

{\small{\sf{\textbf{Continuity.}}}} The matrix-based $\alpha$-cross-entropies are additionally continuous in most instances.
\begin{itemize}
\item[] \-\hspace{0.5cm}{\small{\sf{\textbf{Proposition A.7.}}}} We have that, for $[\hat{\kappa}_1] \!\neq\! [0]_{n \times n}$, $C_\alpha([\hat{\kappa}_1]\|[\hat{\kappa}_2])$ is continuous in $[\hat{\kappa}_1],[\hat{\kappa}_2]$ for $\alpha \!\in\! \mathbb{R}_{+} \!\backslash \{1\}$.\\ \noindent The same result holds for the tripartite case.  It does too for the mirrored case, except that $\alpha \!\in\! \mathbb{R}_{\frac{1}{2},+} \!\backslash \{1\}$.
\end{itemize}
Observe that the ordering of the operators is relevant for continuity for the bipartite measures.  Point discontinuities can arise due to a division by zero if the operators are swapped and hence one of them does not satisfy the set-ordering property.

As a consequence of continuity everywhere, the derivatives of the matrix-based $\alpha$-cross-entropies can be shown to exist everywhere, for appropriate parameter values.

{\small{\sf{\textbf{Convexity\! and\!\, Concavity.}}}} We can also quantify the curvature of the $\alpha$-cross-entropies.
\begin{itemize}
\item[] \-\hspace{0.5cm}{\small{\sf{\textbf{Proposition A.8.}}}} Assume $[\hat{\kappa}_1] \!\neq\! [\hat{\kappa}_2]$, we have that:
\begin{itemize}
   \item[] \-\hspace{0.5cm}(i) The function $([\hat{\kappa}_1],[\hat{\kappa}_2]) \!\mapsto\! \textnormal{tr}([\hat{\kappa}_1]^{\alpha} [\hat{\kappa}_2]^{1 -\alpha})$ is jointly concave for $\alpha \!\in\! \mathbb{R}_{0,1}$ and jointly convex for\\ \noindent $\alpha \!\in\! \mathbb{R}_{1,2}\backslash \{1\}$.  This implies that the non-mirrored, matrix-based R\'{e}nyi's cross-entropy is both jointly\vspace{-0.025cm}\\ \noindent concave and convex, respectively, over the same parameter-value ranges.
   \item[] \-\hspace{0.5cm}(ii) The function $([\hat{\kappa}_1],[\hat{\kappa}_2]) \!\mapsto\! (\textnormal{tr}([\hat{\kappa}_2]^{(1-\alpha)/2\alpha} [\hat{\kappa}_1] [\hat{\kappa}_2]^{(1-\alpha)/2\alpha}))^\alpha$ is jointly concave for $\alpha \!\in\! \mathbb{R}_{\frac{1}{2},1}\hspace{-0.01cm}\backslash \{1\}$\vspace{-0.015cm}\\ \noindent and jointly convex for $\alpha \!\in\! \mathbb{R}_{1,+}\!\backslash \{1\}$.  This implies that the mirrored, matrix-based R\'{e}nyi's cross-entropy is\vspace{-0.015cm}\\ \noindent both jointly concave and convex, respectively, over the same parameter-value ranges.
   \item[] \-\hspace{0.5cm}(iii) The function $([\hat{\kappa}_1],[\hat{\kappa}_{1,2}],[\hat{\kappa}_2]) \!\mapsto\! \mathbb{E}_{(s,s') \sim p_\mathcal{S}}[[\hat{\kappa}_1]] \!+\! \mathbb{E}_{(a,a') \sim p_\mathcal{A}}[[\hat{\kappa}_2]] \!-\! 2\mathbb{E}_{(s,a) \sim \rho_{\mathcal{S} \times \mathcal{A}}}[[\hat{\kappa}_{1,2}]]$ is jointly\\ \noindent convex.  As well, the function $[\hat{\kappa}_1] \!\mapsto\! \textnormal{tr}([\hat{\kappa}_1]^\alpha)$ is Schur-concave for $\alpha \!\in\! \mathbb{R}_{+} \!\backslash \{1\}$.  The tripartite, matrix-based cross-entropy is thus jointly convex for $\alpha \!\in\! \mathbb{R}_{+} \!\backslash \{1\}$.
\end{itemize}
\end{itemize}

{\small{\sf{\textbf{Data-Processing Inequality.}}}} We also obtain a data-processing inequality for divergences.  This inequality states that, for random variables forming a Markov chain, processing the random variables cannot increase the matrix-based R\'{e}nyi's cross-entropies.  That is, any transformation within a particular class cannot increase the distinguishability of the samples.
\begin{itemize}
\item[] \-\hspace{0.5cm}{\small{\sf{\textbf{Proposition A.9.}}}} Let $\theta$ be a completely positive, trace-preserving map between two Hilbert spaces.  We have that $C_\alpha([\hat{\kappa}_1]\|[\hat{\kappa}_2]) \!\geq\! C_\alpha(\theta([\hat{\kappa}_1])\|\theta([\hat{\kappa}_2]))$ for $\alpha \!\in\! \mathbb{R}_{0,2} \backslash \{0\}$.  Likewise, for the mirrored case, we have that\\ \noindent $C_\alpha'([\hat{\kappa}_1]\|[\hat{\kappa}_2]) \!\geq\! C_\alpha'(\theta([\hat{\kappa}_1])\|\theta([\hat{\kappa}_2]))$ for $\alpha \!\in\! \mathbb{R}_+\!\backslash [0,\frac{1}{2})$.
\end{itemize}
For the mirrored case, an additional parameter $\beta \!\in\! \mathbb{R}_+$ can be included in the cross-entropy expression to increase the coverage of the data-processing inequality to $\alpha \!\in\! \mathbb{R}_{0,+}\!\backslash \{1\}$. 
\begin{itemize}
\item[] \-\hspace{0.5cm}{\small{\sf{\textbf{Proposition A.10.}}}} Let
\begin{equation*}
C'_{\alpha,\beta}([\hat{\kappa}_1]\|[\hat{\kappa}_2]) \!=\! \frac{1}{\alpha \!-\! 1}\textnormal{log}\Bigg(\!\textnormal{tr}\Bigg(\![\hat{\kappa}_2]^{\frac{1-\alpha}{2\beta}} [\hat{\kappa}_1]^{\frac{\alpha}{\beta}} [\hat{\kappa}_2]^{\frac{1-\alpha}{2\beta}}\!\Bigg)^{\!\!\beta}\,\Bigg) \!-\! \frac{1}{\alpha \!-\! 1}\textnormal{log}\Bigg(\!\textnormal{tr}(\kappa_1)\!\Bigg).
\end{equation*}
We have that $C'_{\alpha,\beta}([\hat{\kappa}_1]\|[\hat{\kappa}_2]) \!\geq\! C'_{\alpha,\beta}(\theta([\hat{\kappa}_1])\|\theta([\hat{\kappa}_2]))$ for (i) $\alpha \!\in\! \mathbb{R}_{0,1}\backslash \{0,1\}$ with $\beta \!\geq\! \textnormal{max}(\alpha,1\!-\!\alpha)$,\\ \noindent (ii) $\alpha \!\in\! \mathbb{R}_{1,2} \backslash \{1\}$ with $\beta \!=\! 1$ or $\beta \!=\! \alpha/2$, and (iii) $\alpha \!\in\! \mathbb{R}_{1,+} \!\backslash \{1\}$ with $\beta \!=\! \alpha$.
\end{itemize}
Such a functional also satisfies many of the other properties listed in this appendix and thus can be used in lieu of the mirrored matrix-based cross-entropy.

{\small{\sf{\textbf{Monotonicity.}}}} The matrix-based cross-entropies are monotonically increasing with respect to $\alpha$, just as with the classical case.  Such a property is apparent from the general-mean property.  That is, as $\alpha$ rises, increasing preference is given to contributions with a high log-likelihood ratio.
\begin{itemize}
\item[] \-\hspace{0.5cm}{\small{\sf{\textbf{Proposition A.11.}}}} Assume that $[\hat{\kappa}_1] \!\neq\! [0]_{n \times n}$, we have, for $\alpha \!\in\! \mathbb{R}_{+} \!\backslash \{1\}$, that $C_\alpha([\hat{\kappa}_1]\|[\hat{\kappa}_2])$, $C_\alpha'([\hat{\kappa}_1]\|[\hat{\kappa}_2])$, and $C_\alpha''([\hat{\kappa}_1]\|[\hat{\kappa}_{1,2}]\|[\hat{\kappa}_2])$ are monotonically increasing in $\alpha$.
\end{itemize}
Monotonicity occurs regardless of if the operators are commutative or non-commutative.  The latter is important for practical situations, as the Gram matrices will not always be simultaneously diagonalizable.

{\small{\sf{\textbf{Induced Entropies.}}}} The matrix-based cross-entropies can be seen as a parent quantity to corresponding matrix-based R\'{e}nyi's entropies, $S_\alpha([\hat{\kappa}_1]) \!=\! \textnormal{log}(\textnormal{tr}([\hat{\kappa}_1])^\alpha)/(\alpha \!-\! 1)$, that our lab introduced \cite{GiraldoLGS-jour2014a}.  Our matrix-based R\'{e}nyi's\\ \noindent entropies were proven to satisfy standard properties of classical R\'{e}nyi $\alpha$-entropies.
\begin{itemize}
\item[] \-\hspace{0.5cm}{\small{\sf{\textbf{Proposition A.12.}}}} For $\alpha \!\in\! \mathbb{R}_{+} \!\backslash \{1\}$, we have that $S_\alpha([\hat{\kappa}_1])$ is equal to $C_{\alpha}([\hat{\kappa}_1]\|\textnormal{id}_{n \times n})$ and  $C_{\alpha,1}'([\hat{\kappa}_1]\|\textnormal{id}_{n \times n})$, where $\textnormal{id}_{n \times n}$ is the identity matrix.  Moreover, it is equal to $C_{\alpha}''([\hat{\kappa}_1]\|[0]_{n \times n}\|[\hat{\kappa}_1])$.
\end{itemize}
This result does not extend to the mirrored cross-entropy for the case of a single-parameter.

We can view either divergence as a distance measure.  The corresponding matrix-based R\'{e}nyi's entropy can thus be understood as $S_\alpha([\hat{\kappa}_1]) \!=\! \textnormal{log}(n) \!-\! C_\alpha([\hat{\kappa}_1]\|\textnormal{id}_{n \times n}/n)$, that is, the difference between the maximal possible entropy and how far away the Gram matrix is from the normalized identity matrix.  

The set of positive semi-definite matrices, which includes Gram matrices, is closed under the Hadamard product.  This property can be employed to extend the matrix-based divergences from a single-variable entropy to a joint-entropy representation that is given by $S_\alpha([\hat{\kappa}_1],[\hat{\kappa}_2]) \!=\! S_\alpha([\hat{\kappa}_1] \circ [\hat{\kappa}_2]/\textnormal{tr}([\hat{\kappa}_1] \circ [\hat{\kappa}_2]))$.
\begin{itemize}
\item[] \-\hspace{0.5cm}{\small{\sf{\textbf{Proposition A.13.}}}} Let $\alpha \!\in\! \mathbb{R}_{+} \!\backslash \{1\}$. We have that $S_\alpha([\hat{\kappa}_1],[\hat{\kappa}_2]) \!=\! \textnormal{log}(n) \!-\! C_\alpha([\hat{\kappa}_1] \circ [\hat{\kappa}_2] \| [\hat{\kappa}_2] \textnormal{id}_{n \times n}) \hspace{0.015cm}+$\\ \noindent $C_{\alpha}([\hat{\kappa}_1]\|\textnormal{id}_{n \times n})$.  Similar results hold in the two-parameter non-mirrored case.  Likewise, only for univariate Gram matrices that are square, $S_\alpha([\hat{\kappa}_1],[\hat{\kappa}_2]) \!=\! C_{\alpha}''([\hat{\kappa}_1] \circ [\hat{\kappa}_2]/\textnormal{tr}([\hat{\kappa}_1] \circ [\hat{\kappa}_1])\|[0]_{n \times n}\|[\hat{\kappa}_1] \circ [\hat{\kappa}_2]/\textnormal{tr}([\hat{\kappa}_1] \circ [\hat{\kappa}_1]))$.
\end{itemize}
We can interpret the Hadamard product as computing a product kernel.  It is specifying a measure of entropy of a random element defined by a pair of random variables.

We can also use this divergence to specify a matrix-based conditional entropy.  There is, however, no general consensus for a definition of a R\'{e}nyi conditional $\alpha$-entropy.  Here, we take inspiration from Shannon's definition and consider a version, $S_{\alpha}(\kappa_1|\kappa_2) \!=\! S_{\alpha}(\kappa_1,\kappa_2) \!-\! S_{\alpha}(\kappa_2)$, that specifies the uncertainty about one random variable after observing another and taking into account the joint entropy.
\begin{itemize}
\item[] \-\hspace{0.5cm}{\small{\sf{\textbf{Proposition A.14.}}}} Let $\alpha \!\in\! \mathbb{R}_{+} \!\backslash \{1\}$. We have that $S_\alpha([\hat{\kappa}_1]|[\hat{\kappa}_2]) \!=\! \textnormal{log}(n) \!-\! C_\alpha([\hat{\kappa}_1] \circ [\hat{\kappa}_2] \| [\hat{\kappa}_2] \textnormal{id}_{n \times n})\hspace{0.02cm}+$\\ \noindent $C_{\alpha}([\hat{\kappa}_1]\|\textnormal{id}_{n \times n}) \!-\! C_{\alpha}([\hat{\kappa}_2]\|\textnormal{id}_{n \times n})$.  Similar results hold in the two-parameter mirrored and tripartite cases.\\ \noindent For the tripartite measure, the univariate Gram matrices must be square.
\end{itemize}
Likewise, we can define a matrix-based R\'{e}nyi's $\alpha$-mutual-information, as in \cite{GiraldoLGS-jour2014a}, using the notions of the matrix-based R\'{e}nyi's $\alpha$-marginal and $\alpha$-conditional entropies.

Both the matrix-based joint and conditional $\alpha$-entropies derived from our matrix-based $\alpha$-cross-entropies can be extended to arbitrary numbers of matrices \cite{YuS-jour2020a}.  The matrix-based R\'{e}nyi's $\alpha$-mutual-information can too, as a consequence.

\phantomsection\label{secA.2}
\subsection*{\small{\sf{\textbf{A.2.$\;\;\;$Matrix-based R\'{e}nyi's $\alpha$-Divergence Bounds}}}}

It can be shown that the unit-$\alpha$ case of the bipartite, mirrored $\alpha$-cross-entropy is related to a Umegaki relative entropy with an additional weighting factor.

\begin{itemize}
\item[] \-\hspace{0.5cm}{\small{\sf{\textbf{Proposition A.15.}}}} We have that
\begin{equation*}
\textnormal{lim}_{\alpha \to 1}C'_\alpha([\hat{\kappa}_1]\|[\hat{\kappa}_2]) \!=\! \frac{1}{\textnormal{tr}([\hat{\kappa}_1])}\textnormal{tr}\Bigg(\![\hat{\kappa}_1](\textnormal{log}([\hat{\kappa}_1]) \!-\! \textnormal{log}([\hat{\kappa}_2]))\!\Bigg),
\end{equation*}
and $C'_1([\hat{\kappa}_1]\|[\hat{\kappa}_2]) \!=\! \infty$ otherwise.
\end{itemize}
This link permits establishing an upper bound in terms of the trace distance.  We can also obtain a tighter bound via spectral properties of the univariate Gram matrices.

\begin{itemize}
\item[] \-\hspace{0.5cm}{\small{\sf{\textbf{Proposition A.16.}}}} We have that
\begin{itemize}
   \item[] \-\hspace{0.5cm}(i) Trace-distance bound: Let $\lambda_{\kappa_1} \!\!\in\! \mathbb{R}_{+}$ and $\lambda_{\kappa_2} \!\!\in\! \mathbb{R}_{+}$ be the minimal non-zero eigenvalues of the kernel matrices $[\hat{\kappa}_1]$ and $[\hat{\kappa}_2]$, respectively.  Let $\alpha \!\to\! 1$, then
\begin{equation*}
\textnormal{tr}\Bigg(\![\hat{\kappa}_1](\textnormal{log}([\hat{\kappa}_1]) \!-\! \textnormal{log}([\hat{\kappa}_2]))\!\Bigg) \!\leq\! \Bigg(\!(\lambda_{\kappa_2} \!+\! \omega_{\kappa_1,\kappa_2}/2)\textnormal{log}(1 \!+\! \omega_{\kappa_1,\kappa_2}/2\lambda_{\kappa_2}) \!-\! \lambda_{\kappa_1}\textnormal{log}(1 \!+\! \omega_{\kappa_1,\kappa_2}/2\lambda_{\kappa_1})\!\Bigg),
\end{equation*}
where $\omega_{\kappa_1,\kappa_2} \!\in\! \mathbb{R}_{0,+}$ is the $\ell_1$-matrix-distance between $[\hat{\kappa}_1]$ and $[\hat{\kappa}_2]$.
   \item[] \-\hspace{0.5cm}(ii) Tighter trace-distance bound: Let $\lambda_{\kappa_1} \!\!\in\! \mathbb{R}_{+}$ and $\lambda_{\kappa_2} \!\!\in\! \mathbb{R}_{+}$ be the minimal non-zero eigenvalues of the Gram matrices $[\hat{\kappa}_1]$ and $[\hat{\kappa}_2]$, respectively.  Let $\lambda_{\kappa_1}^+ \!\!\in\! \mathbb{R}_+$ be the largest eigenvalue of $[\hat{\kappa}_1]$.  Let $\alpha \!\to\! 1$, then
\begin{equation*}
\textnormal{tr}\Bigg(\![\hat{\kappa}_1](\textnormal{log}([\hat{\kappa}_1]) \!-\! \textnormal{log}([\hat{\kappa}_2]))\!\Bigg) \!\leq\! \Bigg(\frac{\omega_{\kappa_1,\kappa_2}\lambda_{\kappa_1}^+}{\lambda_{\kappa_1} \!-\! \lambda_{\kappa_2}}(\textnormal{log}(\lambda_{\kappa_1}) \!-\! \textnormal{log}(\lambda_{\kappa_2})) \Bigg),
\end{equation*}
which is non-strictly bounded above by $\lambda_{\kappa_1}^+\omega_{\kappa_1,\kappa_2}/\textnormal{min}(\lambda_{\kappa_1},\lambda_{\kappa_2})$.
\end{itemize}
\end{itemize}
These inequalities provide easy tests of the maximal cross-entropies that will be encountered when coupled with spectral-radius bounds.   

Both bipartite matrix-based R\'{e}nyi's cross-entropies can be related to each other.  From the Araki-Lieb-Thirring inequality, we have that $\textnormal{tr}([\hat{\kappa}_1]^\alpha[\hat{\kappa}_2]^\alpha[\hat{\kappa}_1]^\alpha) \!\leq\! \textnormal{tr}([\hat{\kappa}_1][\hat{\kappa}_2][\hat{\kappa}_1])^\alpha$ and $\textnormal{tr}([\hat{\kappa}_1][\hat{\kappa}_2][\hat{\kappa}_1])^\alpha \!\leq\! (\|[\hat{\kappa}_2]\|^\alpha\textnormal{tr}([\hat{\kappa}_1]^{2\alpha}))^{1-\alpha}$, for $\alpha \!\in\! \mathbb{R}_{0,1} \!\backslash \{0,1\}$.  These inequalities hold in the opposite direction for $\alpha \!\in\! \mathbb{R}_{1,+} \!\backslash \{1\}$.   Thus,
\begin{equation*}
\textnormal{tr}([\hat{\kappa}_1]^\alpha [\hat{\kappa}_2]^{1-\alpha}) \leq \|[\hat{\kappa}_2]\|^{(1-\alpha)^2}\Bigg(\!(\textnormal{tr}([\hat{\kappa}_1]^\alpha))^{1 - \alpha}(\textnormal{tr}([\hat{\kappa}_1]^\alpha[\hat{\kappa}_2]^{1-\alpha}))^{\alpha}\!\Bigg),
\end{equation*}
which gives rise to the following claim.

\begin{itemize}
\item[] \-\hspace{0.5cm}{\small{\sf{\textbf{Proposition A.17.}}}} For $\alpha \!\in\! \mathbb{R}_{+} \!\backslash \{1\}$, $C_\alpha([\hat{\kappa}_1]\|[\hat{\kappa}_2]) \!\geq\! C_\alpha'([\hat{\kappa}_1]\|[\hat{\kappa}_2])$.
\end{itemize}
From this, we can see that the magnitudes from the mirrored R\'{e}nyi's $\alpha$-cross-entropy may be more conservative than those from the non-mirrored version.  

\clearpage\newpage

\phantomsection\label{secB}
\subsection*{\small{\sf{\textbf{Appendix B}}}}

\vspace{0.15cm}

\begin{center}
\begin{tabular}{| p{0.85in} | p{3.0in} |}
\hline
{\sc Symbol} & {\sc Description} \\
\hline
\hline

$\mathbb{E}$ & Expected value\\

$\mathbb{R}$ & Real numbers\\

$\mathbb{R}_+$ & Positive real numbers\\

$\mathbb{R}_{0,+}$ & Non-negative real numbers\\

$\mathbb{R}_{a,b}$ & Real numbers in the range $[a,b]$\\

$\mathbb{R}_{a,b}\backslash \{c\}$ & Real numbers in the range $[a,b]$ excluding the set $c$\\

$\mathbb{N}_{a,b}$ & Natural numbers in the range $[a,b]$\\

$\mathbb{L}_2$ & Square-integrable functions\\

$\mathcal{S},\mathcal{A}$ & Topological sample spaces\\

$\mathcal{M}_+^1(\mathcal{S})$ & Space of probability measures on $\mathcal{S}$\\

$\mathcal{M}_+^1(\mathcal{S} \!\times\! \mathcal{A})$ & Space of probability measures on $\mathcal{S} \!\times\! \mathcal{A}$\\

$\mathcal{H}_{\kappa'}$ & Reproducing-kernel Hilbert space for $\kappa'$\\

$\mathcal{H}_{\kappa_1'},\mathcal{H}_{\kappa_2'}$ & Reproducing-kernel Hilbert spaces for $\kappa_1',\kappa_2'$\\

$\mathcal{H}_{\kappa_1'} \!\otimes\! \mathcal{H}_{\kappa_2'}$ & Tensor-product reproducing-kernel Hilbert space\\

$\kappa',\kappa_1',\kappa_2'$ & Reproducing kernels\\

$p_\mathcal{S},q_{\mathcal{S}},p_\mathcal{A}$ & Marginal probability measures\\

$\rho_{\mathcal{S} \times \mathcal{A}}$ & Joint probability measure\\

$\varphi,\psi$ & Measurable functions\\

$\varphi^\otimes$ & Measurable function for tensor-product spaces\\

$\mu_{p_\mathcal{S}}$ & Univariate mean element\\

$\mu_{\rho_{\mathcal{S} \times \mathcal{A}}}$ & Bivariate mean element\\

$s,s',a,a'$ & Topological-space elements\\

$f,g$ & Hilbert-space functions\\

$h,r$ & Hilbert-space dimensionality\\

$\eta_j,\vartheta_k$ & Basis component for $\mathcal{H}_{\kappa'_1},\mathcal{H}_{\kappa'_2}$\\

$\pi_q$ & Basis component for $\mathcal{H}_{\kappa'_1} \!\otimes\! \mathcal{H}_{\kappa'_2}$\\

$\kappa,\kappa_1,\kappa_2$ & Univariate Gram operators\\

$\hat{\kappa},\hat{\kappa}_1,\hat{\kappa}_2$ & Approximate univariate Gram operators\\

$[\hat{\kappa}],[\hat{\kappa}_1],[\hat{\kappa}_2]$ & Univariate Gram matrices\\

$\gamma$ & Bivariate Gram operator\\

$\hat{\gamma}$ & Approximate bivariate Gram operator\\

$[\hat{\gamma}]$ & Bivariate Gram matrix\\

$\lambda$ & Joint, bivariate Gram operator\\

$\hat{\lambda}$ & Approximate joint, bivariate Gram operator\\

$[\hat{\lambda}],[\hat{\kappa}_{1,2}]$ & Joint, bivariate Gram matrices\\

$\mathcal{K}$ & Multilinear form specified by $\kappa$\\

$\mathcal{G}$ & Multilinear form specified by $\gamma$\\

$\mathcal{L}$ & Multilinear form specified by $\lambda$\\

$\tau_i$ & Gram operator eigenvalue\\

$\beta_i$ & Gram operator eigenvector\\

$\theta$ & Completely positive, trace-preserving mapping\\

$n,m$ & Number of samples\\

$i,j,k,q$ & Indexes\\

$c_1,c_2,c_3,c_4,c_5$ & Real-valued constants\\

$\textnormal{supp}(\kappa)$ & Support of $\kappa$\\

$\textnormal{tr}(\cdot)$ & Trace\\

$\langle \cdot,\cdot \rangle_{\mathcal{H}_{\kappa'}}$ & Inner product for $\mathcal{H}_{\kappa'}$\\

$\| \cdot \|_{\mathcal{H}_{\kappa'}}$ & Norm for $\mathcal{H}_{\kappa'}$\\

$H_\alpha(\cdot\|\cdot)$ & R\'{e}nyi's $\alpha$-order cross-entropy\\

$C_\alpha(\cdot\|\cdot)$ & Bipartite, non-mirrored R\'{e}nyi's $\alpha$-order cross-entropy\\

$C_\alpha'(\cdot\|\cdot)$ & Bipartite, mirrored R\'{e}nyi's $\alpha$-order cross-entropy\\

$C_{\alpha,\beta}'(\cdot\|\cdot)$ & Bipartite, mirrored R\'{e}nyi's $\alpha$-$\beta$-order cross-entropy\\

$C_\alpha''(\cdot\|\cdot\|\cdot)$ & Tripartite R\'{e}nyi's $\alpha$-order cross-entropy\vspace{0.045cm}\\

\hline

\end{tabular}
\end{center}

\end{document}